\documentclass[final,5p,authoryear,times]{elsarticle} 
\pdfoutput=1
\usepackage{natbib}
\usepackage{caption}
\usepackage{amsmath,xparse,mleftright}
\usepackage{lineno}
\usepackage{subcaption}
\newcommand{\rthis}[1]{\textcolor{black}{#1}}
\usepackage[plainpages=false, colorlinks=true, anchorcolor=blue, linkcolor=blue, citecolor=blue, bookmarks=false]{hyperref}
\modulolinenumbers[5]
\journal{Astronomy \& Computing}
\author[1]{S.R. Bhavanam}\ead{ee19resch01008@iith.ac.in}
\author[1]{Sumohana S. Channappayya}
\ead{sumohana@ee.iith.ac.in} 
\author[2]{P.K. Srijith}\ead{srijith@cse.iith.ac.in}
\author[3]{Shantanu Desai}\ead{shantanud@phy.iith.ac.in}
\address[1]{Department of Electrical Engineering, IIT Hyderabad, Kandi,  Telangana-502285, India}
\address[2]{Department of Computer Science and Engineering, IIT Hyderabad, Kandi,  Telangana-502285, India}
\address[3]{Department of Physics, IIT Hyderabad, Kandi,  Telangana-502285, India}

\begin{document}

\begin{frontmatter}


\title{Cosmic Ray Rejection with Attention Augmented Deep Learning}

\begin{abstract}

Cosmic Ray (CR) hits are the major contaminants in astronomical imaging and spectroscopic observations involving solid-state detectors. Correctly identifying and masking them is a crucial part of the  image processing pipeline, since it may otherwise lead to spurious detections. For this purpose, we have developed and tested a novel Deep Learning based framework for the automatic detection of CR hits from astronomical imaging data from two different imagers: Dark Energy Camera (DECam) and Las Cumbres Observatory Global Telescope (LCOGT). We considered two baseline models namely deepCR and Cosmic-CoNN, which are the current state-of-the-art learning based algorithms that were trained using Hubble Space Telescope (HST) ACS/WFC and LCOGT Network images respectively. We have experimented with the idea of augmenting the baseline models using Attention Gates (AGs) to improve the CR detection performance. We have trained our models on DECam data and demonstrate a consistent \rthis{marginal} improvement by adding AGs in True Positive Rate (TPR) at 0.01\% False Positive Rate (FPR) and Precision at 95\% TPR over the aforementioned baseline models for the DECam dataset. \rthis{We demonstrate that the proposed AG augmented models provide significant gain in TPR at 0.01\% FPR when tested on previously unseen LCO test data having images from three distinct telescope classes.} Furthermore, we demonstrate that the proposed baseline models with and without attention augmentation outperform state-of-the-art models such as Astro-SCRAPPY, Maximask (that is trained natively on DECam data) and pre-trained ground-based Cosmic-CoNN. \rthis{This study demonstrates that the AG module augmentation enables us to get a better deepCR and Cosmic-CoNN models and to improve their generalization capability on unseen data.} 



\end{abstract}


\end{frontmatter}


\section{Introduction}

In astronomical imaging involving solid-state detectors such as Charge-Coupled Devices (CCDs), Cosmic Ray (CR) hits are the primary source of charged particle contamination. The cosmic detritus is caused due to the energy transfer from the CR particles to the CCD electrons in the valence band~\citep{popowicz2016efficiency}. As the particles enter from different directions and move differently within the internal detector structure, the CR hits impact the observations in numerous patterns, ranging from dots affecting one or two pixels to long wandering tracks, known as worms. The detector thickness and incidence angle also impact the sensitivity of the imager to CR hits. The CRs in observational data often masquerade as astronomical sources, contaminating observations and causing false detection, thus degrading the sensitivity to detect faint or transient sources. Hence, the detection and flagging of CR hits should be done before further analysis to ensure top quality data is used for any science analysis. However, a visual inspection is usually impossible due to the sheer volume of data generated by modern wide-field photometric or spectroscopic surveys. As a result, developing fully automated techniques to separate contaminants such as CR hits from the actual astrophysical sources in modern astronomical survey pipelines is a critical challenge.


The CR hits in astronomical observations are transient, implying that the likelihood of a specific pixel being affected by CRs when the same sky region is captured multiple times is relatively low. As a result, getting multiple exposures of the same sky region is one straightforward approach. This is because CRs will  contaminate image pixels  in only one exposure, and the remaining exposures help to detect the CR hits. This is achieved by comparing each pixel's deviation from the mean or median value from the stack of aligned exposures~\citep{windhorst1994removing, freudling1995image, fruchter2002drizzle, Gruen14, desai2016detection}. However, getting multiple exposures are not always possible for various reasons, including constraints on the observing strategy and the targeting of moving or transient objects. When the exposures are displaced by a non-integer number of pixels about each other, or if the seeing varies significantly between the images~\citep{rhoads2000cosmic}, CR rejection using only single exposures is necessary.


Numerous methods for CR rejection from single exposure imaging  and spectroscopic data have been proposed in the literature. Most are based on the fact that the CR hits appear sharper and brighter than the actual astronomical sources present in the image. This is because the CR hits are not blurred by telescope optics or the atmosphere, unlike astronomical sources. Furthermore, they exhibit less symmetrical morphologies than the sources present in the image because they can have any incidence angle. Convolution using an adaptive Point Spread Function (PSF)~\citep{rhoads2000cosmic}, histogram analysis ~\citep{pych2003fast}, fuzzy logic-based algorithm~\citep{shamir2005fuzzy}, and Laplacian edge detection (LACosmic)~\citep{van2001cosmic} are a few techniques that have been developed using these properties. A comparison of some of these algorithms is discussed in ~\citet{farage2005evaluation}. Removal of cosmic rays using single-exposure spectroscopic fiber images is discussed in ~\citet{Bai17}. These techniques, as well as the IRAF tasks like {\tt xzap} by M. Dickinson, frequently involve the tuning of one or more hyperparameters to produce the optimal CR mask per image ~\citep{farage2005evaluation}. Machine-Learning (ML) algorithms in which the classification rules are learned from labelled data such as K-nearest neighbours, multi-layer perceptrons~\citep{murtagh1991detecting}, and decision-tree classifiers \citep{salzberg1995decision} were also applied for CR detection and yielded encouraging results on small HST data sets. However, in terms of generalization, these algorithms fall short of image-filtering techniques such as LACosmic.

In recent years, ML techniques, particularly Deep-Learning (DL) algorithms, have become increasingly ubiquitous throughout  astrophysics~\citep{Ball,baron2019machine} and masking of cosmic ray artifacts is no exception. Recently, deepCR ~\citep{zhang2020deepcr} began employing DL algorithms to automate the CR rejection, demonstrating the potential of deep learning for processing CR hits in the observational data. The main algorithm used here is based on Convolutional Neural Networks (CNNs, hereafter)~\citep{o2015introduction}, which are well-suited for identifying patterns in images. The convolutional kernels in deep learning are learned automatically through back-propagation. This is in contrast to the Laplacian-like edge detection kernel used in LACosmic, which has to be tuned by hand. In the context of deep learning and computer vision, CR detection can be framed as an image semantic segmentation problem, which refers to the process of classifying each pixel in an image as belonging to one of several categories. The deepCR~\citep{zhang2020deepcr} is an UNet~\citep{ronneberger2015u} based model and was initially trained and tested on HST ACS/WFC F606W images of sparse extragalactic fields, globular clusters, and resolved galaxies. It achieved 92.8\%, 95.5\%, and 73.3\% CR detection rates, respectively, with a 0.1\% false detection rate and surpassed the state-of-the-art approach, LACosmic. Similarly, MaxiMask~\citep{paillassa2020maximask} is another CNN-based model inspired by SegNet~\citep{badrinarayanan2017segnet} that attempted to produce a versatile and robust tool for the community at large, avoiding the trap inherent in software specialized to a single or large number of instruments while maintaining high performance. 
Unlike deepCR, which was trained on data from space-based instruments, MaxiMask has been trained using images that encompass a wide range of telescopes (CTIO Blanco, CFHT, Subaru, VST, and others), detector types (modern CCDs and near-infrared cameras), and ground-based observation sites to ensure that the model covers most recent astronomical wide-field surveys. The DL algorithms are also employed to detect other image artifacts, such as spurious reflections (commonly referred to as "ghosts") and the scattering of light off the surfaces of a camera and/or telescope ~\citep{tanoglidis2021deepghostbusters, chang2021machine}.

The network architectures and training procedures adopted in deepCR make it susceptible to variations in the CR rates and morphologies, restricting their application to other unseen instruments. Furthermore, the low CR density in ground-based data makes the training of CNN-based models more complex. The class-imbalance issue~\citep{buda2018systematic} occurs because for  ground-based data, the  ratio of CR affected pixels to non-CR pixels  is usually quite small for nominal exposure times.
For DECam images with 90 seconds exposure, this ratio has been calculated using data from the Science verification phase, and is  approximately equal to 0.027\%~\citep{desai2016detection}. 
This results  in too few CR pixels for spatial convolution, making learning inefficient. We note however that  this ratio could  also increase with  the exposure time.
A novel CNN-based algorithm called Cosmic-CoNN~\citep{xu2021cosmic} was proposed to develop generic CR detection models for all ground and space-based instruments separately, while explicitly addressing the class imbalance problem. They also developed novel strategies for optimizing the neural network for astronomical images' unique spatial and numerical properties. Cosmic-CoNN achieved a CR detection rate of 99.9\% with 0.1\% false detection when trained on LCO Global Telescope Network imaging data. The same model achieved 99.8\% detection on other ground-based unseen data (Gemini GMOS-N/S), demonstrating its generalization capacity. This framework was also evaluated on spatial data from the HST ACS/WFC images similar to deepCR. It achieved 93.4\%, 96.5\%, and 80.3\% CR detection rates on sparse extragalactic fields, globular clusters, and resolved galaxies, respectively at 0.1\% false detection. The Cosmic-CoNN was also trained on LCO spectroscopic data and demonstrated a CR detection rate of 99.8\%.



CNNs have become the de-facto standard for image semantic segmentation due to their high representational power, quick inference, and filter-sharing properties. Fully Convolutional Networks (FCNs)~\citep{long2015fully}, SegNet~\citep{badrinarayanan2017segnet} and U-Net~\citep{ronneberger2015u} are the most commonly employed architectures. Despite their high representational power, these architectures rely on multi-stage cascaded CNNs to accommodate large variations in the shape and size of the target. Cascaded frameworks take a region of interest (ROI) and create dense predictions. On the other hand, this technique results in unnecessary and redundant use of computational resources and model parameters; for example, all models in the cascade extract similar low-level features again. Attention Gates (AGs)~\citep{oktay2018attention} offer a basic but effective solution to this general problem. CNN models with AGs are trained from scratch, similar to FCN models, and AGs automatically learn to focus on target structures without further supervision. These gates create soft region proposals implicitly on the fly during testing and emphasise significant features relevant to a specific task. By suppressing feature activations in irrelevant regions, the AGs increase model sensitivity and accuracy for dense label predictions. We demonstrate the implementation of AGs in standard U-Net architectures, for eg., Attention U-Net~\citep{oktay2018attention}, and apply them for the CR identification problem.

In this work, we test and apply a deep learning framework to photometric data from the DECam  and LCO Global Telescope Network imaging data. MaxiMask employed DECam images in its training process to detect CR hits and outperformed LACosmic. However, this was at the cost of significantly increased processing resources since it tries to detect multiple contaminants at once in a given image. Hence, developing more robust CR-alone detection models for DECam and other imaging data is essential. Our primary goals in this work are described as follows:
\begin{itemize}
    \item Developing a paired data set of CR contaminated images and the corresponding CR masks from the DECam observations.
    \item Training two baseline models, deepCR and Cosmic-CoNN using DECam data. 
    \item Studying the CR detection performance by adding attention gates to the baseline models and allowing them to focus on CR structures of various sizes and shapes.
    \item Evaluating the generalization performance of both the baseline and attention augmented models using the unseen LCO imaging data.
    \item Evaluating the performance of the pre-trained ground based Cosmic-CoNN model using unseen DECam data.
\end{itemize}

The remainder of this paper is organized as follows. In Section~\ref{sec:data}, details on the instrument, a summary of the data collection and data synthesis are presented. The baseline models deepCR and Cosmic-CoNN, as well as the proposed attention gate added variants, are described in Section~\ref{sec:models} along with their neural network architectures. Section~\ref{sec:results} summarizes the performance on DECam and LCO imaging data, which were conducted to demonstrate the potential of the proposed approaches. Finally, conclusions and future work are presented in Section~\ref{sec:conclusion}.
\label{sec:intro}

\section{Data}
This work considers images from the DECam instrument and Las Cumbres Observatory Global Telescope Network. This section gives an overview of the instruments and describes the dataset used for our analysis.
\label{sec:data}

\subsection{DECam}
The DECam is a 570 megapixel imaging camera~\citep{Flaugher}, with a 2.2-degree diameter field of view and a pixel scale of 0.263 arcsecond/pixel. The imaging is done by 62 2K $\times$ 4K CCDs. The thickness of the CCDs is 250 $\mu$ m with 15 $\mu$ m pixels~\citep{Flaugher}. It is mounted on the Victor M. Blanco 4-meter Telescope at the Cerro Tololo Inter-American Observatory (CTIO). In this work, 56 raw science images from four different photometric bands of DECam, namely $g$, $r$, $i$, and $z$, each with 90 sec exposure times, are considered for training and testing the CR detection models. These images  come from the  detrended DECam data taken during the Science Verification phase from Nov 2012 - Feb 2013, which have been processed  using the {\tt CosmoDM} pipeline~\citep{Desai12,cosmoDM}, where the cosmic rays are removed according to the algorithm described in ~\citet{desai2016detection}. The median seeing of these images is equal to 1.1" in $g$-band and about 1" in $r$, $i$, $z$ bands. Although a direct estimate of the sky brightness is not possible for these images, since they are not photometrically calibrated, these images were taken during similar observing months as those  used for Dark Energy Survey data. Therefore, we would expect the sky brightness to be about the same as in DES DR1 dataset, which is $\sim$ 22, 21.1, 19.9, and 18.7 mag/arcsec$^2$ in $g,r,i,z$ bands respectively~\citep{DR1}. The expected object density is approximately equal to 10 galaxies/sq. arcminute~\citep{Sevilla}. 

\subsubsection{The DECam Dataset}
Recent works, including deepCR~\citep{zhang2020deepcr}, MaxiMask~\citep{paillassa2020maximask} and Cosmic-CoNN~\citep{xu2021cosmic} presented the CR detection as a supervised image segmentation problem and demonstrated the superiority of deep learning models on segmentation tasks. However, training deep learning algorithms in a supervised setting requires large amounts of annotated data. In the case of CR detection, annotated data correspond to paired images having CR contamination with the corresponding CR mask. However, getting accurate CR masks from single exposures like DECam data is complex and requires manual effort. Hence, we chose to use the dark frame data (which were downloaded from the  NOAO NVO portal) similar to MaxiMask~\citep{paillassa2020maximask} to generate CR contamination synthetically. The objective for synthetically generating CR contaminated images is that the added CR hits should follow the dynamics of the corresponding image to which they are added so as to closely mimic the real data. The paired dataset reconstructed in this way could also be used to train other CR detection algorithms on DECam data. The key stpdf involved in this procedure are described as follows:
\begin{figure}[!tbp]
  \centering
  \subfloat[Image with CR hits]{\includegraphics[width=0.23\textwidth,
  height=0.24\textwidth, keepaspectratio,]{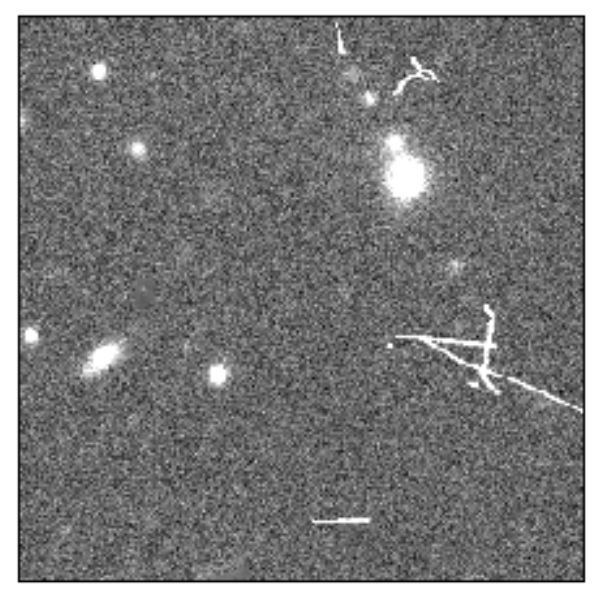}\label{fig:f1}}
  \hfill
  \subfloat[Groundtruth CR mask]{\includegraphics[width=0.23\textwidth,
  height=0.24\textwidth, keepaspectratio,]{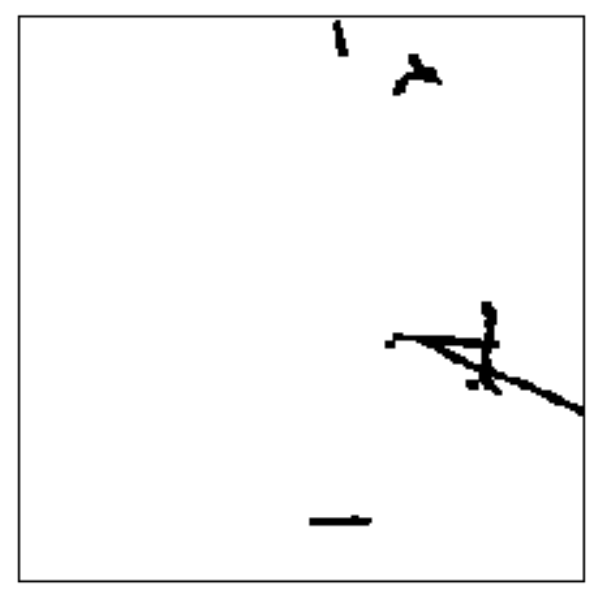}\label{fig:f2}}
  \caption{Synthetic data with CR contamination on DECam-g band image.}
  \label{fig:conta}
\end{figure}

\begin{description}
  \item[$\bullet$ CR \rthis{Identification}:] Using Astro-SCRAPPY~\citep{mccully2019astro}, a multi-core optimized implementation of LACosmic, the vast majority of CR hits are detected and replaced using $5\times5$ median filter sampling, which allows us to get the uncontaminated image without CR artifacts.
  \item[$\bullet$ Dark Frame Extraction:] The dark frames are images obtained when no light is incident on the image sensor, and so the only contributors to the content of undamaged pixels are the offset, dark current, noise and CR hits. We used 42 dark frames with medium to long exposures ($\geq$ 90 seconds) with maximum exposure time of 1200 seconds,  to create a library of real CR hits from DECam. We used the dark frames post the Science Verification phase. Note that we did not consider any cosmic rays near the edges, and also visually scanned the cosmic rays in the dark frames in order to ensure that they do not intersect any bad columns or bad pixels.
  \item[$\bullet$ Dark Data Addition for CR Contamination:] Given a dark data frame $D$, the corresponding CR mask $M$ can be easily generated by applying a simple detection threshold. We empirically choose this threshold to be 3$\sigma_{D}$ above the median value ($m_D$) of $D$. Thus we obtain the CR mask $M$ as follows:
  
    \[
       \forall \text{pixels\ }p,\ M_{p} = 
    \begin{cases}
        1,& \text{if} \ D_{p} > m_{D} + 3\sigma_{D}\\
        0,              & \text{otherwise}
    \end{cases}
    \]
  
  Then the mask $M$ is dilated using a $3\times3$ pixel kernel to create the final mask $M^D$, which is then used as both the ground truth for the classifier, as well as to generate the final "CR contaminated" image $C$ by adding CR pixels from the dark data $D$ with re-scaled values to the uncontaminated image $U$:
  
  \[
       C = U + k_{c}\frac{\sigma_{U}}{\sigma_{D}} \ D \odot M^{D}
    \]
    
  where $\sigma_{D}$ is the standard deviation of the non-CR pixels from the dark data $D$, the estimated standard deviation of the non-source pixels (background) of the uncontaminated image $U$ is denoted by $\sigma_{U}$, $\odot$ is the element-wise multiplication operation and $k_c$ is a scaling factor that needs to be adjusted to make the CR hits look more realistic. We experimented with multiple values for $k_c$ and empirically choose the factor $1/8$. We found through visual scanning that this threshold value of  $k_c$  does a better job in generating realistic CR hits. An example image with added CR hits and the corresponding ground truth CR mask are illustrated in Fig~\ref{fig:conta}. 
  The CR mask dilation helps in two ways: to make the CR hits more realistic when added to the uncontaminated image. The other avoids the under segmentation problem, which made for missing the CR peripherals, especially at the CR edges.
  
  

\end{description}



The images and masks (originally with $4K \times 2K$ pixel resolution) are then divided into image chunks of size $256 \times 256$ before feeding to the neural network to facilitate batch training. 

\subsection{LCO CR data set}
We also considered the Las Cumbres Observatory (LCO) CR data set~\citep{xu_chengyuan_2021_5034763}, that has been made publicly available from~\citet{xu2021cosmic} to validate our CR detection models on previously unseen data. This data set has been constructed by leveraging the data from LCO's {\tt BANZAI} data pipeline~\citep{mccully2018real}. A variety of CCD imagers with different pixel scales,  field of views, and filters were used in each telescope class of LCO's global telescopes network. It consists of over 4,500 scientific images from LCO's 23 globally distributed telescopes. About half of the images have pixel resolutions of  $4K \times 4K$ pixels, whereas the rest have size equal to  $3K \times 2K$ or $2K \times 2K$. The dataset is constructed such that each sequence has at least three consecutive observations with the same exposure time and separated by a few minutes. All the images are with an exposure time of 100 seconds or longer. For our analysis, we have used the test data from LCO CR dataset from three distinct imagers, with specifications specified in Table 1 of ~\citet{xu2021cosmic}. The CCDs used in these imagers are described in ~\citet{brown2013cumbres} and have pixel sizes in the range 9-15 $\mu$ m. The median seeing for these imagers is  equal to 2.07", 2.04", and 1.44" for SBIG 6303, Sinistro, and Spectral imagers, respectively.

\subsection{Data Augmentation}


All the DECam images considered in this work for training the CR detection models are with 90 sec exposure time. To facilitate the models to work on other extremely short or long exposures, we used the idea of augmenting the image sky background level using the same procedure as in deepCR~\citep{zhang2020deepcr}. Note that data augmentation was done only for DECam as the LCO data already had images with varying exposure times. Note however that the expected pixel values observed in images are realizations of Gaussian and Poisson processes and an exact deterministic scaling may not be completely realistic. However, this is sufficient to build up our training dataset. \rthis{Different exposure times and sky background levels change the contrast of CR artifacts and astronomical objects against the background, thereby affecting model prediction. The sky background level is adjusted by scaling by a factor of three,  followed by  subtracting it by up to 0.9 times its original level. Since an image's original pixel value can be expressed as:}
\[
  {n = (f_{star} + f_{sky}) \ . \ t_{exp} + n_{CR}}
\]
\rthis{where $n$ is in units of $e^-$ and flux ($f$) in units of $e^-$/s, the pixel value after augmentation is:}

\[
  {n^l = n + \alpha . f_{sky} . t_{exp} = \Big(\frac{f_{star}}{1+\alpha} + f_{sky} \Big) . (1+\alpha).t_{exp}+n_{CR}}
\]

\rthis{Adding or subtracting a multiplicative factor of the sky level, i.e., $\alpha$.$f_{sky}$, is equivalent to simulating an exposure time of $(1+\alpha)$.$t_{exp}$, with the flux from astronomical objects scaled down by $1+\alpha$, which is only a minor concern given that astronomical fluxes already span many orders of magnitude. Alternatively, by simply scaling the image with a multiplicative factor, one could emulate different exposure times. As a result, various CR statistics would be generated, each of which would contribute to the observed pixel value, independent of the integration time. While this augmentation technique will invariably affect the image noise properties, homogeneous noise in the training set should have little effect on model performance, according to similar reasoning in~\citep{lehtinen2018noise2noise}.}

\section{Deep Learning Framework}
\label{sec:models}
As described earlier, the U-Net~\citep{ronneberger2015u} model is a popular encoder-decoder based deep learning approach for image segmentation and was originally proposed for biomedical images. U-Net based architectures have been adopted for the CR detection task as well. Specifically, deepCR and Cosmic-CoNN are two popular CR detection algorithms based on this architecture. Given an image with CR contamination as input, these models predict a probabilistic map of each pixel affected by the CR hits. Segmentation is a simple binary classification between CR and Non-CR pixels in the context of CR detection. Finally, setting a threshold transforms the predicted probability map to a binary CR mask, with 1 indicating CR and 0 indicating Non-CR pixels. 


\begin{figure*}
    \centering
    \includegraphics[width=12cm, height=25cm, keepaspectratio,]{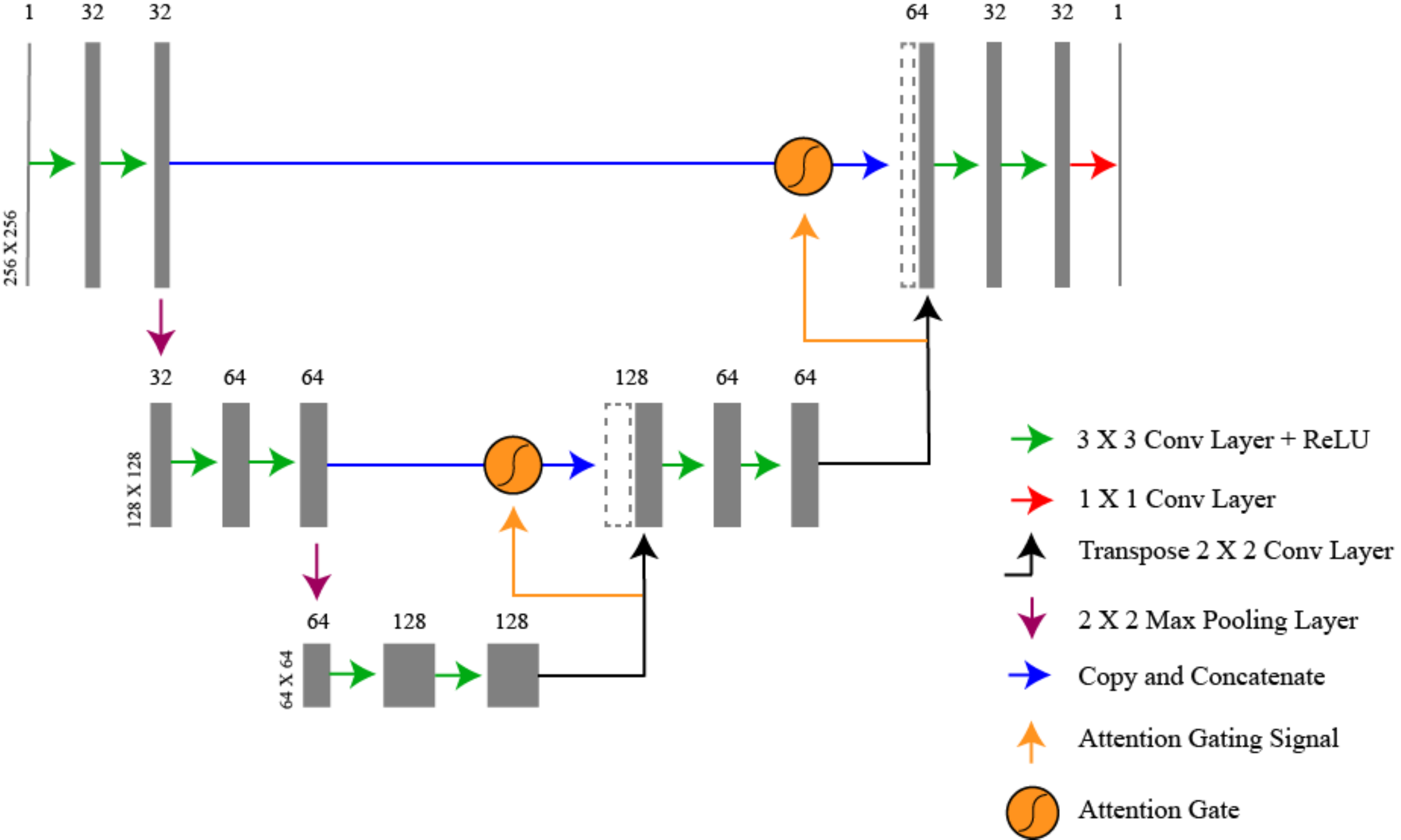}
    \caption{The neural network architecture of deepCR and Cosmic-CoNN. Feature maps are represented by grey boxes. The number of channels and the feature map size are given on the top and left of each feature map, respectively. In the legend on the lower left, different computational procedures are marked. The rectifier activation unit, ReLU is written as $f(x) = \text{max}(0, x)$ and is used here. Unfilled boxes to the right of blue arrows represent feature maps that have been copied from the left and are to be combined with the adjacent feature map. The attention gate is represented by the orange color circle at each skip connection.}
    \label{fig:network}
\end{figure*}

In order to optimize the neural networks trained on astronomical images, their unique features such as high dynamic range and spatial fluctuations of high magnitude need to be addressed explicitly. Also, the high CR rates in space-based data do not represent the acute class imbalance problem that has been reported in ground-based imaging data. CNN models find it challenging to train and converge on ground-based imaging data due to low CR density. A two-phase training is designed in the deepCR~\citep{zhang2020deepcr} framework to overcome some of these limitations. In order to converge, the model freezes the  feature normalization parameters in the second phase, assuming correct data statistics were learned in the first phase. This method works well when the inference and training data have similar statistics and allows an instrument-specific model to be learned. However, this approach is not suited for designing a generic CR detection model that can be used with a wide range of ground or space-based instruments and data statistics. Cosmic-CoNN~\citep{xu2021cosmic} presented a symbiotic combination of three enhancements to address the difficulties mentioned above. Based on the effective U-shaped architecture, Cosmic-CoNN proposed a novel loss function that explicitly handles the problem of class disparity, improved data sampling methods that have proven to be critical for training efficiency, and a feature normalization technique that uses group normalisation~\citep{wu2018group} to improve model generalization ability. Thus, the Cosmic-CoNN has emerged as the community's first deep learning-based generalized CR detection model.


This work uses deepCR and Cosmic-CoNN (both adopted from the U-Net architecture) as baseline models and augments them with an attention gate (AG) module at each decoder block of the U-Net. We describe our proposed methodology next.


\subsection{U-Net}
The architectures of the baseline models are simple modifications of U-Net, an encoder-decoder-based CNN model with skip connections between each encoder and decoder depth. The U-Net convolves the image at multiple scales and concatenates features of the same scale with skip connections, allowing the network to propagate context information to higher resolution layers, thereby producing pixel-level classification predictions on large images. A deep CNN model optimizes millions of kernel parameters in its hierarchical convolution layers instead of the hand-crafted kernels used in traditional image-filtering methods such as LACosmic. Both methods apply convolution to extract the features from a small portion of the input image, known as the receptive field. LACosmic finds a CR streak by the sharpness of its outermost pixel and proceeds inwards via an iterative method, making the CR detection computationally expensive and inefficient due to the limited receptive field ($3 \times 3$, extent of the Laplacian kernel). On the other hand, a deep CNN model uses deeper layers of its hierarchical network to produce a larger receptive field to capture the morphological characteristics (edges, corners, or sharpness) of CR hits, along with the contextual cues from surrounding pixels. 

A reference model to depict the network architectures of both the baseline and the attention augmented baseline models are presented in Fig.~\ref{fig:network}. The baseline models do not contain the attention gates that are represented with the orange color circles at each decoder layer of the U-Net. Each model represents a UNet-3-32 architecture which is a depth-3 network with 32 channels at the first convolution layer, as illustrated in Fig.~\ref{fig:network}. The major distinction between the two baseline models is the feature-normalization technique that expedites training and makes deep CNNs more amenable to optimize. deepCR uses batch normalization, whereas Cosmic-CoNN uses group normalization after every convolution layer.


\subsection{Attention U-Net}
The attention mechanism was initially introduced in \citet{vaswani2017attention}, which helps in improving the performance of deep-learning models by allowing them to focus more on target structures and even provides an explanation for the model behaviour. It is used for several applications, including image captioning~\citep{anderson2018bottom}, image classification and semantic segmentation~\citep{wang2017residual, ypsilantis2017learning}. While addressing the semantic segmentation problem, the spatial information obtained during upsampling in the expanding path of encoder-decoder architecture is imprecise. U-Net employs skip connections to address this issue and combine spatial data from the downsampling and upsampling paths. However, feature representation in the initial layers is inadequate, resulting in many redundant low-level feature extractions, wasting computational resources and model parameters. Attention U-Net~\citep{oktay2018attention} employs a novel self Attention Gating (AG) mechanism that learns to suppress the unimportant regions in an input image while emphasising essential features for a specific task. Thus it allows the U-Net to more focus on target structures of varying size and shape. Specifically, implementing AGs using soft attention at the skip connections actively suppresses activation in irrelevant regions, minimizing the number of redundant features carried across. Soft attention works by giving different sections of the image distinct weights. Those regions with high significance are given a higher weight, whereas those with low relevance are given a lower weight. As the model is trained, the weighting is improved, enabling the model to make more informed decisions about which parts to pay more attention to. The AGs can be easily integrated with standard CNN topologies such as the U-Net model (shown in Fig.~\ref{fig:network}), with minimal computational overhead and increased model sensitivity and prediction accuracy. 

\subsubsection{Attention Gate (AG) Module}
\begin{figure}
    \centering
    \includegraphics[width=70mm, height=40mm]{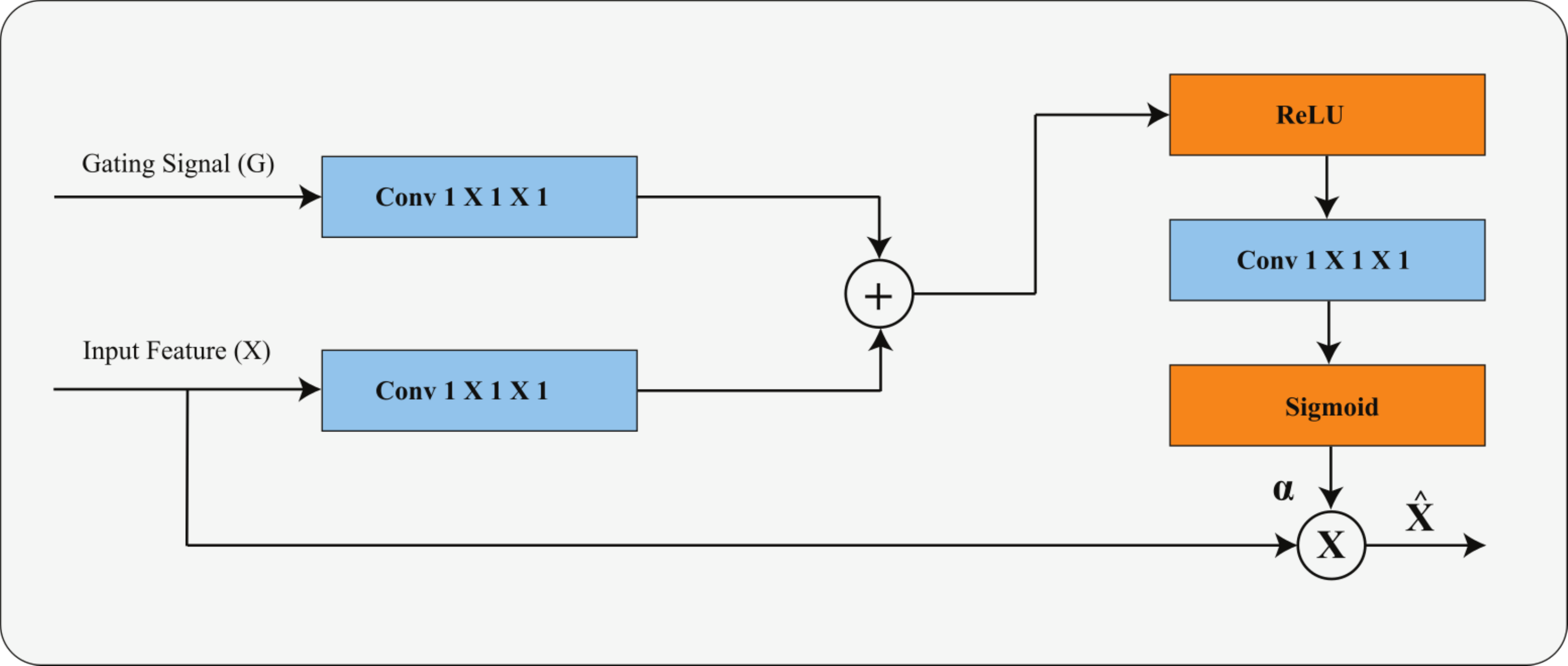}
    \caption{The architecture of additive Attention Gate (AG) module represented in the orange colour circle at each decoder of U-Net in Figure \ref{fig:network}. Unlike base models in which the feature maps are directly copied from the encoder via skip connection, they are first passed through the attention gate in attention augmented models. This allows U-Net to automatically focus on target structures in an image and then concatenate with the adjacent feature map.}
    \label{fig:att1}
\end{figure}

Attention U-Net ~\citep{oktay2018attention} provided a grid-based gating that allows attention coefficients to be more specific to local regions, building on the attention paradigm introduced in \citet{jetley2018learn}. The architecture of a stand-alone AG module is depicted in Fig~\ref{fig:att1}. It has two inputs: the first being the attention gating signal, which represents the feature maps from the previous layer, and the other is the activation map from the corresponding encoder layer that are transferred via a skip connection. Fig.~\ref{fig:att1} illustrates the step-by-step procedure involved in implementing the AGs at every skip connection and are described as follows. Specifically, we explain the operations using tensors from the second skip connection as an example fed to the attention gate.

\begin{itemize}
    \item The vectors \textit{\textbf{X}} and \textit{\textbf{G}} of dimensions $64\times 128\times 128$ and $64\times 128\times 128$ are the two inputs to the AG module from the encoder and decoder at the first skip connection.
    \item The vector \textit{\textbf{G}} is the upsampled version of the previous layer's feature maps performed via transposed convolution.
   \item Both the vectors \textit{\textbf{X}} and \textit{\textbf{G}} undergo a $1\times 1$ convolution, resulting in tensors of dimension $32\times 128 \times 128$ and $32\times 128 \times 128$ respectively. Here, 32 is the number of multi-dimensional attention coefficients, whose values need to be learned.
   \item The two vectors are then added element by element, resulting in aligned weights getting larger and the non-aligned weights becoming smaller.
   \item The resulting vector is processed with a ReLU activation layer and a $1\times 1$ convolution, reducing the dimensions to $1\times 128\times 128$.
   \item A sigmoid layer scales this vector between [0,1], yielding the attention coefficients (or weights, represented with \boldmath$\alpha$), where the coefficients closer to 1 indicate more significant features and those closer to 0 represent less significant ones.
   \item The original vector \textit{\textbf{X}} is multiplied element-by-element with the attention coefficients, scaling the vector according to relevance.
   \item Finally, the refined feature maps \textit{$\hat{\textbf{X}}$} are passed along normally in the skip connection.
   
\end{itemize}

Since the AG module is differentiable, it is trained during back-propagation, making the attention coefficients better at highlighting relevant regions. Thus, the attention gate enhances the U-Net performance by allowing it to automatically focus on varying target structures in an image.

In all these frameworks, training the CR detection models require a set of labeled data that includes CR contaminated images ($X$) and the corresponding ground truth CR masks ($M$). Our training data is constructed using DECam images, and more details on how the dataset is constructed can be found in Section \ref{sec:data}. We used LCOGT images for testing our models and so no training is needed. Similar to deepCR \citep{zhang2020deepcr}, we also chose to use the Binary Cross-Entropy (BCE) loss while training the CR-mask detection models. The BCE loss between the model prediction (\textit{P}, the probability map) and the ground-truth CR mask (\textit{Y}, with '1' indicating CR pixels and '0' indicating Non-CR pixels) can be defined as follows: 
\[ BCE(P, Y) = -(Y_{ij} \text{log}(P_{ij}) + (1-Y_{ij})\text{log}(1-P_{ij}))  \]

The ground-truth CR mask, $Y$ is defined as $Y_{ij}$ = 1 for CR pixels and $Y_{ij}$ = 0 for Non-CR pixels. When $Y_{ij}$ = 1, the first term in BCE loss function $Y_{ij}$ log($P_{ij}$) measures the loss for CR pixels, while the second term becomes 0. Similarly, when $Y_{ij}$ = 0 the second term measures the loss for Non-CR pixels while the first term becomes 0. The optimization objective is to minimize the total loss, which is the sum of the two loss terms to account for both CR and Non-CR classes.


\section{Results and Discussion}
\label{sec:results}
We have trained and evaluated the baseline deepCR and Cosmic-CoNN models with and without attention gate module augmentation. The models are trained using 20,480 image patches, each with $256 \times 256$ pixel resolution with 40 images from each of the {\em{'griz'}} DECam photometric bands. From the training dataset, 1\% of the images, which is 2048 patches are reserved for validation. While we trained our models using only images from the DECam imager, testing was performed on both DECam and LCOGT Network images. The DECam test set consists of 8192 image patches with 16 images from each DECam band. On the other hand, the LCO CR test dataset consists of 119 images from three different telescope classes. CR detection using image-filtering based algorithms such as LACosmic \citep{van2001cosmic}, Astro-SCRAPPY \citep{mccully2019astro} and other recent learning-based algorithms including Maximask \citep{paillassa2020maximask} and Cosmic-CoNN (pre-trained on LCO) \citep{xu2021cosmic} are also evaluated using the test data. We use various methods to assess the performance of the models both quantitatively and qualitatively. LACosmic and Astro-SCRAPPY are image-filtering based methods and can be applied for CR detection on any optical or spectroscopic image obtained either from ground-based or space-based telescopes. In contrast, the other learning-based models are specific to the data (or data with similar statistics) that they have seen during training.

The network is a binary classifier for every class in the proposed deep learning framework while detecting the CR induced pixels. Hence, we can compute the Receiver Operating Characteristic (ROC) curve and use it as an evaluation metric to compare the performance of different detectors at varying thresholds ({\em{t}}). ROC curves represent the False Positive Rate (FPR) versus the True Positive Rate (TPR) and in the context of CR detection:
\[
  TPR = \frac{CR \ pixels \ correctly \ classified}{Total \ CR \ pixels} = \frac{TP}{TP \ + \ FN}
\]
\[
  FPR = \frac{Non-CR \ pixels \ mistaken \ as \ CR}{Total \ Non-CR \ pixels} = \frac{FP}{TN \ + \ FP}
\]

where TP is the number of true positives (CR induced pixels successfully recovered as CRs), FN is the number of false negatives (CR pixels wrongly classified as Non-CRs), FP is the number of false positives (Non-CR pixels wrongly classified as CRs), and TN is the number of true negatives (Non-CR pixels successfully recovered as Non-CRs).

Ideally, the network should deliver the highest possible TPR for a fixed FPR. Standard ROC plots on ground-based imaging data, such as DECam and LCOGT, can be misleading for imbalanced datasets \citep{saito2015precision} with far fewer CR pixels than Non-CR pixels. Even if the ROC plot provides a model-wide evaluation at all possible thresholds, this is possible. The main reason for this optimistic picture is because of the use of True Negatives in the False Positive Rate in the ROC Curve. Even though evaluating TPR at fixed and lower FPR is a decent metric and enough for the CR detection problem, further caution can help better interpret the performance of these models on imbalanced data. For imbalanced datasets, the Precision-Recall curve, on the other hand, \rthis{could also provide an independent diagnostic} and is used in our work \rthis{as a supplement to the ROC curve}. Because while computing the PRC plots, we can carefully avoid using True Negatives, the dominant class in the CR detection problem. Also, PRC plots can provide the viewer with an accurate prediction of future classification performance since they evaluate the fraction of true positives among positive predictions. While the recall is equivalent to TPR, in the context of CR detection, Precision (or Purity) is defined as:

\[
  Precision = \frac{CR \ pixels \ correctly \ classified}{Total \ CR \ pixels \ predicted \ by\ model} = \frac{TP}{TP \ + \ FP}
\]

Unlike FPR, precision is evaluated by the model's proportion of correct CR predictions, which is less susceptible to the ratio of CR and Non-CR pixels in an image and varying CR rates between datasets. Given a fixed proportion of actual CRs correctly discovered (e.g., 95$\%$ recall), the better model should make fewer mistakes and thus result in higher precision. It also assists in determining how well a model works on two different datasets with the same recall and vice versa.

In addition to the ROC and PRC curves, we use three other segmentation metrics helpful for assessing the performance of each CR detector quantitatively. These include F1-score, which is the harmonic mean of the Precision and Recall, Intersection-Over-Union (IOU), and False Discovery Rate (FDR) and are defined  as follows~\citep{Bethapudi}: 

\[
  F1-score = 2 \ . \ \frac{Precision \ * \ Recall}{Precision \ + \ Recall} = \frac{2TP}{2TP \ + \ FP + \ FN}
\]

\[
  IOU = \frac{Area \ of \ Overlap}{Area \ of \ Union} =  \frac{TP}{TP \ + \ FP \ + \ FN}
\]

\[
  FDR =  \frac{FP}{TP \ + \ FP}
\]

Considering the binary classification between CR and Non-CR pixels, the metrics F1-score and IOU should be as high as possible, while FDR should be as low as possible. The ROC and PRC plots are threshold-independent measures. However, the F1-score, IOU, and FDR metrics are computed at a specific threshold at which the model is performing well.

\subsection{Performance on DECam data}
The DECam test set is first analyzed with the image-filtering based CR detectors, LACosmic \citep{van2001cosmic} and it's optimized version, Astro-SCRAPPY \citep{mccully2019astro} for reference. We chose \texttt{objlim=1.0} and \texttt{sigfrac=0.1} for optimal performance across the DECam test dataset and held it constant for both LACosmic and Astro-SCRAPPY. Then we produce the ROC curves varying the \texttt{sigclip} parameter between [1, 20]. Next, we evaluated the previously trained deep-learning-based algorithms (MaxiMask and Cosmic-CoNN (pre-trained)) along with the proposed baselines with and without attention augmentation. The ROC and PRC plots for deep learning models are plotted by varying the threshold between [0, 1] on the probability map obtained from the output of each deep learning model. Varying the threshold changes the probability map for CR hits and thus the final binary CR mask. The CR masks obtained from LACosmic, Astro-SCRAPPY and the Cosmic-CoNN (pre-trained) models are dilated using a $3 \times 3$ dilation kernel for fair comparisons with the DECam test data constructed using the method described in Section \ref{sec:data}. However, the output CR mask from MaxiMask is not dilated as it has also used dilation for the CR mask while training. \rthis{We note that the efficacy of our CR masking algorithms is evaluated by comparing how many CR induced pixels are correctly identified. However, one caveat when applying some of our algorithms on external tools that were trained with different ``ground truths'', is that a slightly larger margin around CR events could have been defined in the Astro-SCRAPPY or MaxiMask footprints compared to the choices made in this paper. This would show Astro-SCRAPPY or MaxiMask in a bad light even if these algorithms were as reliable. One possible solution to this would be to compare the number of distinct cosmic rays, which are currently identified instead of cosmic ray pixels. However, the statistics would be noisier and its also not trivial to count the number of distinct cosmic rays if they are of arbitrary shape. With this caveat in mind, we now present our results.}

\begin{figure*}[!tbp]
  \centering
  \subfloat[LACosmic - ROC]{\includegraphics[width=0.25\textwidth,
  height=0.25\textwidth, keepaspectratio,]{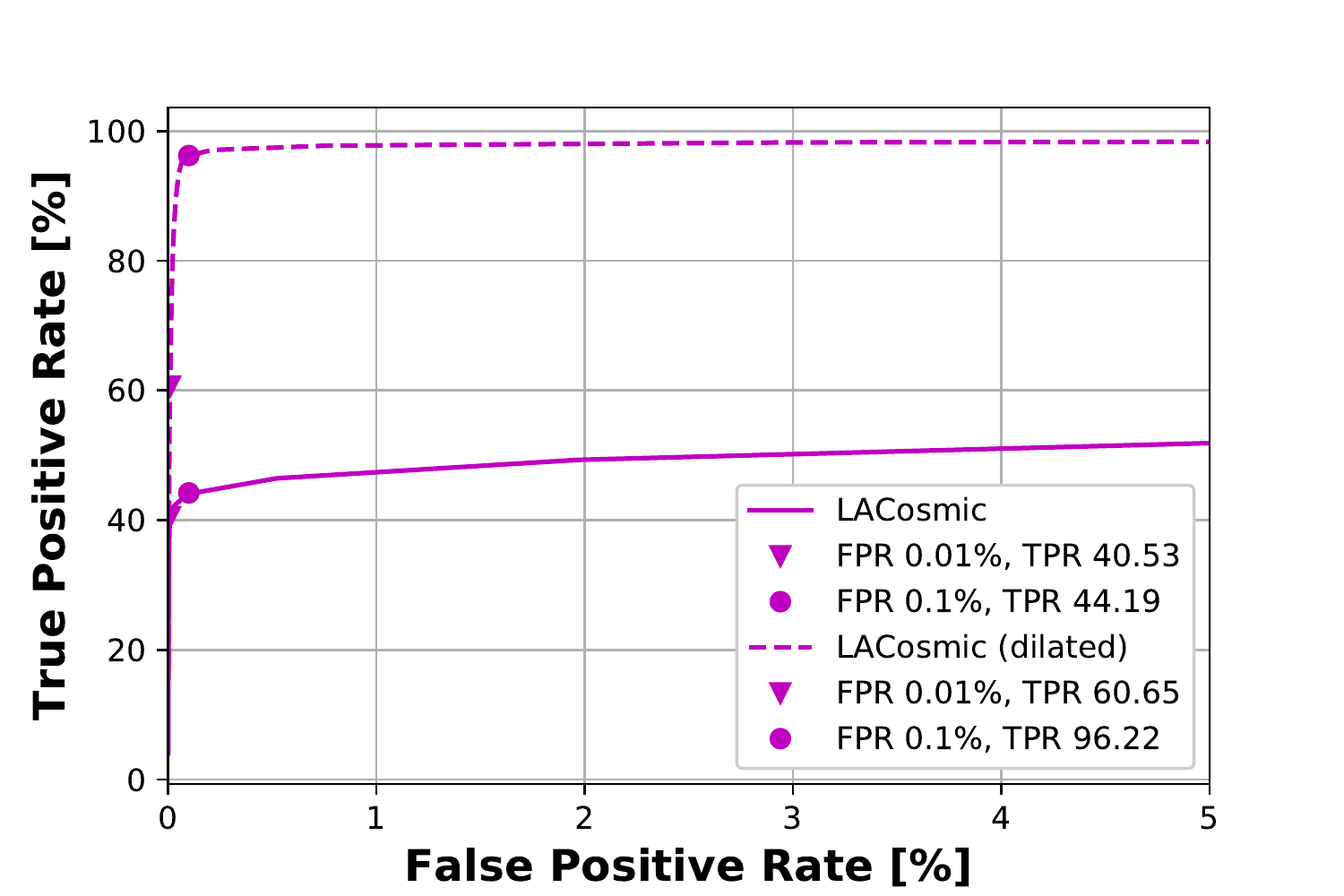}\label{fig:f11}}
  \hfill
  \subfloat[Astro-SCRAPPY - ROC]{\includegraphics[width=0.25\textwidth,
  height=0.25\textwidth, keepaspectratio,]{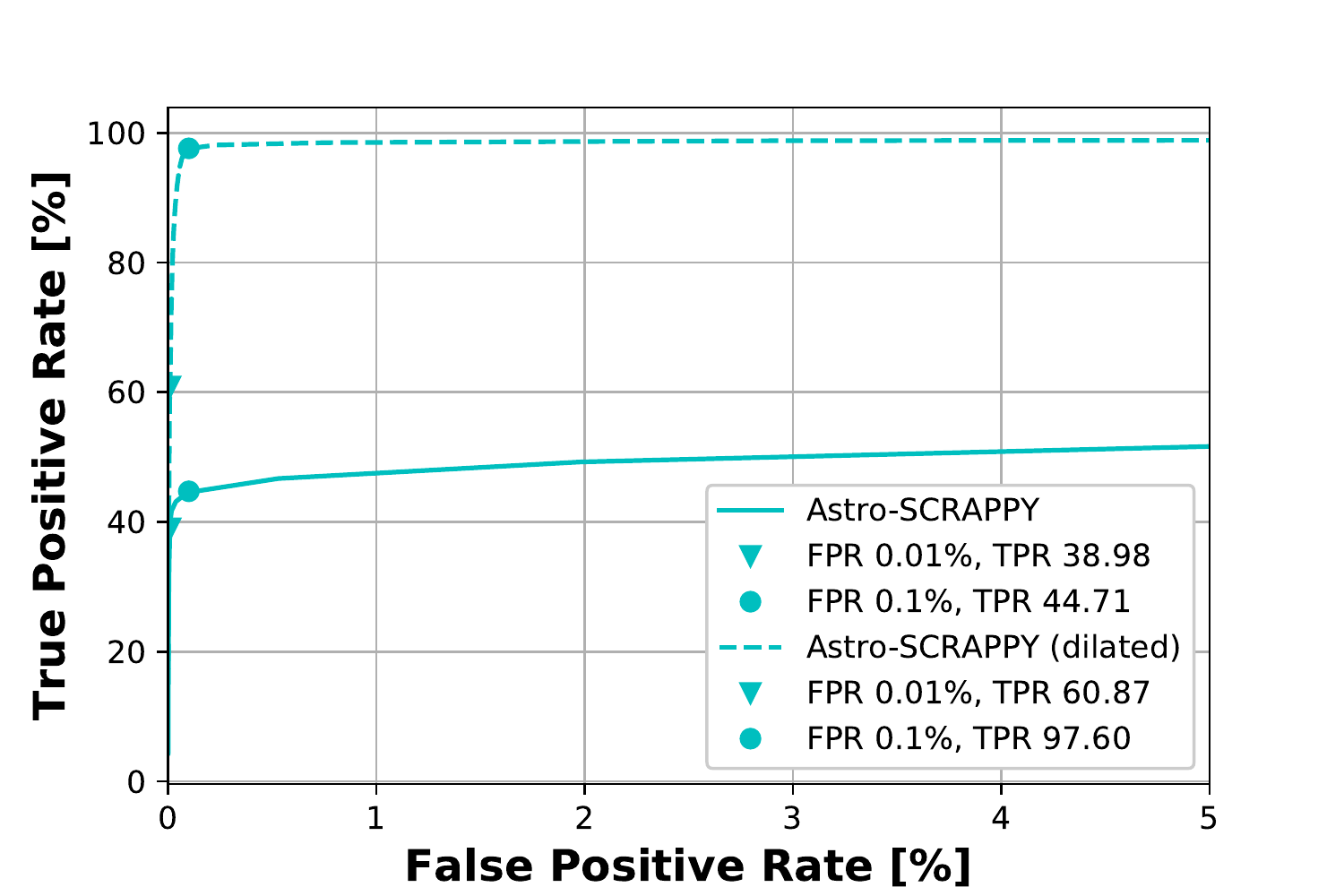}\label{fig:f12}}
  \hfill
  \subfloat[MaxiMask and Cosmic-CoNN - ROC \\ (pre-trained models)]{\includegraphics[width=0.25\textwidth,
  height=0.25\textwidth, keepaspectratio,]{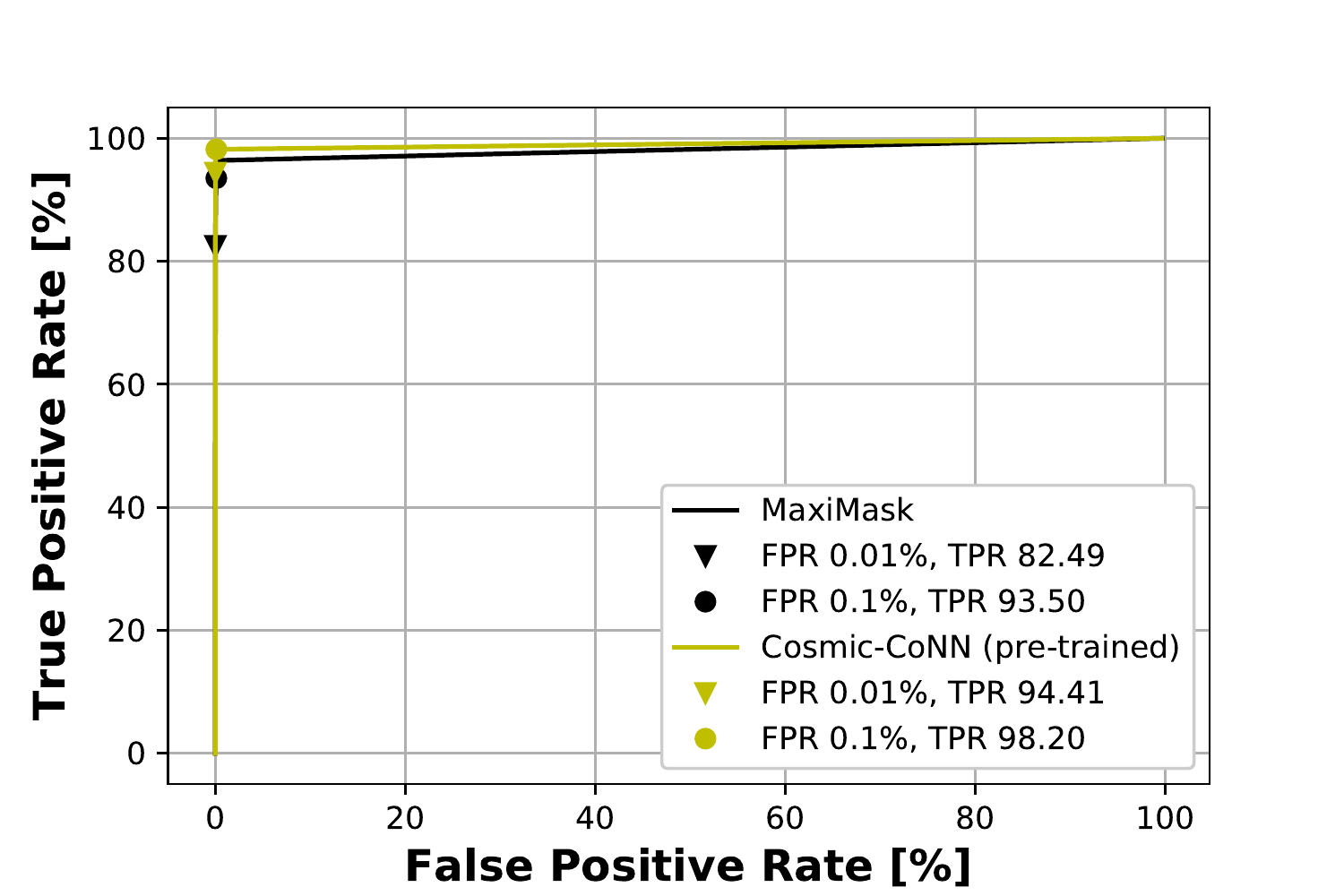}\label{fig:f13}}
  \hfill
  \subfloat[MaxiMask and Cosmic-CoNN - PRC \\ (pre-trained models)]{\includegraphics[width=0.25\textwidth,
  height=0.25\textwidth, keepaspectratio,]{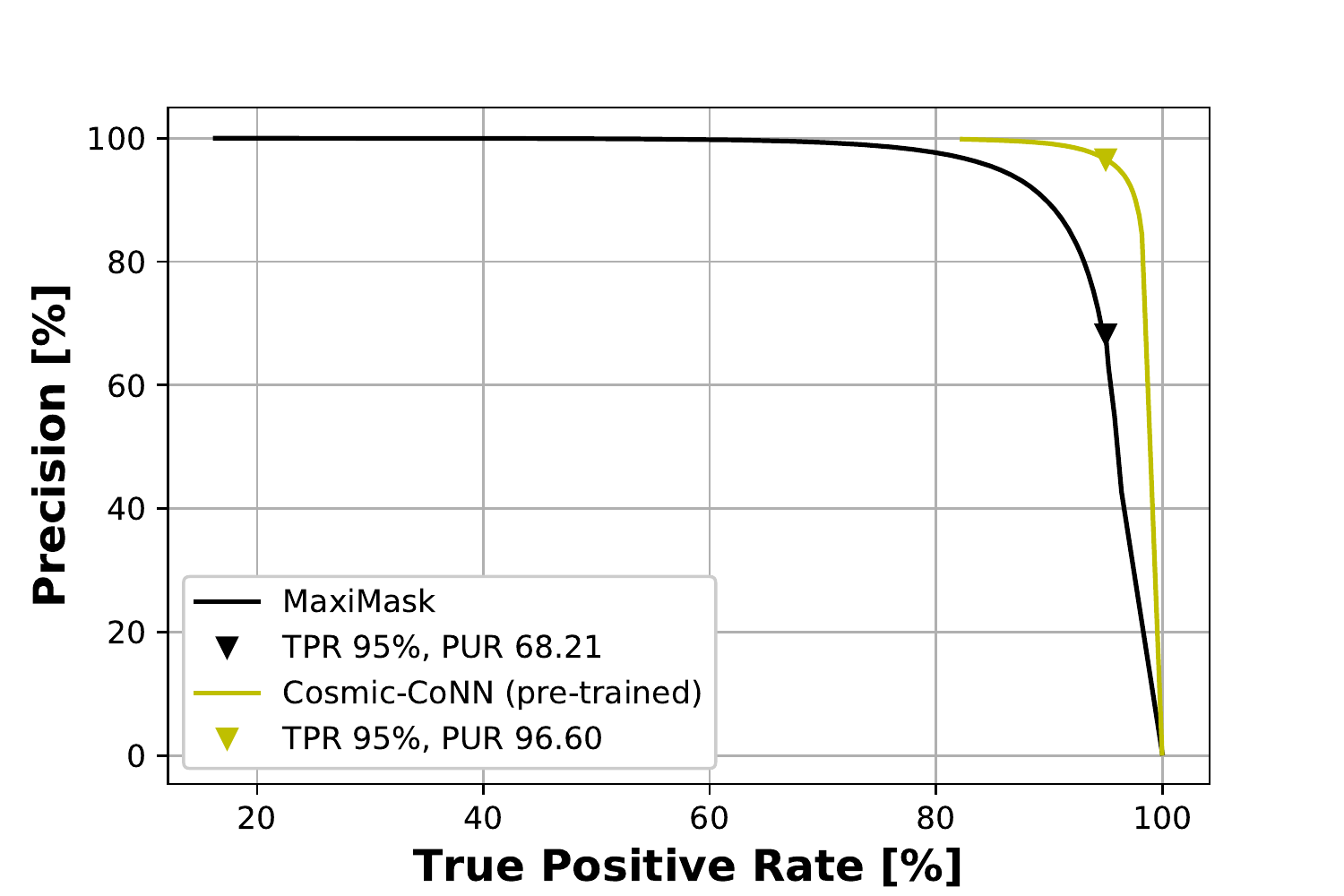}\label{fig:f14}}
  \caption{The ROC curves obtained on DECam test data with conventional algorithms (a) LACosmic and (b) Astro-SCRAPPY. Previously trained algorithms such as MaxiMask and Cosmic-CoNN (pre-trained on LCO) are also evaluated on DECam test data and the correponding ROC and PRC plots are presented in (c) and (d) respectively. The CR mask is dilated using a $3 \times 3$ kernel for fair comparisons with the groundtruth while using LACosmic, Astro-SCRAPPY and pre-trained Cosmic-CoNN models.}
  \label{fig:decam1}
\end{figure*}

\begin{figure*}[!tbp]
  \centering
  \subfloat[LACosmic - ROC]{\includegraphics[width=0.25\textwidth,
  height=0.25\textwidth, keepaspectratio,]{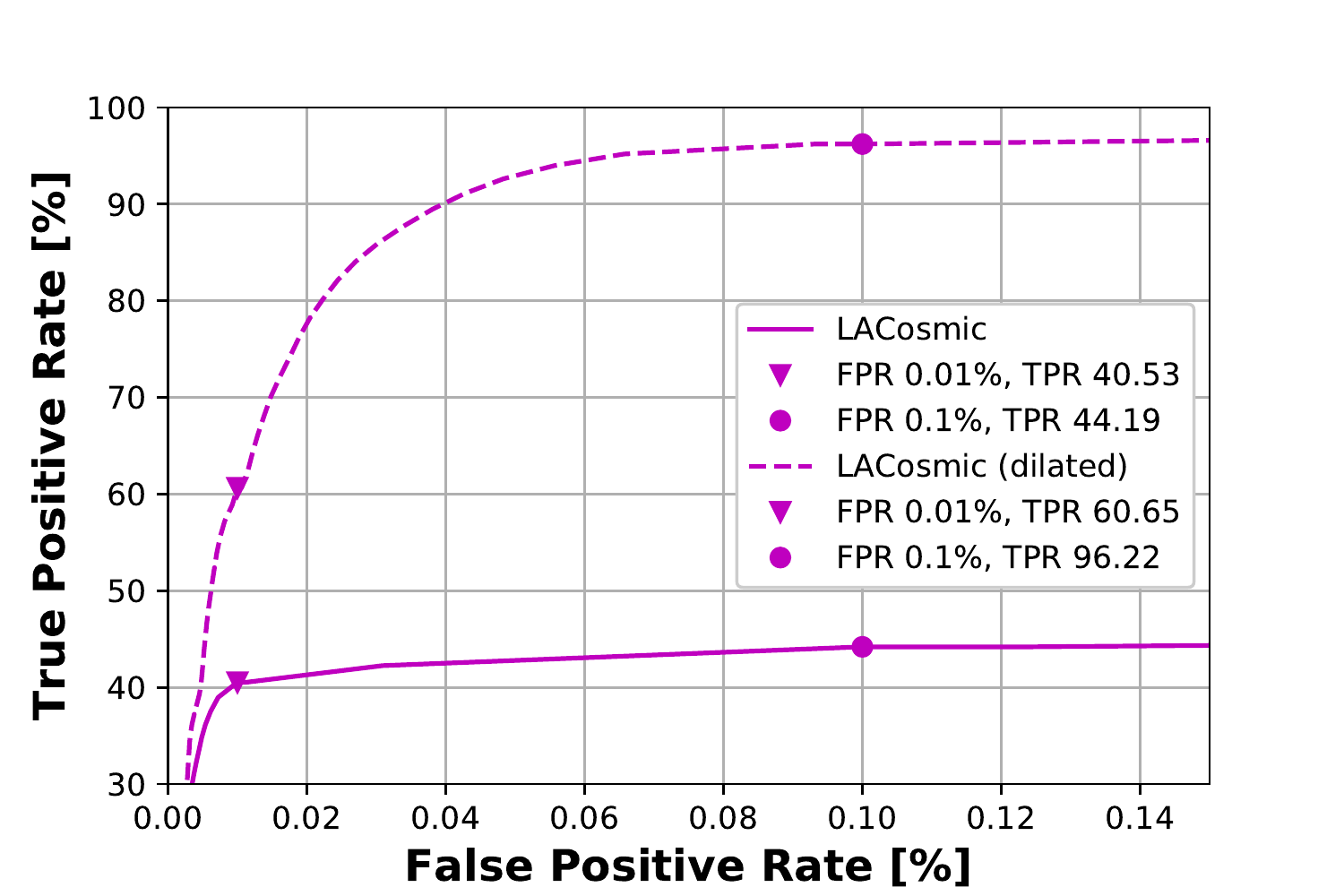}\label{fig:f11_ext}}
  \hfill
  \subfloat[Astro-SCRAPPY - ROC]{\includegraphics[width=0.25\textwidth,
  height=0.25\textwidth, keepaspectratio,]{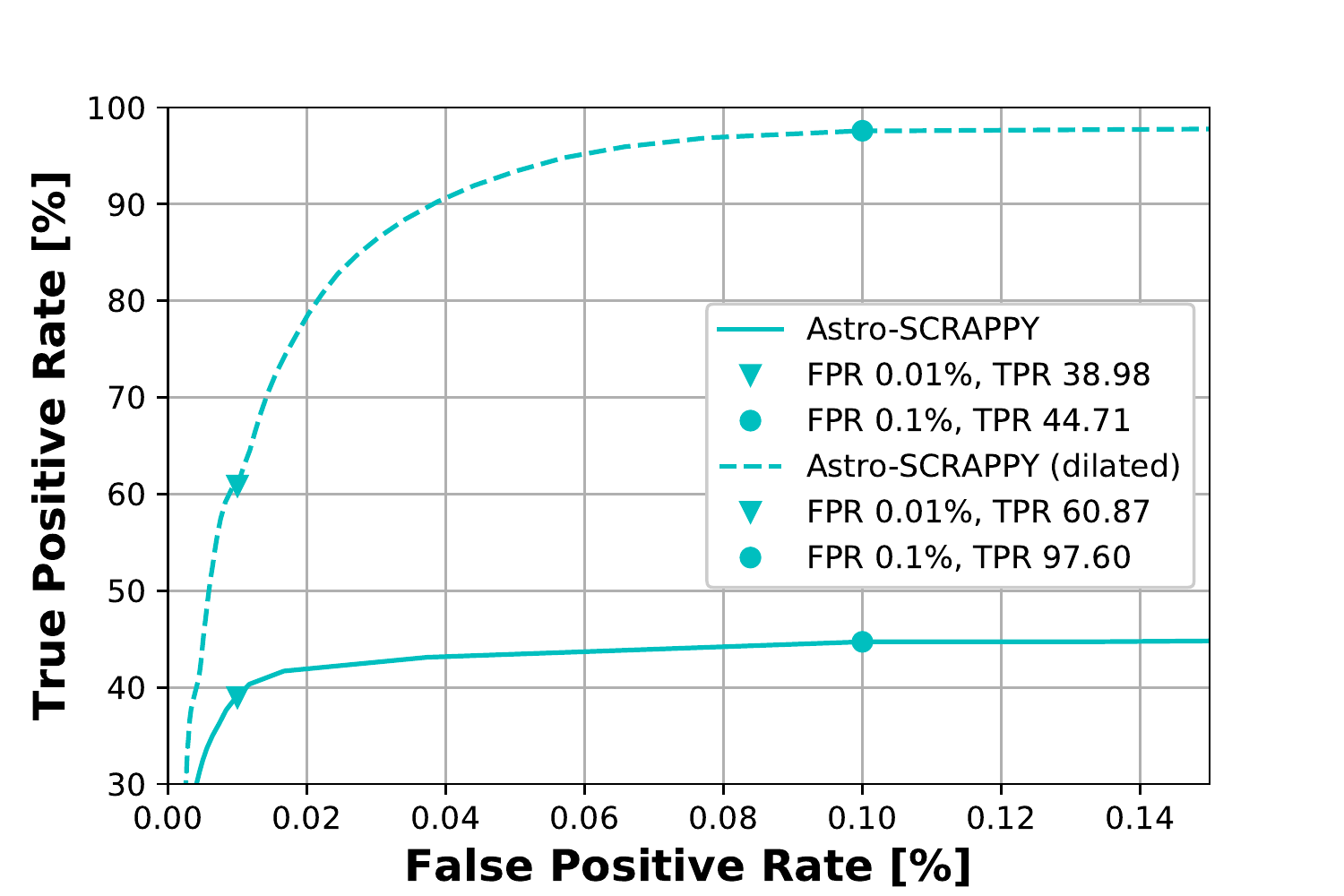}\label{fig:f12_ext}}
  \hfill
  \subfloat[MaxiMask and Cosmic-CoNN - ROC \\ (pre-trained models)]{\includegraphics[width=0.25\textwidth,
  height=0.25\textwidth, keepaspectratio,]{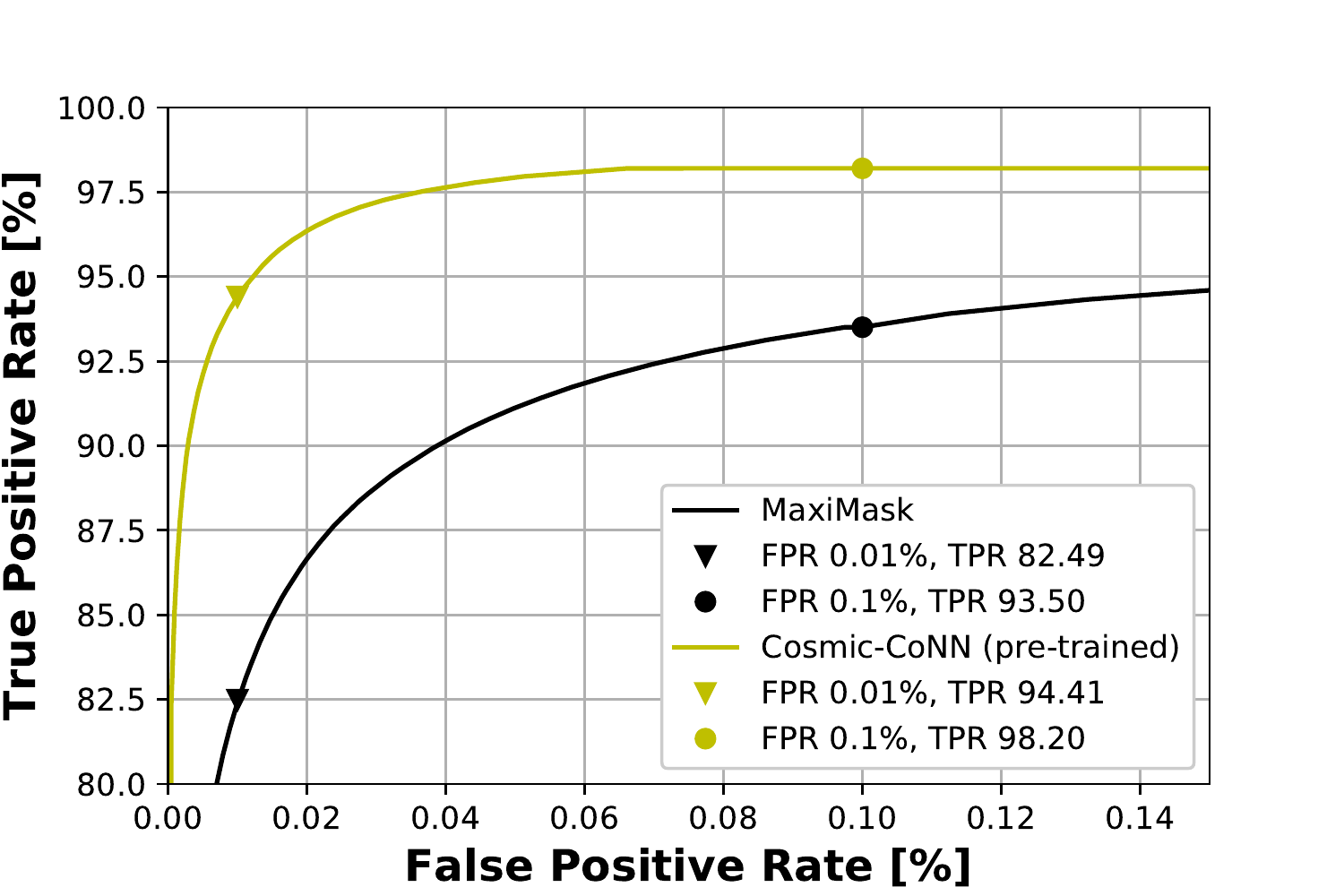}\label{fig:f13_ext}}
  \hfill
  \subfloat[MaxiMask and Cosmic-CoNN - PRC \\ (pre-trained models)]{\includegraphics[width=0.25\textwidth,
  height=0.25\textwidth, keepaspectratio,]{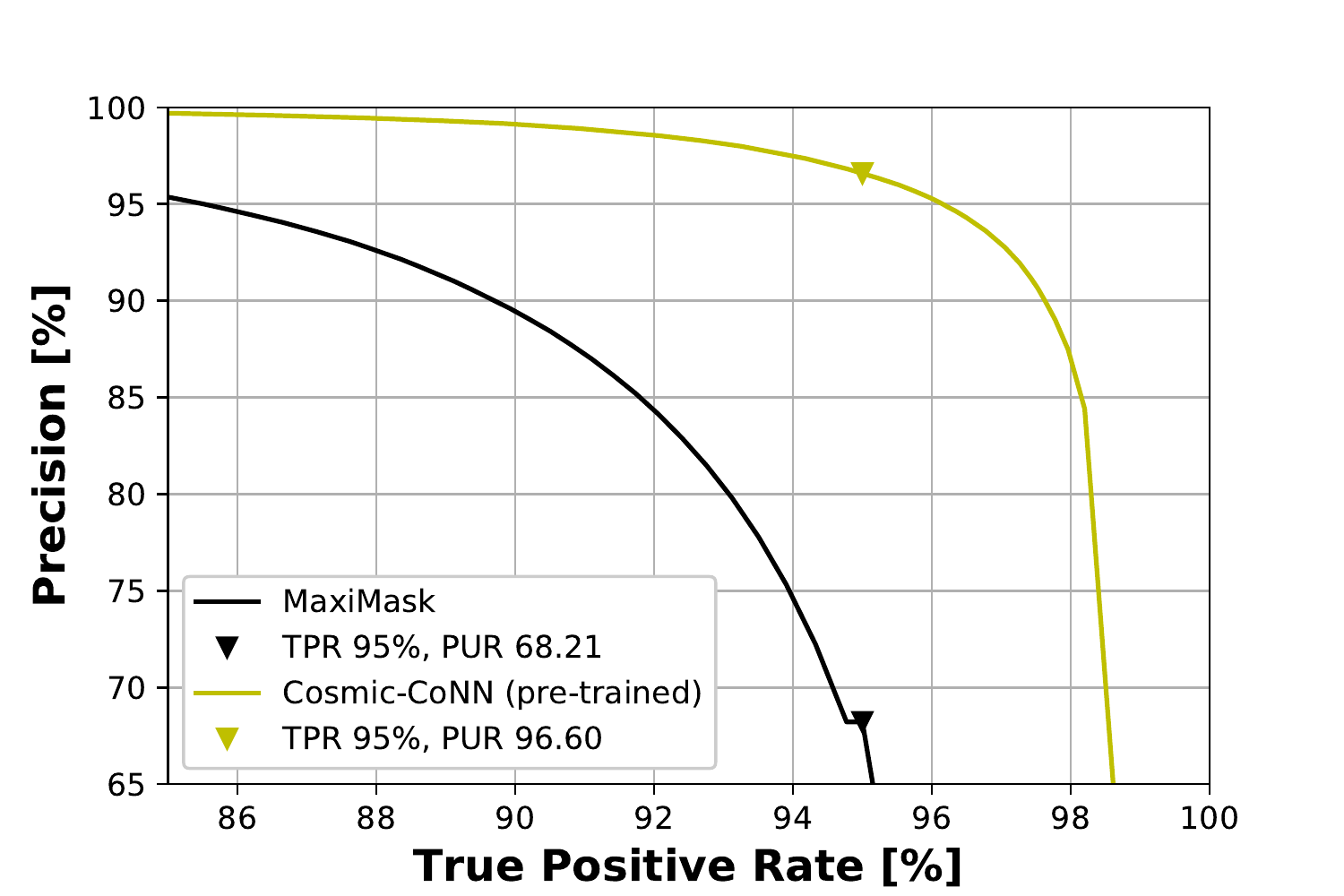}\label{fig:f14_ext}}
  \caption{The extended ROC and PRC plots for a better understanding of the model performances from Figure \ref{fig:decam1}. Both MaxiMask and pre-trained Cosmic-CoNN models outperform LACosmic and Astro-SCRAPPY. However, the pre-trained Cosmic-CoNN provides the highest TPR of 97.96 \% at 0.1 \% FPR and 96.6 \% Precision at 95 \% TPR.}
  \label{fig:decam1_ext}
\end{figure*}

\begin{figure*}[!tbp]
  \centering
\subfloat[deepCR - ROC]{\includegraphics[width=0.25\textwidth,
  height=0.25\textwidth, keepaspectratio,]{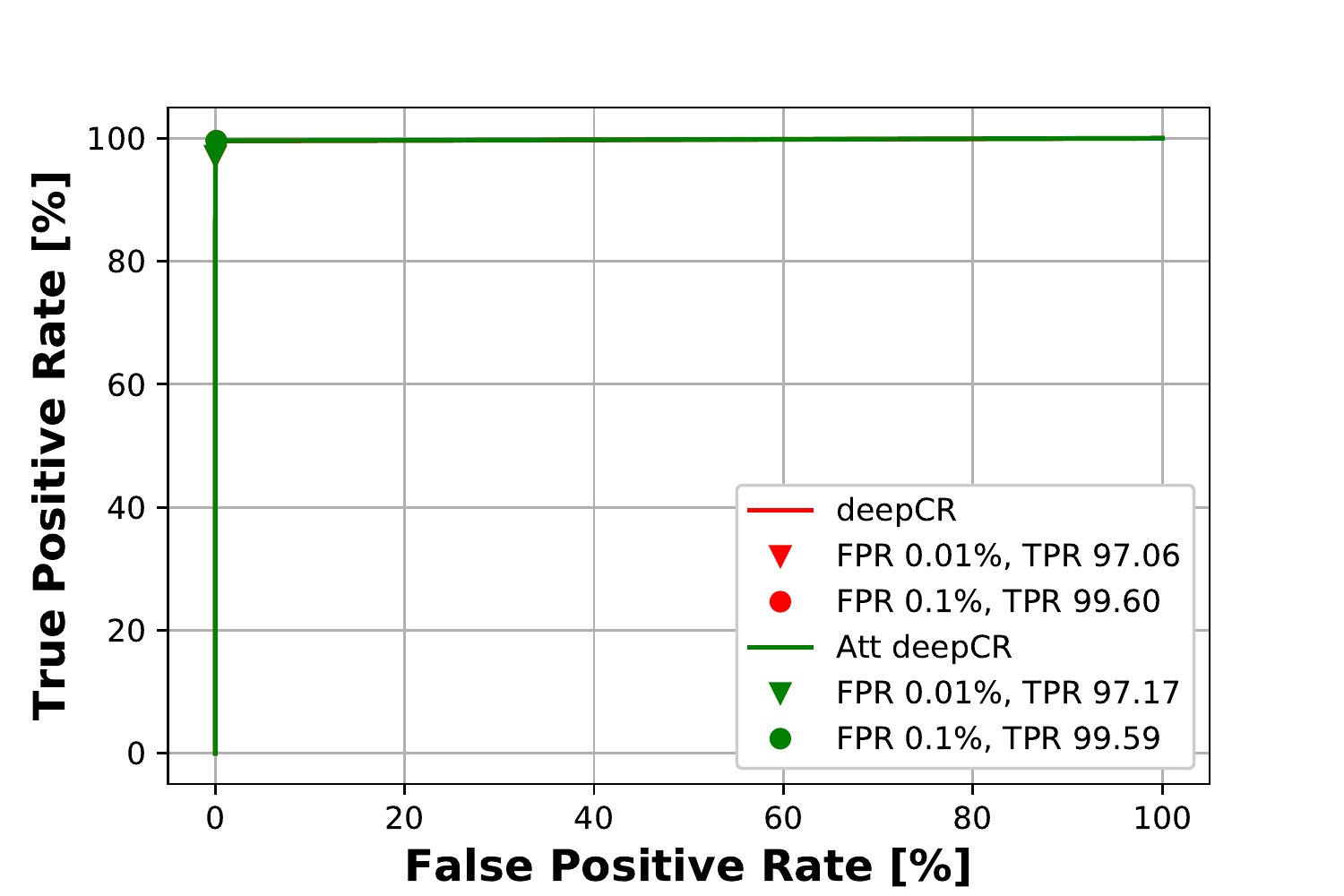}\label{fig:f21}}
  \hfill
  \subfloat[deepCR - PRC]{\includegraphics[width=0.25\textwidth,
  height=0.25\textwidth, keepaspectratio,]{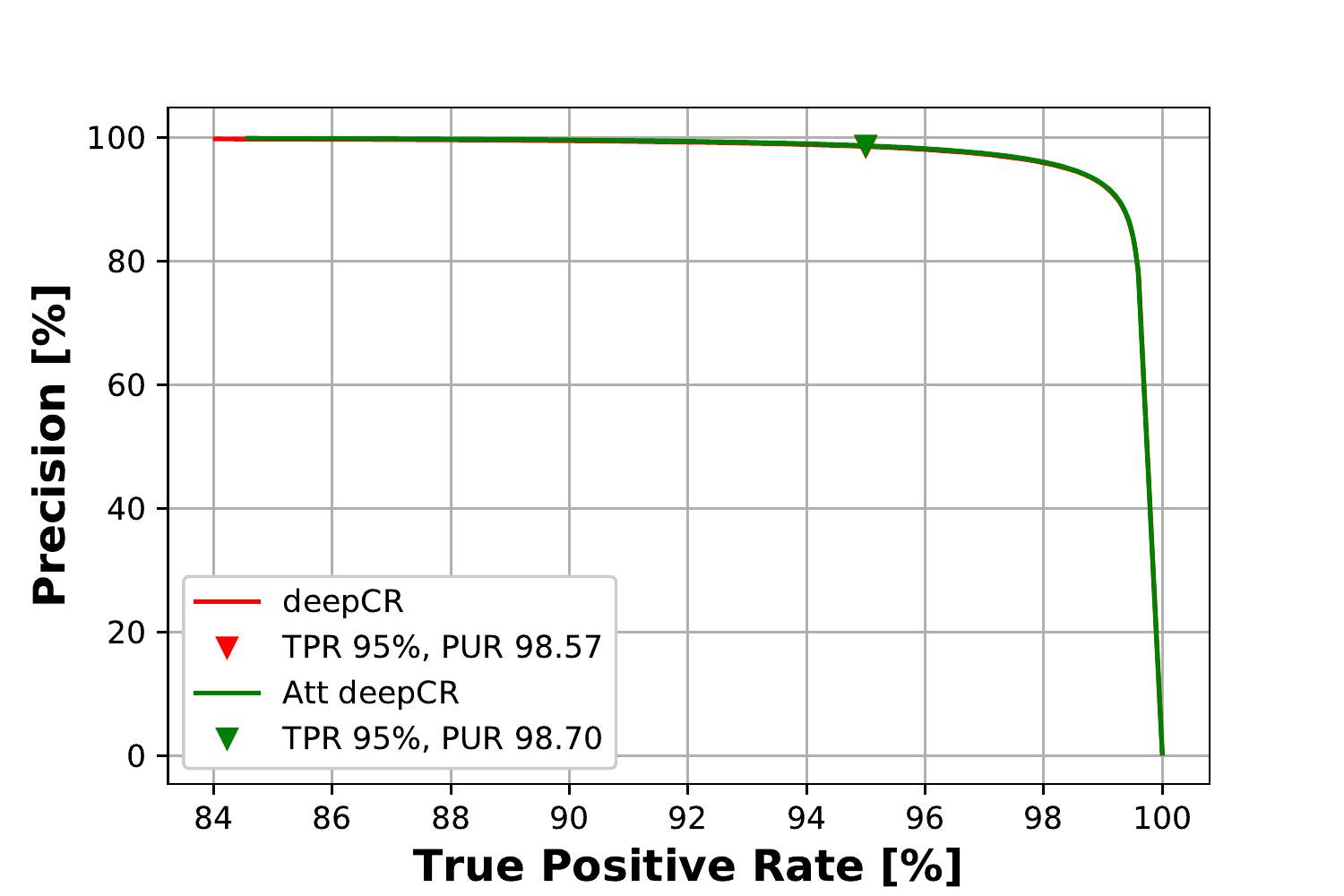}\label{fig:f22}}
  \hfill
  \subfloat[Cosmic-CoNN - ROC]{\includegraphics[width=0.25\textwidth,
  height=0.25\textwidth, keepaspectratio,]{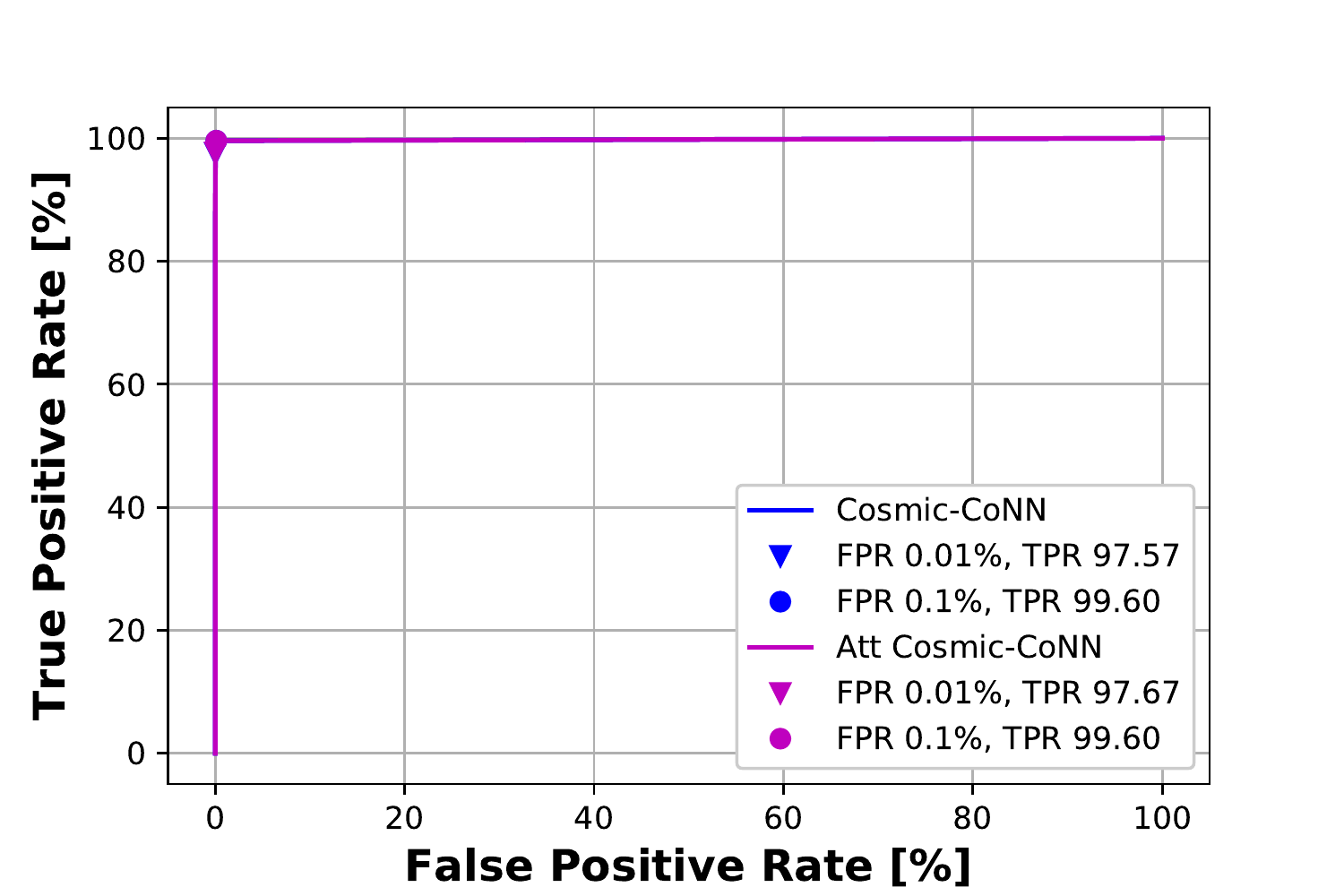}\label{fig:f23}}
  \hfill
  \subfloat[Cosmic-CoNN - PRC]{\includegraphics[width=0.25\textwidth,
  height=0.25\textwidth, keepaspectratio,]{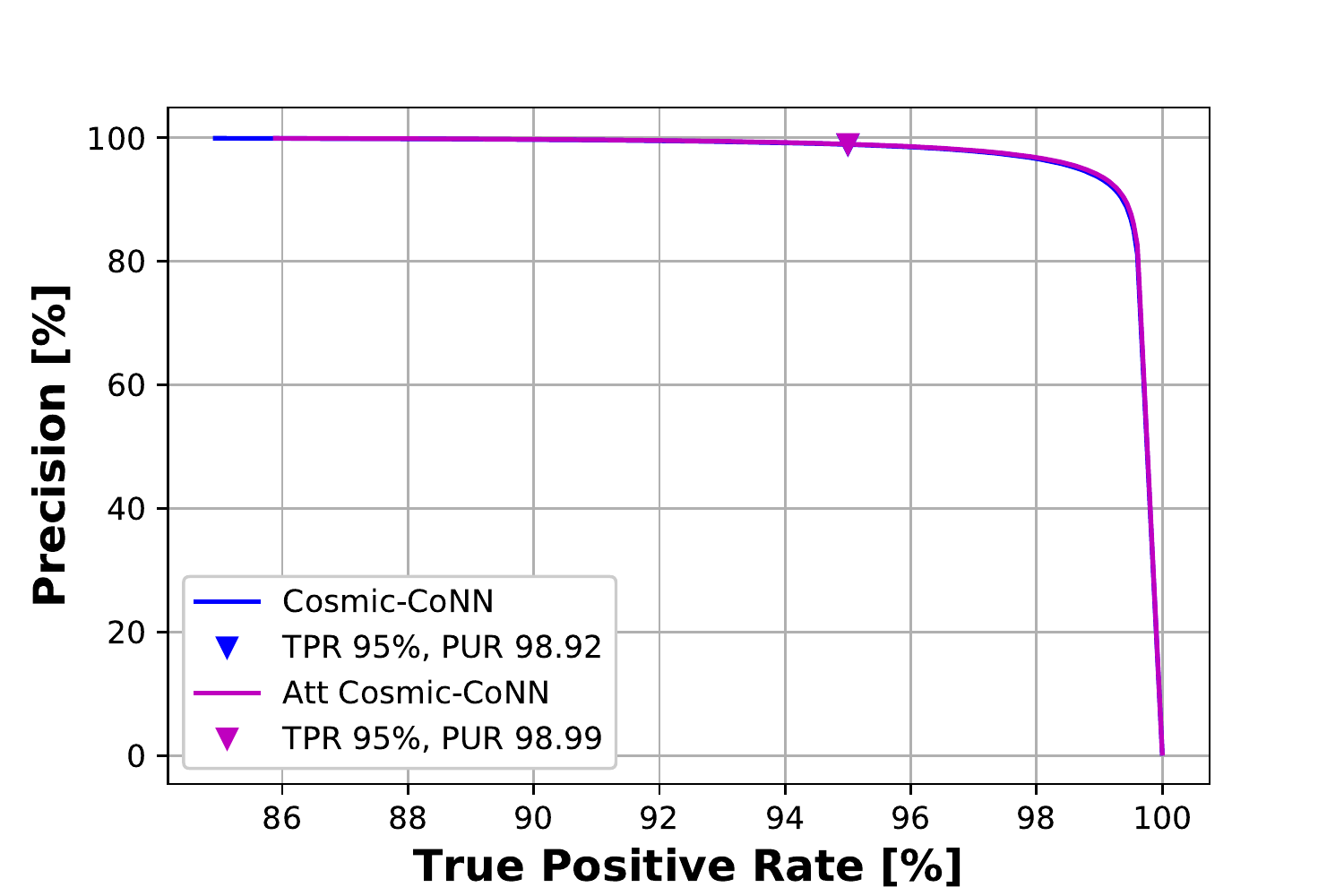}\label{fig:f24}}
  \caption{The ROC and PRC plots on DECam test data with the proposed deepCR based model with and with out attention gate module insertion are in (a) and (b) respectively. The figures (c) and (d) present similar plots with Cosmic-CoNN model. We can notice marginal improvement with AG models in most cases.}
  \label{fig:decam2}
\end{figure*}

\begin{figure*}[!tbp]
  \centering
\subfloat[deepCR - ROC]{\includegraphics[width=0.25\textwidth,
  height=0.25\textwidth, keepaspectratio,]{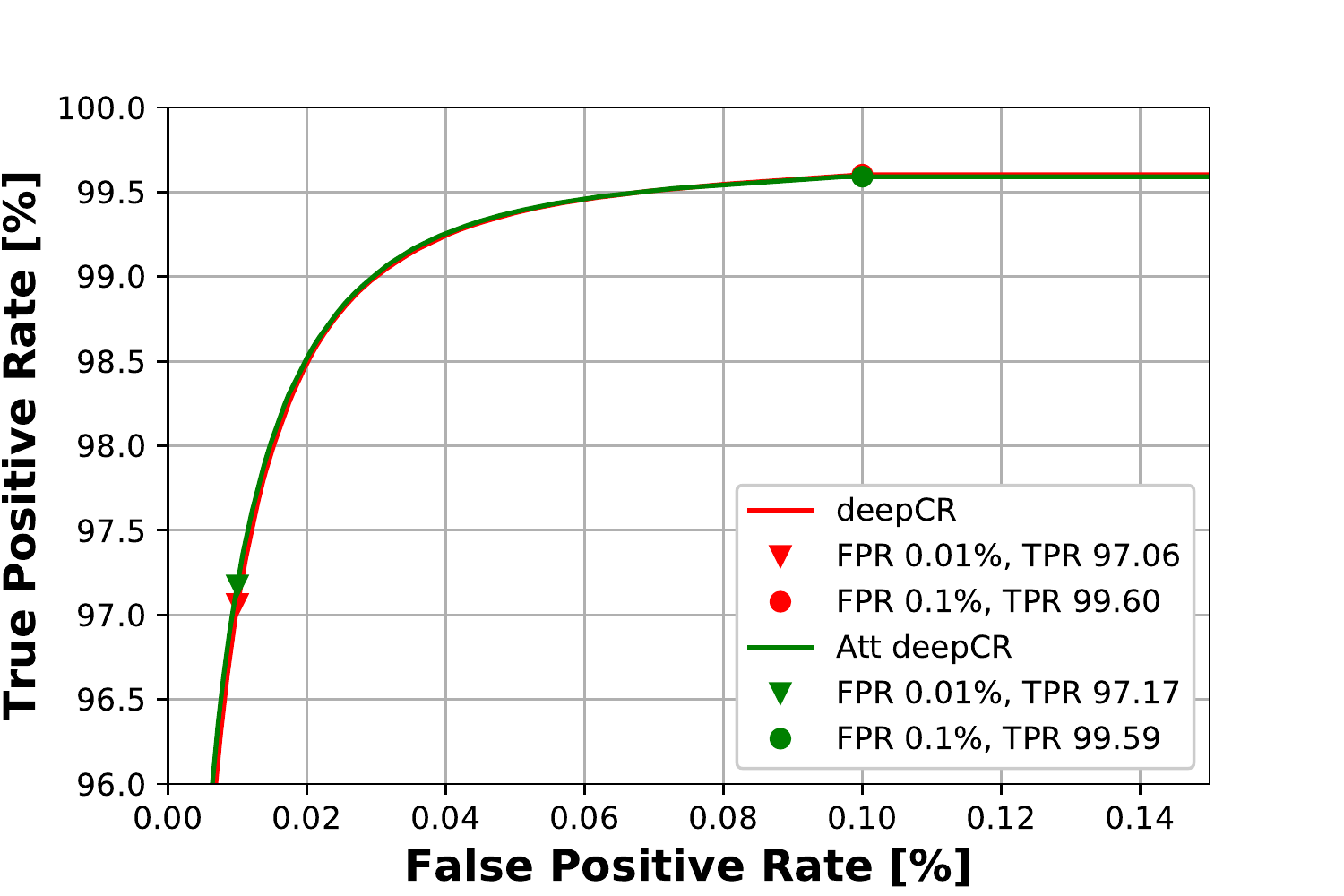}\label{fig:f21_ext}}
  \hfill
  \subfloat[deepCR - PRC]{\includegraphics[width=0.25\textwidth,
  height=0.25\textwidth, keepaspectratio,]{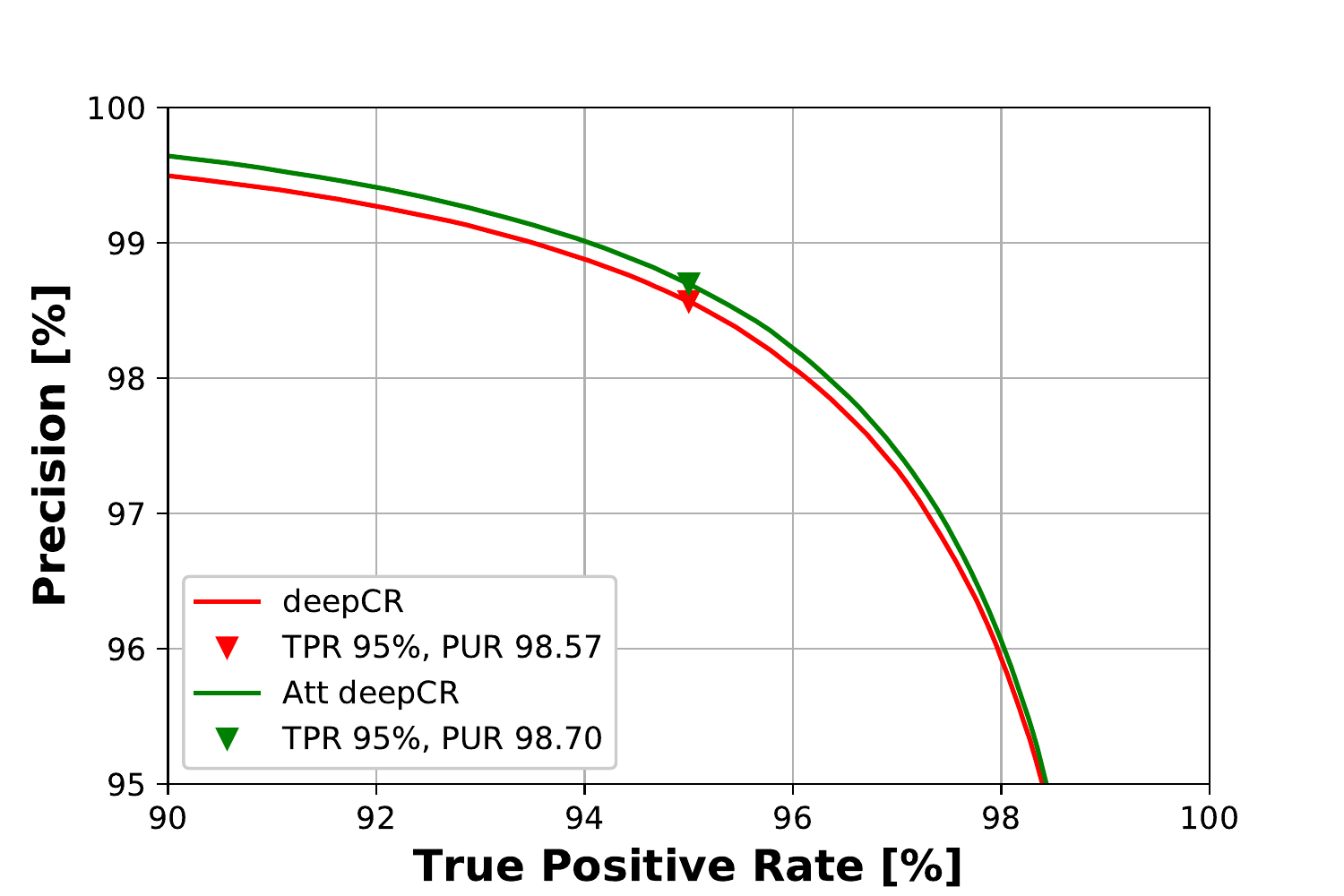}\label{fig:f22_ext}}
  \hfill
  \subfloat[Cosmic-CoNN - ROC]{\includegraphics[width=0.25\textwidth,
  height=0.25\textwidth, keepaspectratio,]{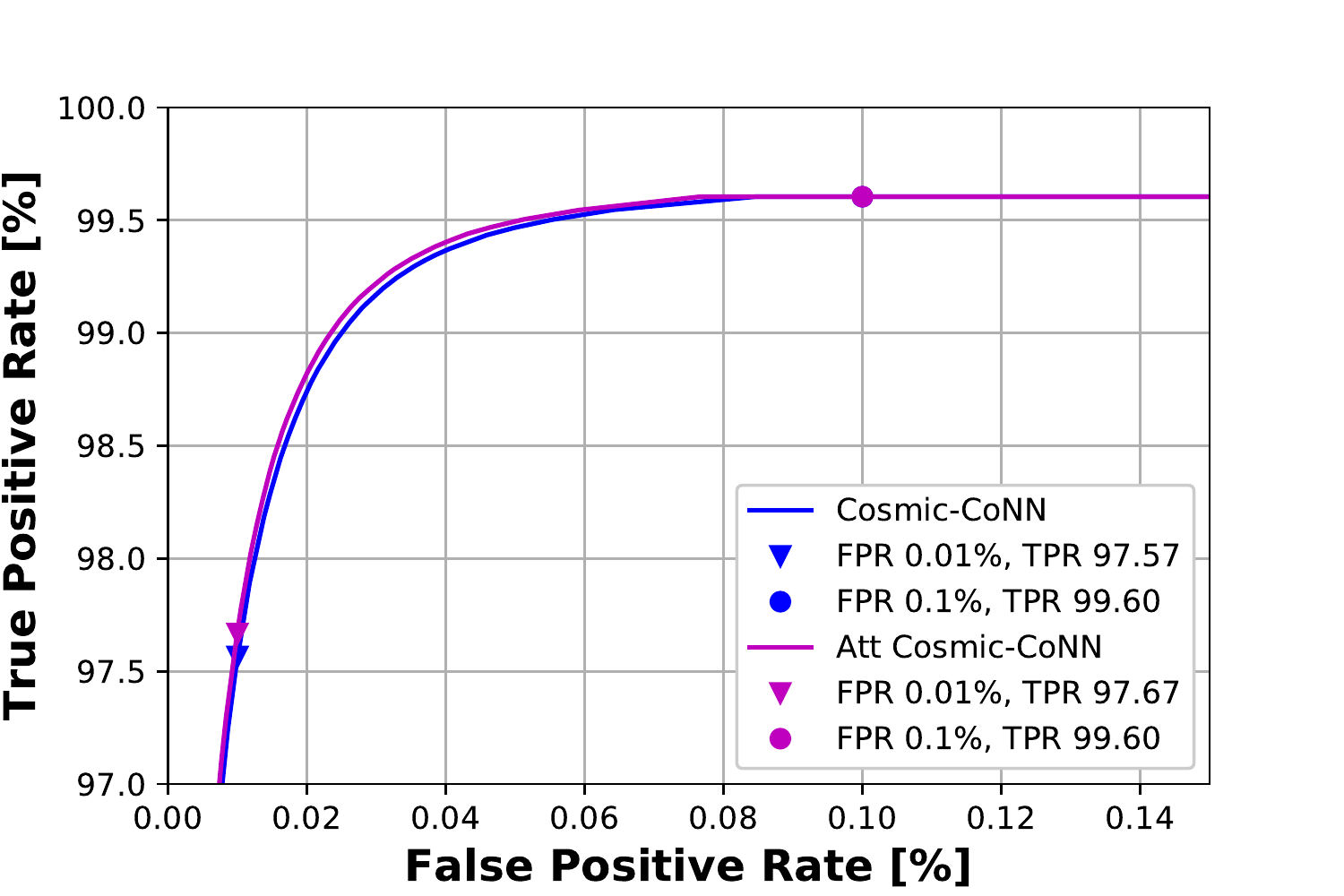}\label{fig:f23_ext}}
  \hfill
  \subfloat[Cosmic-CoNN - PRC]{\includegraphics[width=0.25\textwidth,
  height=0.25\textwidth, keepaspectratio,]{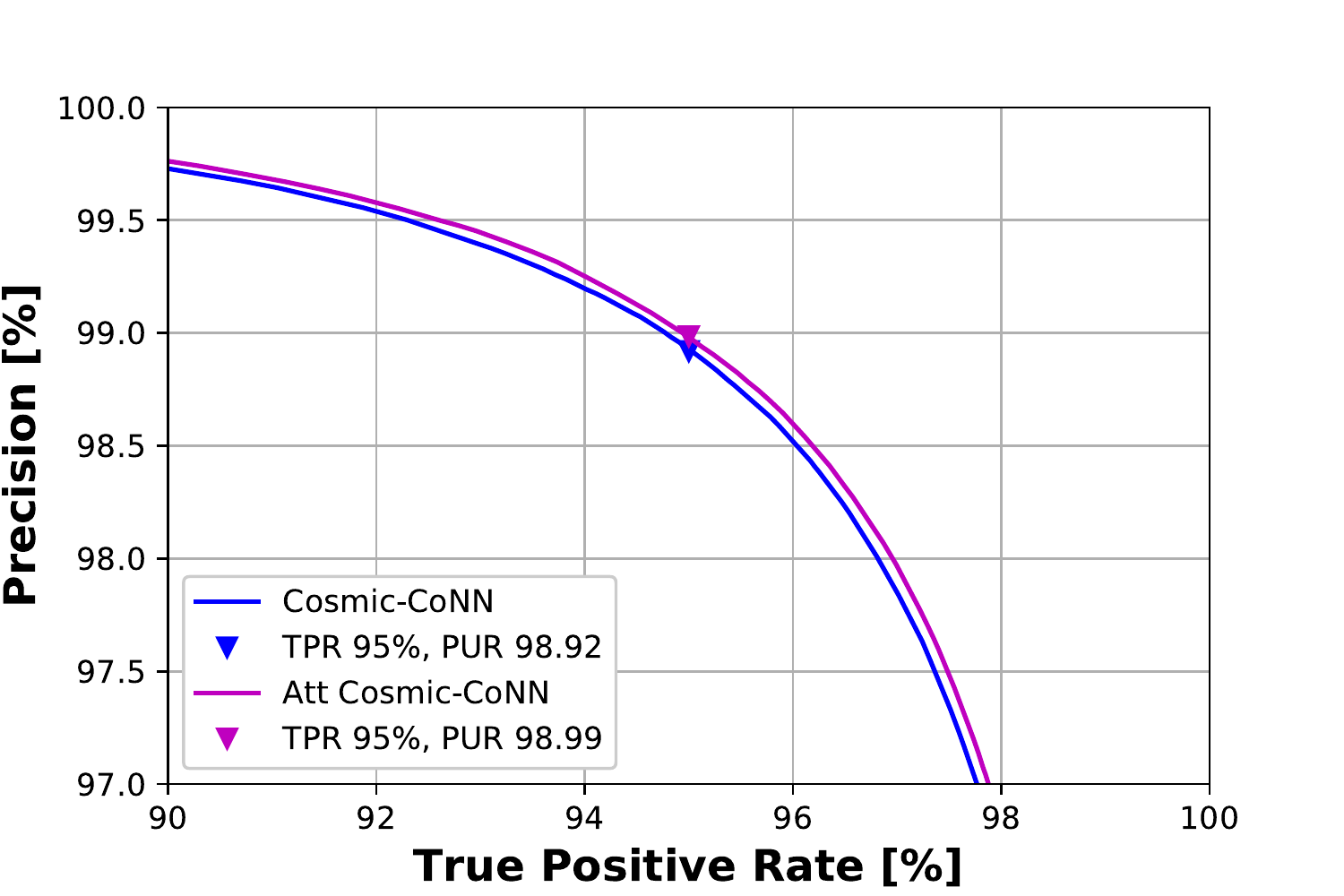}\label{fig:f24_ext}}
  \caption{The extended ROC and PRC plots for a better understanding of the performance of the proposed models from Figure \ref{fig:decam2}. For both deepCR and Cosmic-CoNN models, the attention augmentation is helping in improving the CR detection performance slightly. Moreover, the Cosmic-CoNN model is marginally better than the deepCR model.}
  \label{fig:decam2_ext}
\end{figure*}

\begin{table*}
\centering
\caption{Quantitative findings on CR detection performance with DECam and LCO test datasets using conventional and deep-learning-based models. The True-Positive Rate (TPR) is evaluated at a fixed False-Positive Rate (FPR) of 0.01\%, 0.1\% and Precision (or Purity) is evaluated at a fixed TPR of 95\% to access the model performance. The proposed baseline models with and without AG augmentation outperform all other models with the DECam test data. The marginal gains with adding attention gates to the baseline models can be noticed across most of the metrics listed. For the DECam test data, the Cosmic-CoNN model outperforms the deepCR model with marginal performance gain. The proposed models trained using DECam (mounted on the Victor M. Blanco 4-meter Telescope) data are validated on previously unseen LCO test datasets with images from three different telescope classes separately. These telescope classes include 0.4-meter, 1-meter and 2-meter.
Given that both MaxiMask and our proposed models did not see the LCO data while training, both these models are doing well on the LCO data. However, there are many false positives and false negatives with these models. The proposed models and MaxiMask are better with 1-meter and 2-meter telescope data than 0.4-meter. The classical LACosmic and Astro-SCRAPPY are also doing well on the LCO data across all the telescope classes. These classical algorithms perform better than the deep-learning-based algorithms for the 1-meter and 2-meter data. Almost all the models showed the least performance on the 0.4-meter data for all metrics. Further, the attention models achieve significant performance gain in all data classes with over 50\% TPR at 0.01 \% FPR compared to their corresponding baseline models. \\  \\ \textbf{Note:} The Cosmic-CoNN model was initially trained using the LCO CR dataset, and whenever it is used, we mentioned in the bracket that it is pre-trained. The Cosmic-CoNN and Att Cosmic-CoNN represent our baseline, and attention added models trained using the DECam dataset.}
\begin{tabular*}{0.98\textwidth}{l l l l l}
  \hline\hline    
  Data & Algorithm & TPR at 0.01\% FPR & TPR at 0.1\% FPR & Purity at 95\% TPR \\
   &  & (loss in performance) &  & (loss in performance) \\
  \hline 
  {\em{DECam}} & LACosmic & 60.65 & 96.22 & - \\ 
    & Astro-SCRAPPY & 60.87 & 97.60 & - \\ 
    & MaxiMask & 82.49 & 93.50 & 68.21 \\ 
    & Cosmic-CoNN (pre-trained on LCO) & 94.41 & 98.20 & 96.60 \\ 
    &  deepCR  & 97.06  & 99.60 & 98.57 \\
    & Att deepCR  & 97.17  & 99.59 & 98.70 \\
    & Cosmic-CoNN  & 97.57  & 99.60 & 98.92 \\
    & Att Cosmic-CoNN &  97.69 & 99.60 & 98.99 \\
  \hline\hline
  {\em{LCO-data (0m4)}} & LACosmic & 55.69 & 80.83 & - \\
    & Astro-SCRAPPY & 59.70 & 78.37 & - \\
    & MaxiMask & 37.38 & 62.38 & 0.04 \\
    & deepCR  & 46.53 & 95.24 & 31.13 \\
    & Att deepCR  & 51.56 & 95.48 & 30.00 \\
    & Cosmic-CoNN  & 46.73 & 95.69 & 31.23 \\
    & Att Cosmic-CoNN & 50.09 & 92.89 & 19.31 \\
  \hline\hline
  {\em{LCO-data (1m0)}} & LACosmic & 77.53 & 97.71 & - \\
    & Astro-SCRAPPY & 95.32 & 99.18 & - \\
    & MaxiMask & 86.94 & 99.09 & 54.59 \\
    & deepCR  & 71.24 & 99.32 & 37.00 \\
    & Att deepCR & 74.17 & 99.38 & 41.92 \\
    & Cosmic-CoNN & 83.64 & 99.40 & 63.02 \\
    & Att Cosmic-CoNN & 86.83 & 99.21 & 64.14 \\
  \hline\hline
  {\em{LCO-data (2m0)}} & LACosmic & 86.10 & 99.38 & - \\
    & Astro-SCRAPPY & 86.72 & 99.49 & - \\
    & MaxiMask & 82.34 & 99.46 & 72.84 \\
    & deepCR  & 75.56 & 99.12 & 66.31 \\
    & Att deepCR  & 78.66 & 99.15 & 66.32 \\
    & Cosmic-CoNN  & 80.85 & 99.37 & 68.51 \\
    & Att Cosmic-CoNN & 84.71 & 99.22 & 68.54 \\
  \hline\hline
\end{tabular*}
\label{tab:perf1}
\end{table*}

\begin{table*}
\centering
\caption{\rthis{Quantitative results on CR detection performance using conventional methods, proposed and pre-trained deep-learning models on DECam and LCO test datasets. The first row corresponds to various CR detection models with the DECam data. The second, third, and fourth rows have LCO data for 0.4-meter, 1-meter, and 2-meter telescope classes. The proposed baseline models with and without AG module augmentation perform better on DECam test data than all the other models. Further, the proposed attention-based models show marginal gains across all the metrics listed over the corresponding baselines. The proposed models also perform well on previously unseen LCO data and can generalize well. However, the metrics are not as good as with DECam data. The gain with AG models also can be noticed for each telescope class. For the 1-meter telescope data, the gains with attention models are consistent for all the metrics. The lowest FDR is achieved with the MaxiMask model for DECam and LCO datasets. \\   \textbf{Note:} The Cosmic-CoNN model was initially trained using the LCO CR dataset, and whenever it is used, we have indicated in the parenthesis  that it is pre-trained. The Cosmic-CoNN and Att Cosmic-CoNN represent our baseline, and attention added models trained using the DECam dataset.}}
\label{tab:perf2}  
\begin{tabular*}{0.702\textwidth}{|l | l | l | l | l |}
  \hline    
  Data & Algorithm & F1-Score & IOU & FDR \\
  \hline 
    {\em{DECam}} & LACosmic & 0.8999 & 0.8180 & 0.1249 \\
    & Astro-SCRAPPY & 0.9011 & 0.8200 & 0.1412 \\
    & MaxiMask & 0.8544 & 0.7459 & 0.0129 \\
    & Cosmic-CoNN (pre-trained on LCO) &  0.9574 & 0.9184  & 0.0267 \\
    & deepCR  & 0.9711 & 0.9438 & 0.0329 \\
    & Att deepCR  & 0.9721  & 0.9458  & 0.0281 \\
    & Cosmic-CoNN & 0.9743 & 0.9499  & 0.0250 \\
    & Att Cosmic-CoNN & 0.9749 & 0.9511 & 0.0246 \\
  \hline
    {\em{LCO-data (0m4)}} & LACosmic & 0.4458 & 0.2868 & 0.6711 \\
    & Astro-SCRAPPY & 0.3909 & 0.2429 & 0.7327 \\
    & MaxiMask & 0.4115 & 0.2590 & 0.2463 \\
    & deepCR  &  0.6452 & 0.4763 & 0.4634 \\
    & Att deepCR  & 0.6468 & 0.4780 & 0.4716 \\
    & Cosmic-CoNN  & 0.6423 & 0.4731 & 0.4835 \\
    & Att Cosmic-CoNN & 0.6217 & 0.4511 & 0.5012 \\
  \hline
    {\em{LCO-data (1m0)}} & LACosmic & 0.7656 & 0.6202 & 0.3113 \\
    & Astro-SCRAPPY & 0.8647 & 0.7616 & 0.1607 \\
    & MaxiMask & 0.7258 & 0.5696 & 0.1183 \\
    & deepCR  &  0.5317 & 0.3621 & 0.6309 \\
    & Att deepCR  & 0.5765 & 0.4050 & 0.5871 \\
    & Cosmic-CoNN  & 0.7237 & 0.5670 & 0.4230 \\
    & Att Cosmic-CoNN & 0.7317 & 0.5769 & 0.4127 \\
  \hline
    {\em{LCO-data (2m0)}} & LACosmic & 0.7201 & 0.5627 & 0.4154 \\
    & Astro-SCRAPPY & 0.7642 & 0.6184 & 0.3567 \\
    & MaxiMask & 0.7385 & 0.5854 & 0.1943 \\
    & deepCR  &  0.7737 & 0.6309 & 0.3545 \\
    & Att deepCR  & 0.7717 & 0.6282 & 0.3569 \\
    & Cosmic-CoNN  & 0.7907 & 0.6538 & 0.3372 \\
    & Att Cosmic-CoNN & 0.7856 & 0.6470 & 0.3450 \\
  \hline
\end{tabular*}
\end{table*}

\begin{figure*}
    \centering
    \includegraphics[width=18cm, height=30cm, keepaspectratio,]{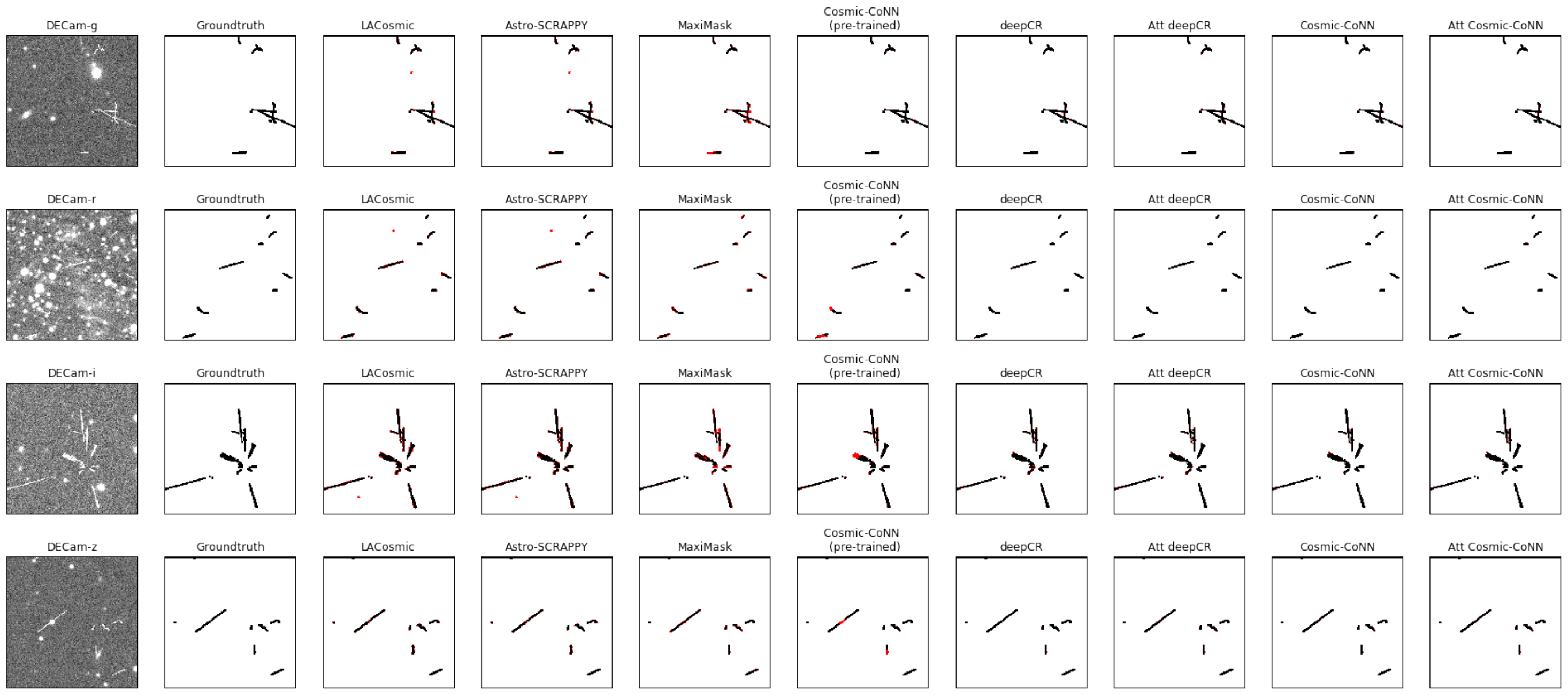}
    \caption{CR Detection discrepancy on DECam imaging data (one image from each of the {\em{'griz'}} DECam photometric bands) with different CR detection algorithms. Incorrect or missing CR pixels are marked in red color. For LACosmic, Astro-SCRAPPY and Cosmic-CoNN (pre-trained on LCO), the output CR masks are dilated using a 3 $\times$ 3 kernel for a fair comparison with the ground truth. Both baseline models with and without attention detect more peripheral CR pixels than other methods. We notice that the LACosmic and Astro-SCRAPPY algorithms provide more False Positives, whereas the MaxiMask and Cosmic-CoNN (pre-trained) models provide more False Negatives. Compared to all other CR detection models, the proposed model's predictions are more faithful to the ground-truth labels.}
    \label{fig:decam_cr}
\end{figure*}

\begin{figure*}
    \centering
    \includegraphics[width=18cm, height=20cm, keepaspectratio,]{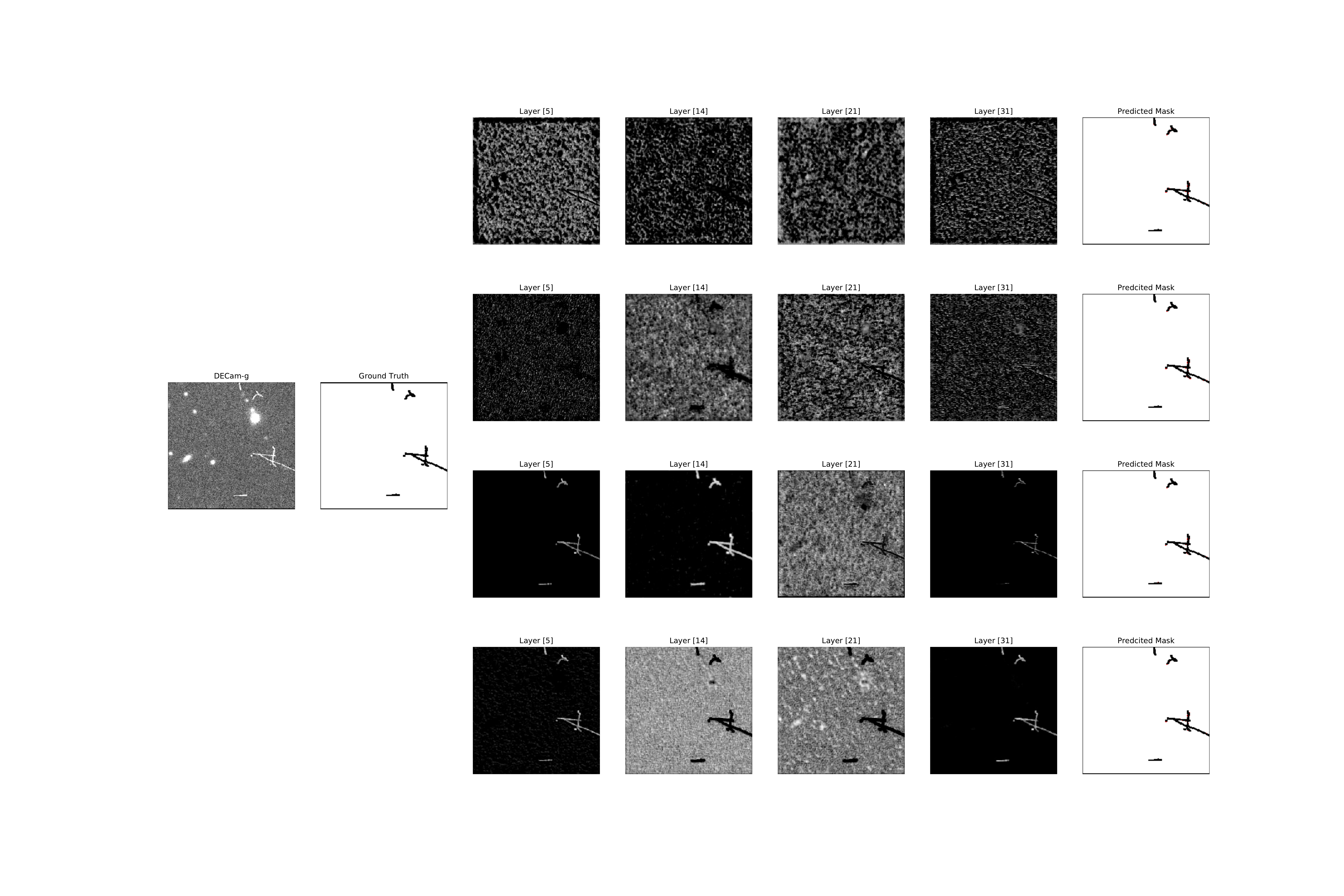}
    \caption{Feature maps obtained on DECam image (DECam-{\em{g}}) at different channels of baseline and attention augmented baseline models. The first row of each image is deepCR without AG augmentation, while the second row is deepCR with augmenting the AG module. Similarly, the third and fourth rows correspond to the Cosmic-CoNN models with and without attention augmentation. The AG models help in better highlighting the image regions contaminated with CR hits than the corresponding baselines.}
    \label{fig:Feature_dc}
\end{figure*}

The ROC and PRC plots obtained on DECam test data with LACosmic, Astro-SCRAPPY and previously trained deep-learning models such as MaxiMask and Cosmic-CoNN (pre-trained) are shown in Fig.~\ref{fig:decam1} along with the extended plots in Fig.~\ref{fig:decam1_ext}, for better understanding the model's performance. Similar plots with the proposed baseline and AG augmented baseline models are shown in Fig.~\ref{fig:decam2} and Fig.~\ref{fig:decam2_ext}. The quantitative evaluation of the CR detection models is presented in the first rows of Table \ref{tab:perf1} and Table \ref{tab:perf2}. Specifically, Table \ref{tab:perf1} shows the TPR at 0.01\% and 0.1\% FPR, and the Purity at 95\% TPR. From Table~\ref{tab:perf1}, the proposed models provide better performance in all the cases compared to previous CR detection algorithms. Also, the \rthis{marginal} improvement offered by the attention augmented models are clear and consistent for most cases across the DECam test data. As with TPR and Purity, the attention augmented models are consistently better than the corresponding baselines and for most cases. Between the two baseline models considered, the Cosmic-CoNN model performs slightly better than the deepCR model. The pre-trained MaxiMask and Cosmic-CoNN models perform well on the DECam test data. Given that the pre-trained Cosmic-CoNN model did not see the DECam data while training, it provides better performance on DECam test data when compared to the conventional LACosmic and Astro-SCRAPPY and serves as a generic model for CR detection. However, both MaxiMask and Cosmic-CoNN (pre-trained) underperform compared to the proposed models. 

Similarly, Table \ref{tab:perf2} lists the F1 score, IOU and FDR values for the models considered in this work on DECam test data. These metrics are evaluated at \texttt{threshold=0.5} for all deep-learning based models except for the MaxiMask, which uses \texttt{threshold=0.43} (this threshold is doing better with the MaxiMask model through their experiments). Similarly, for LACosmic and Astro-SCRAPPY, we choose the following parameters \texttt{sigclip=3.0}, \texttt{objlim=1.0} and \texttt{sigfrac=0.1}. These numbers show that the proposed models perform better than the previous algorithms. Further, attention augmentation offers consistent \rthis{marginal} gains in this scenario over the corresponding baselines. However, the lowest FDR is achieved through the MaxiMask model on DECam test data. 

The qualitative analysis of the performance of each CR detection model is compared and illustrated in Fig.~\ref{fig:decam_cr}. To understand the performance better, we used different images from each photometric band {\em{'griz'}} of the DECam test data. From Fig.~\ref{fig:decam_cr}, the proposed baseline models with and without attention augmentation detect more CR peripherals than the other algorithms. We can notice that the image-filtering-based algorithms (such as LACosmic and Astro-SCRAPPY) provide more False Positives. In contrast, the other learning-based algorithms (pre-trained MaxiMask and Cosmic-CoNN) provide more False Negatives. Further, we provide a qualitative explanation for the performance improvement due to attention augmentation. Specifically, we would like to draw attention to the model from Fig.~\ref{fig:network}. The cost of these gains is a nominal 1\% increase in the number of trainable parameters due to the attention gates introduced into the model architecture. The feature maps obtained on DECam images at different layers of the baseline and AG augmented baseline models are illustrated in Fig.~\ref{fig:Feature_dc}. The AG models give the pixels higher weightage with CR hits than the Non-CR pixels. \rthis{Specifically, this can be noticed from the last layer of the network (layer 31) from Fig.~\ref{fig:Feature_dc}.}


\subsection{Performance on LCOGT Network Images}
Next, we evaluated the performance of the proposed models on previously unseen data using the LCO CR test dataset. \rthis{Test data from the LCO CR dataset are from three distinct telescope imagers with different diameters, namely 0.4-meter, 1-meter, and 2-meter telescopes.} First, we evaluated the CR detection performance on the LCO test data with previously developed CR detection models such as LACosmic, Astro-SCRAPPY, and MaxiMask. The Cosmic-CoNN (pre-trained) model was initially trained using LCOGT Network images, and further details on model performance can be found in ~\citet{xu2021cosmic}. \rthis{The LCO CR test data consists of 119 images with 55, 51, and 13 images from 0.4-meter, 1-meter and 2-meter telescopes, respectively.} All these images consist of  three consecutive exposures  with the same exposure time, which is 100 seconds or longer. These images are of different sizes including, $4K \times 4K$, $3K \times 2K$ or $2K \times 2K$. \rthis{We have used all the available sequences from the LCO test data in our experiments.} For LACosmic and Astro-SCRAPPY, we used \texttt{objlim=2.0} for LCO 1.0-meter and 2.0-meter telescopes’ data and \texttt{objlim=0.5} for 0.4-meter telescope data for optimal performance in different telescope classes. \texttt{sigfrac=0.1} is held constant for all telescope classes, and we produce the ROC curves by varying the sigclip parameter between [1, 20]. While evaluating Astro-SCRAPPY and LACosmic, we adopted similar parameters from ~\citet{xu2021cosmic}.


\begin{figure*}[!tbp]
  \centering
  \subfloat[LACosmic - ROC]{\includegraphics[width=0.25\textwidth,
  height=0.25\textwidth, keepaspectratio,]{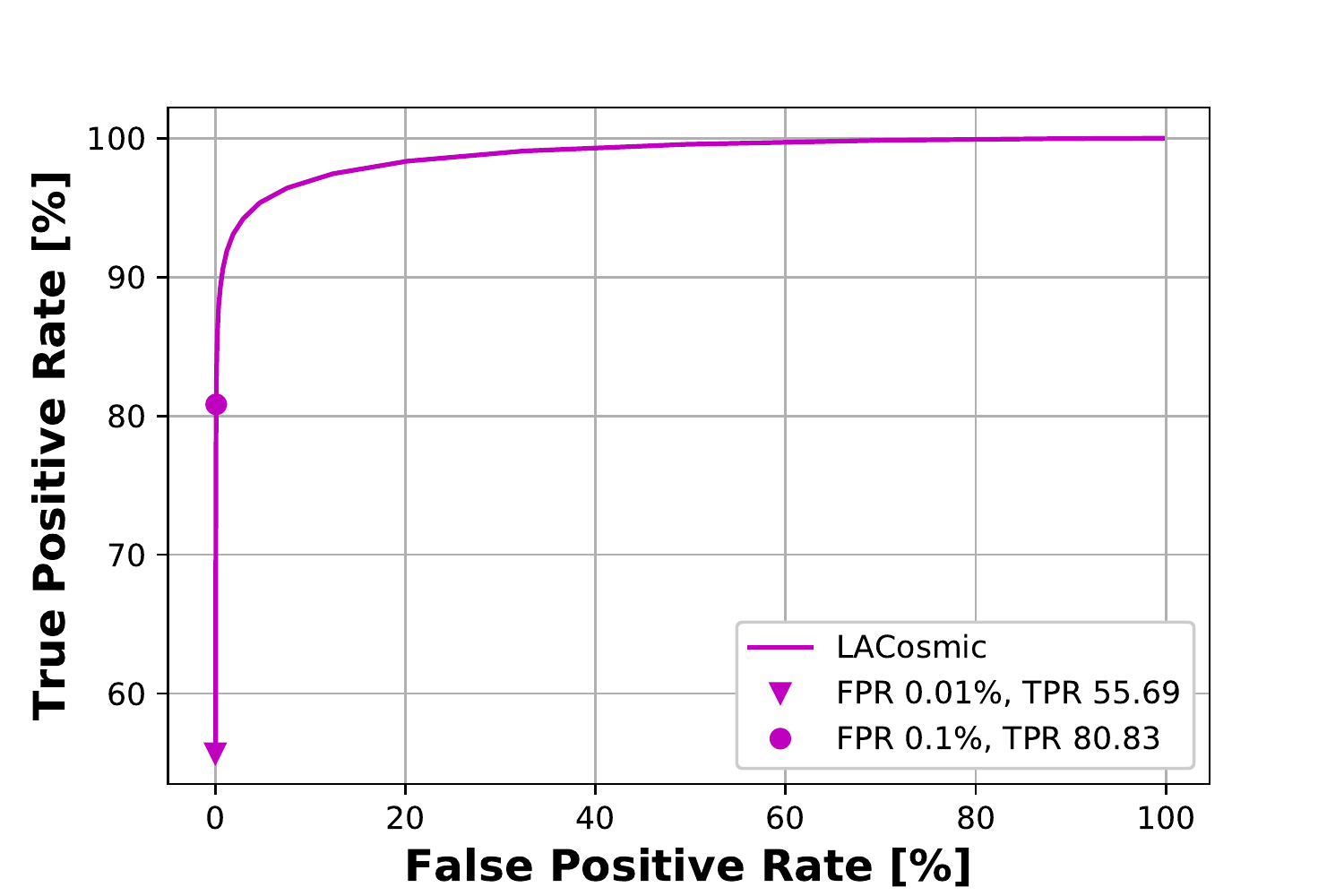}\label{fig:f71}}
  \hfill
  \subfloat[Astro-SCRAPPY - ROC]{\includegraphics[width=0.25\textwidth,
  height=0.25\textwidth, keepaspectratio,]{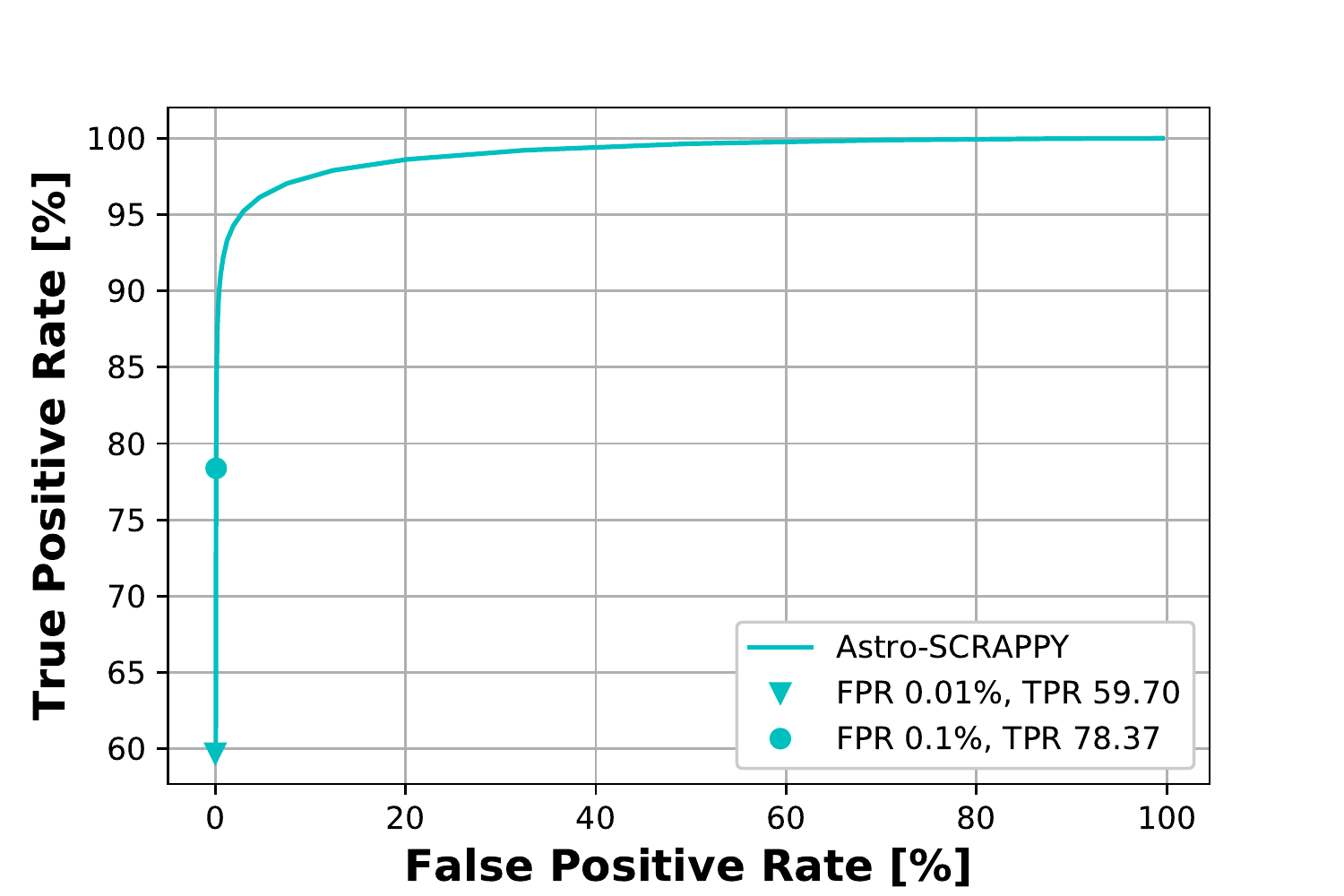}\label{fig:f72}}
  \hfill
  \subfloat[Pre-trained models - ROC]{\includegraphics[width=0.25\textwidth,
  height=0.25\textwidth, keepaspectratio,]{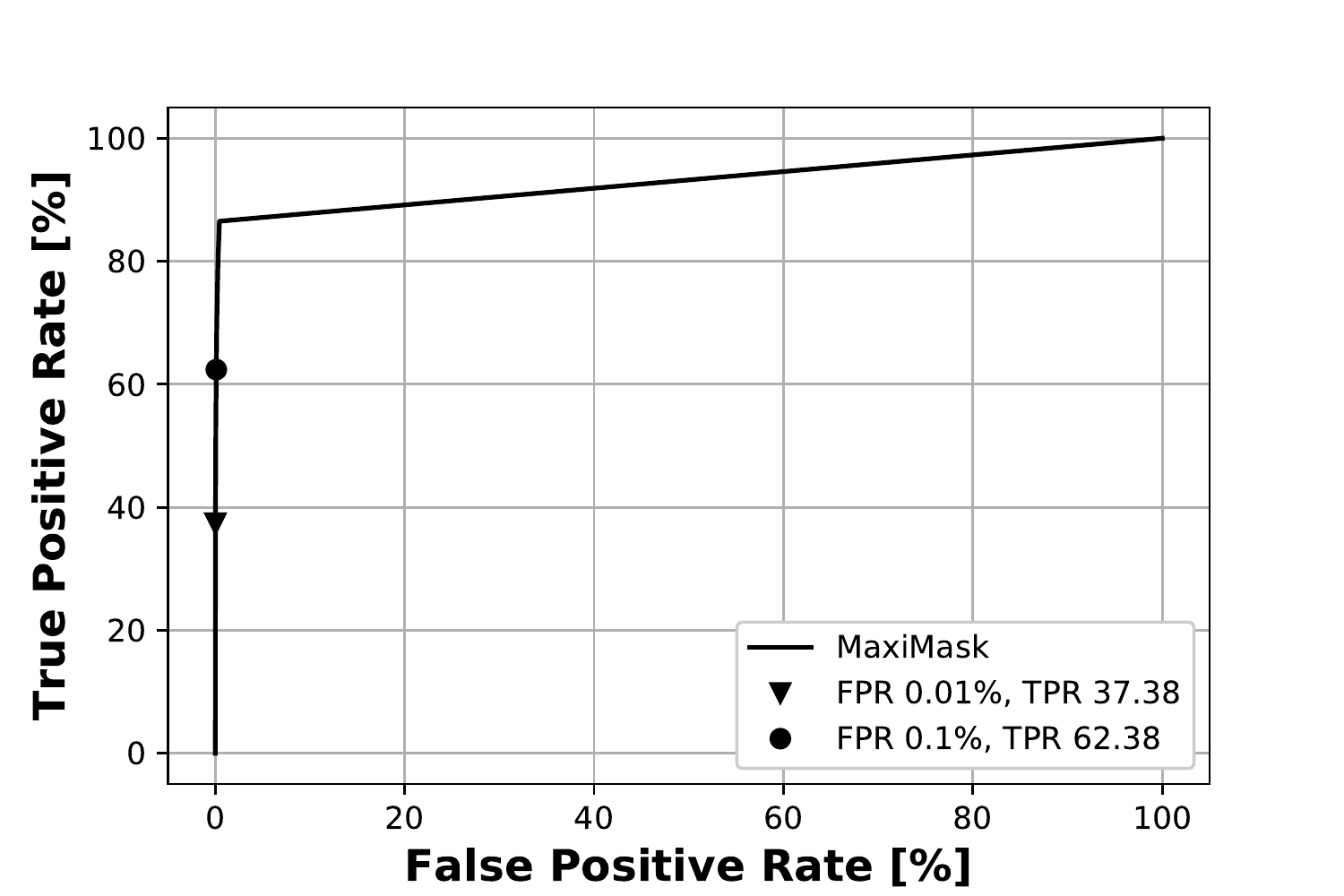}\label{fig:f73}}
  \hfill
  \subfloat[Pre-trained - PRC]{\includegraphics[width=0.25\textwidth,
  height=0.25\textwidth, keepaspectratio,]{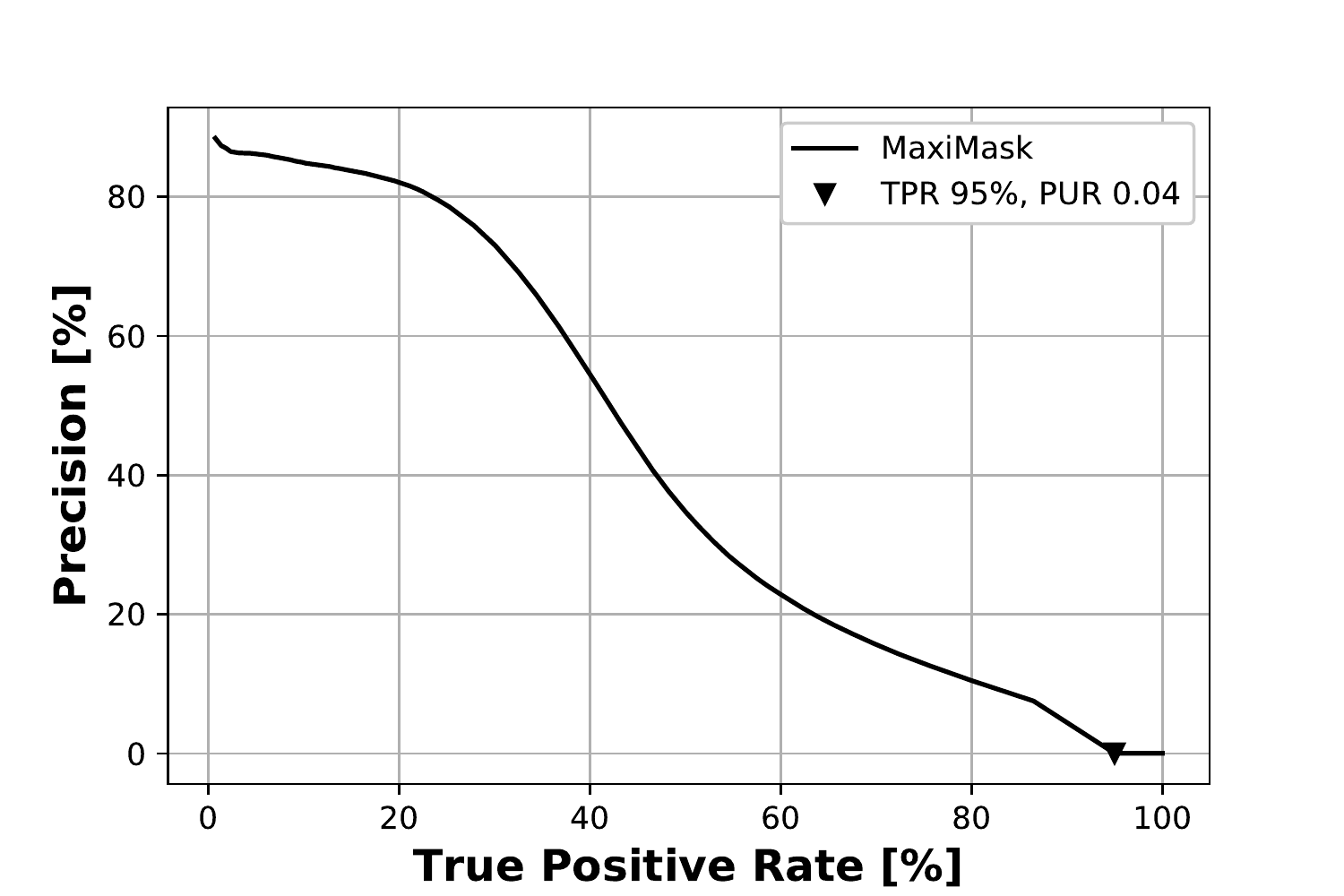}\label{fig:f74}}
  \caption{The ROC and PRC plots obtained with traditional CR detection algorithms, LACosmic and Astro-SCRAPPY on data with 0.4-meter telescope from LCO CR test dataset are presented in (a) and (b). The performance on the same data with pre-trained MaxiMask algorithm is presented with ROC and PRC plots in (c) and (d) respectively. The MaxiMask model has least performance than the LACosmic and Astro-SCRAPPY.}
  \label{fig:lco04m_prev}
\end{figure*}

\begin{figure*}[!tbp]
  \centering
  \subfloat[LACosmic - ROC]{\includegraphics[width=0.25\textwidth,
  height=0.25\textwidth, keepaspectratio,]{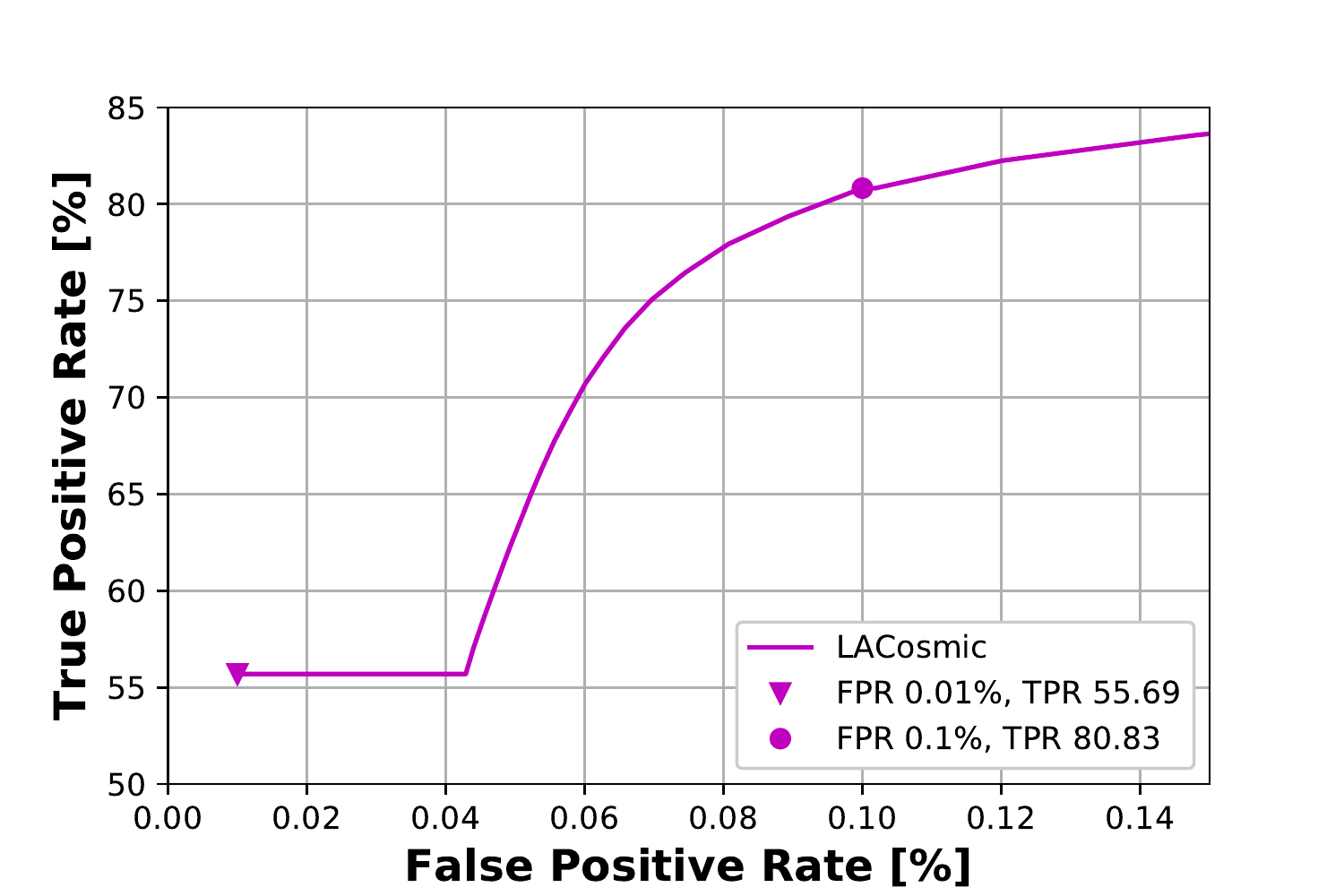}\label{fig:f71_ext}}
  \hfill
  \subfloat[Astro-SCRAPPY - ROC]{\includegraphics[width=0.25\textwidth,
  height=0.25\textwidth, keepaspectratio,]{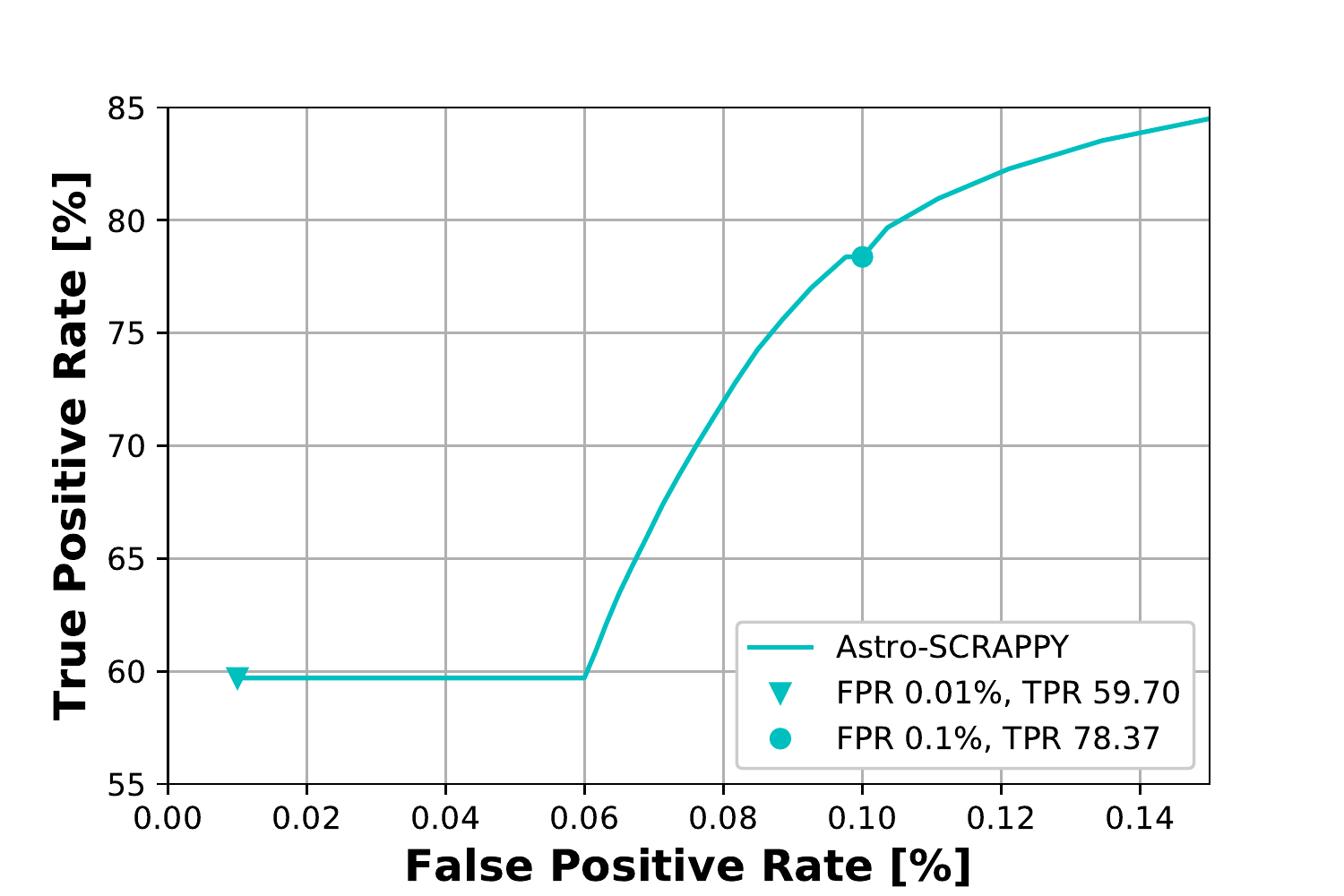}\label{fig:f72_ext}}
  \hfill
  \subfloat[Pre-trained models - ROC]{\includegraphics[width=0.25\textwidth,
  height=0.25\textwidth, keepaspectratio,]{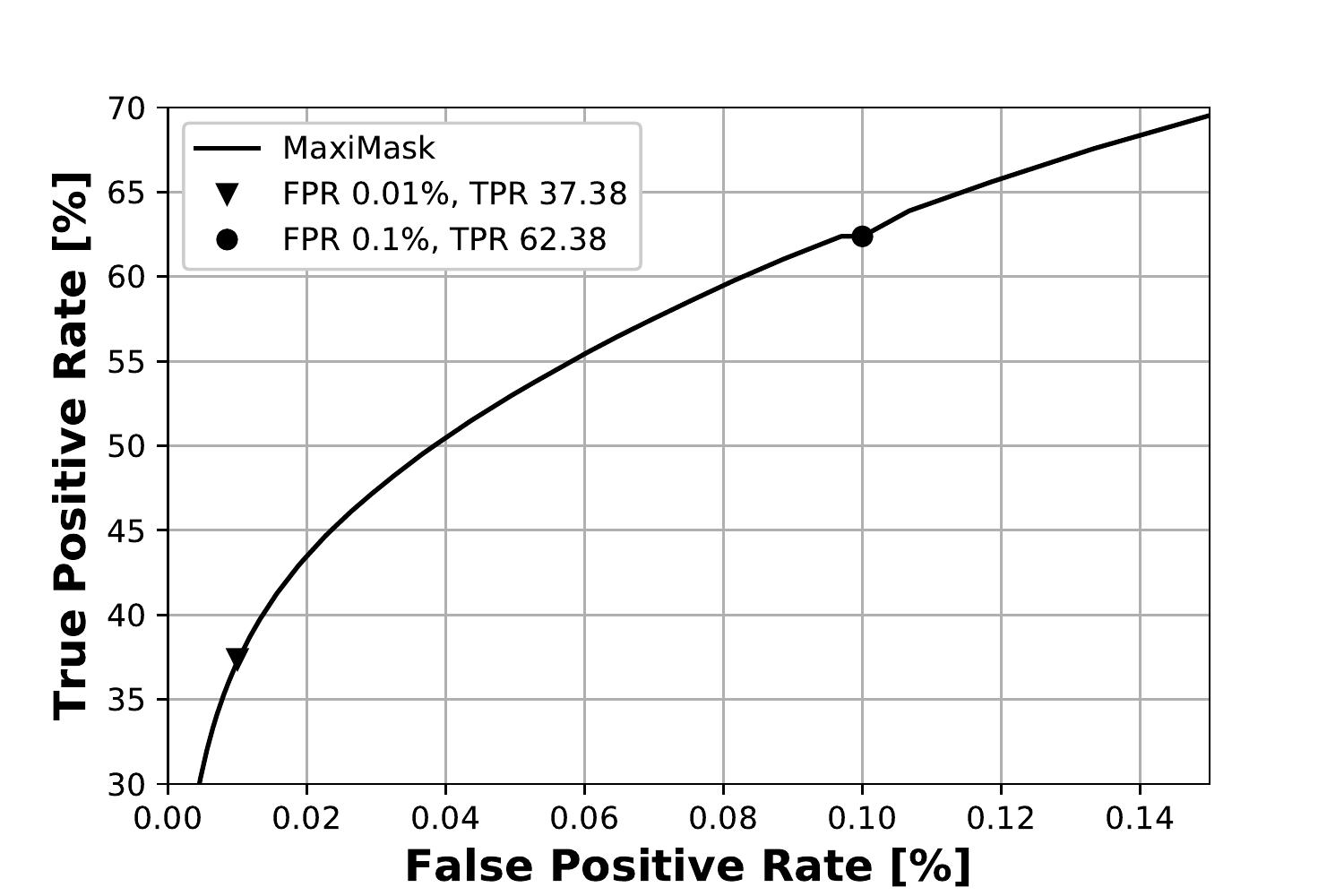}\label{fig:f73_ext}}
  \hfill
  \subfloat[Pre-trained - PRC]{\includegraphics[width=0.25\textwidth,
  height=0.25\textwidth, keepaspectratio,]{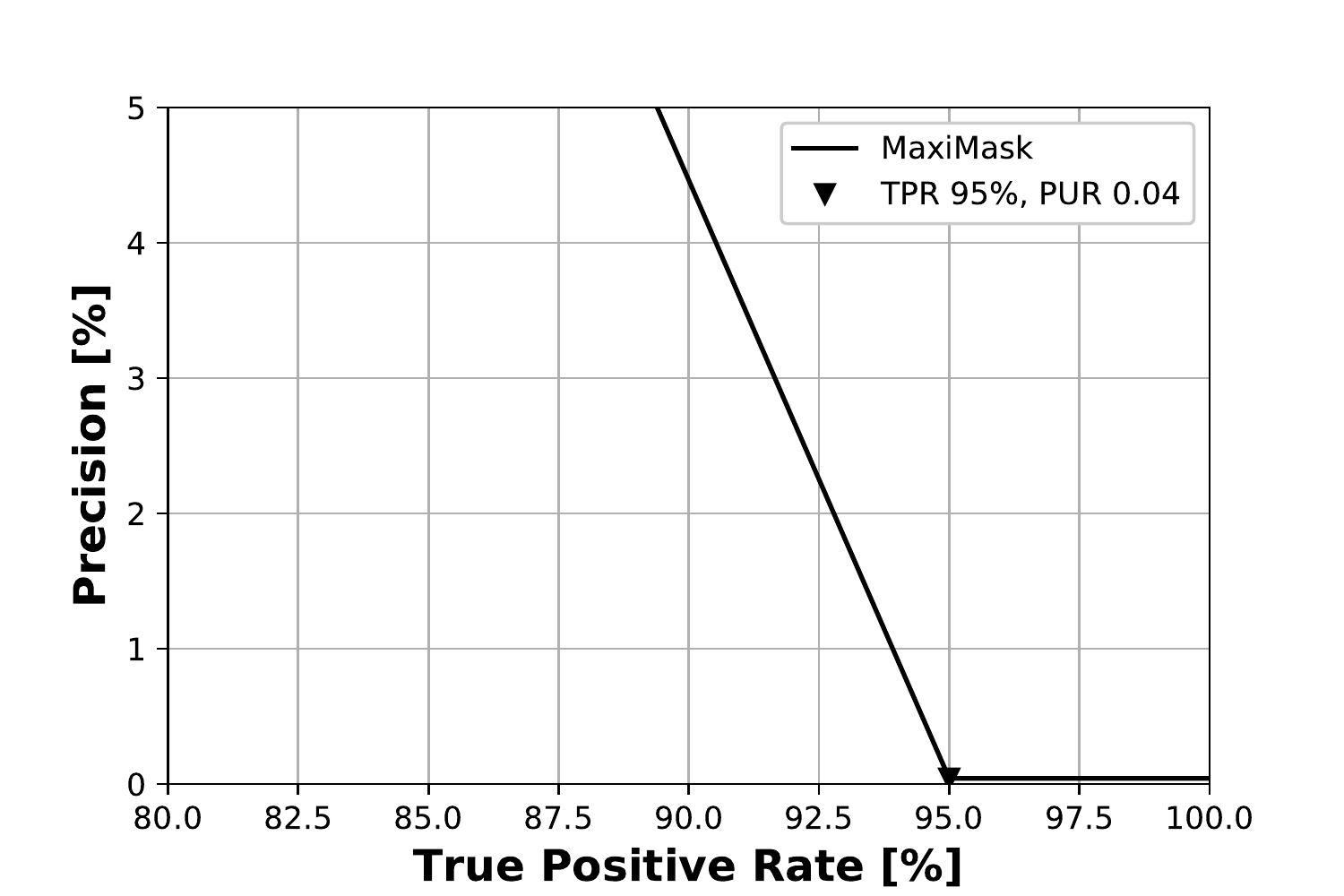}\label{fig:f74_ext}}
  \caption{The extended ROC and PRC plots for a better understanding of the model performances from Fig.~\ref{fig:lco04m_prev}. The high performance in CR detection is noticed with the classical CR detection models. The performance is poor on LCO 0.4-meter telescope data with the MaxiMask model.}
  \label{fig:lco04m_prev_ext}
\end{figure*}

\begin{figure*}[!tbp]
  \centering
  \subfloat[deepCR - ROC]{\includegraphics[width=0.25\textwidth,
  height=0.25\textwidth, keepaspectratio,]{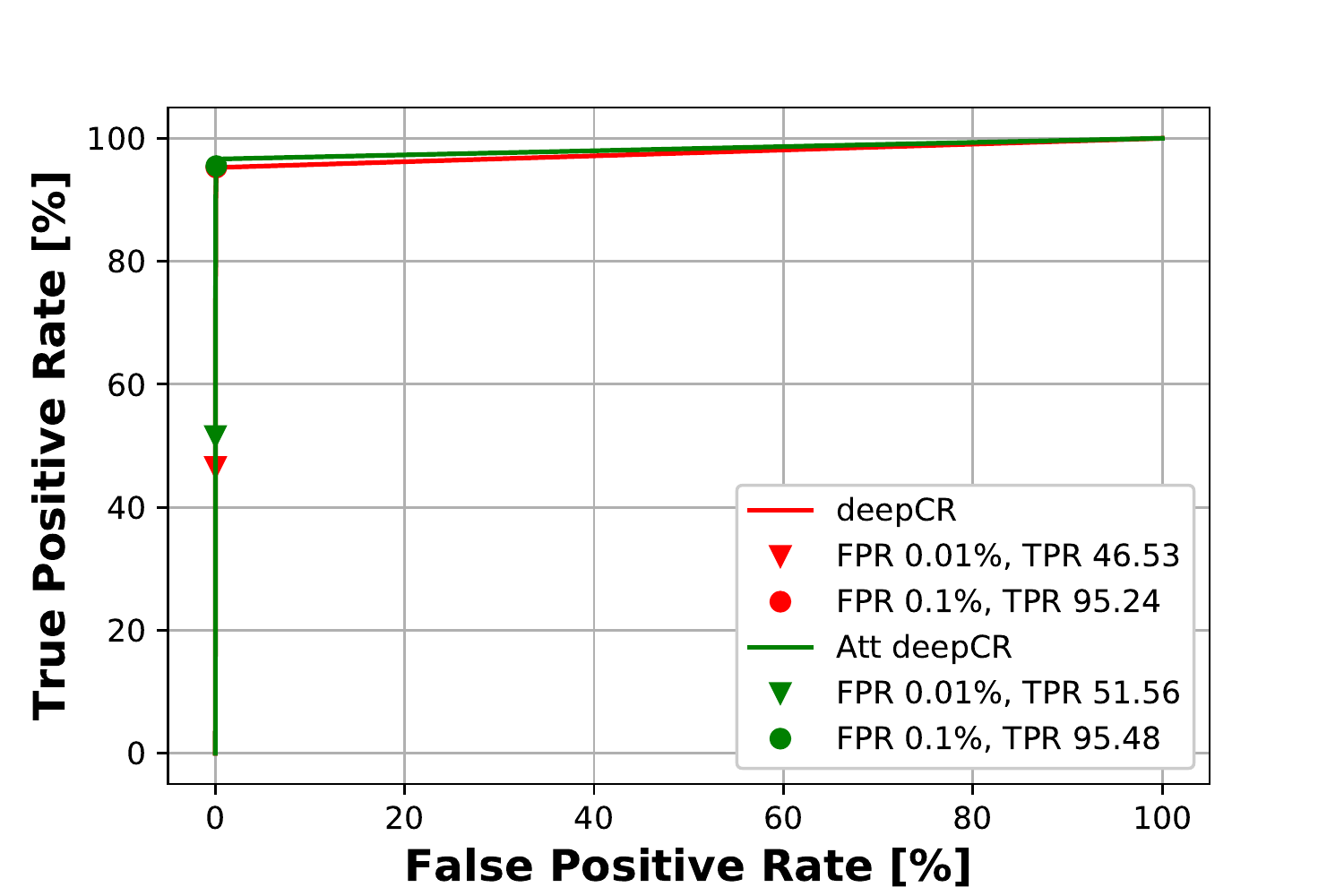}\label{fig:f81}}
  \hfill
  \subfloat[deepCR - PRC]{\includegraphics[width=0.25\textwidth,
  height=0.25\textwidth, keepaspectratio,]{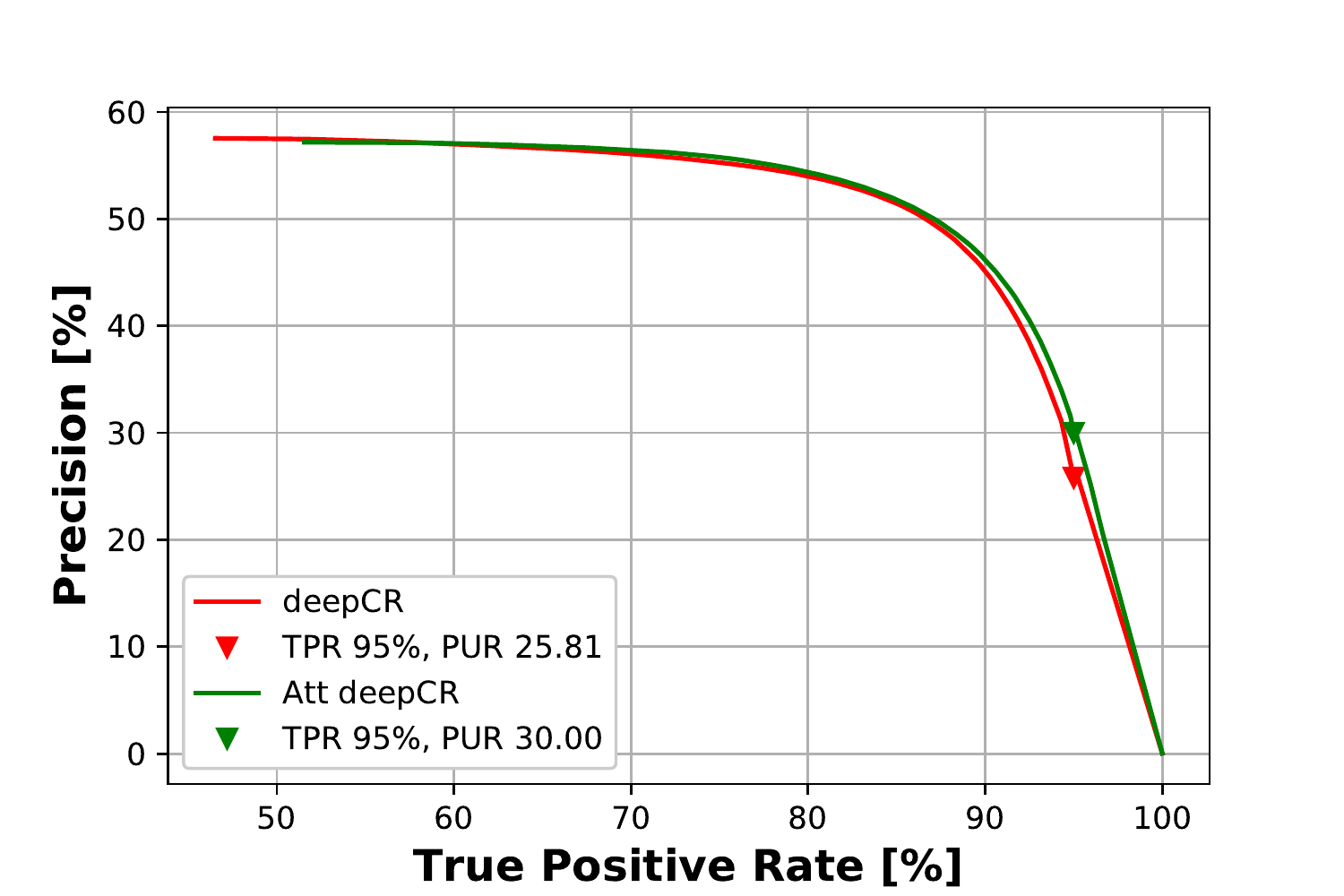}\label{fig:f82}}
  \hfill
  \subfloat[cosmic-CoNN - ROC]{\includegraphics[width=0.25\textwidth,
  height=0.25\textwidth, keepaspectratio,]{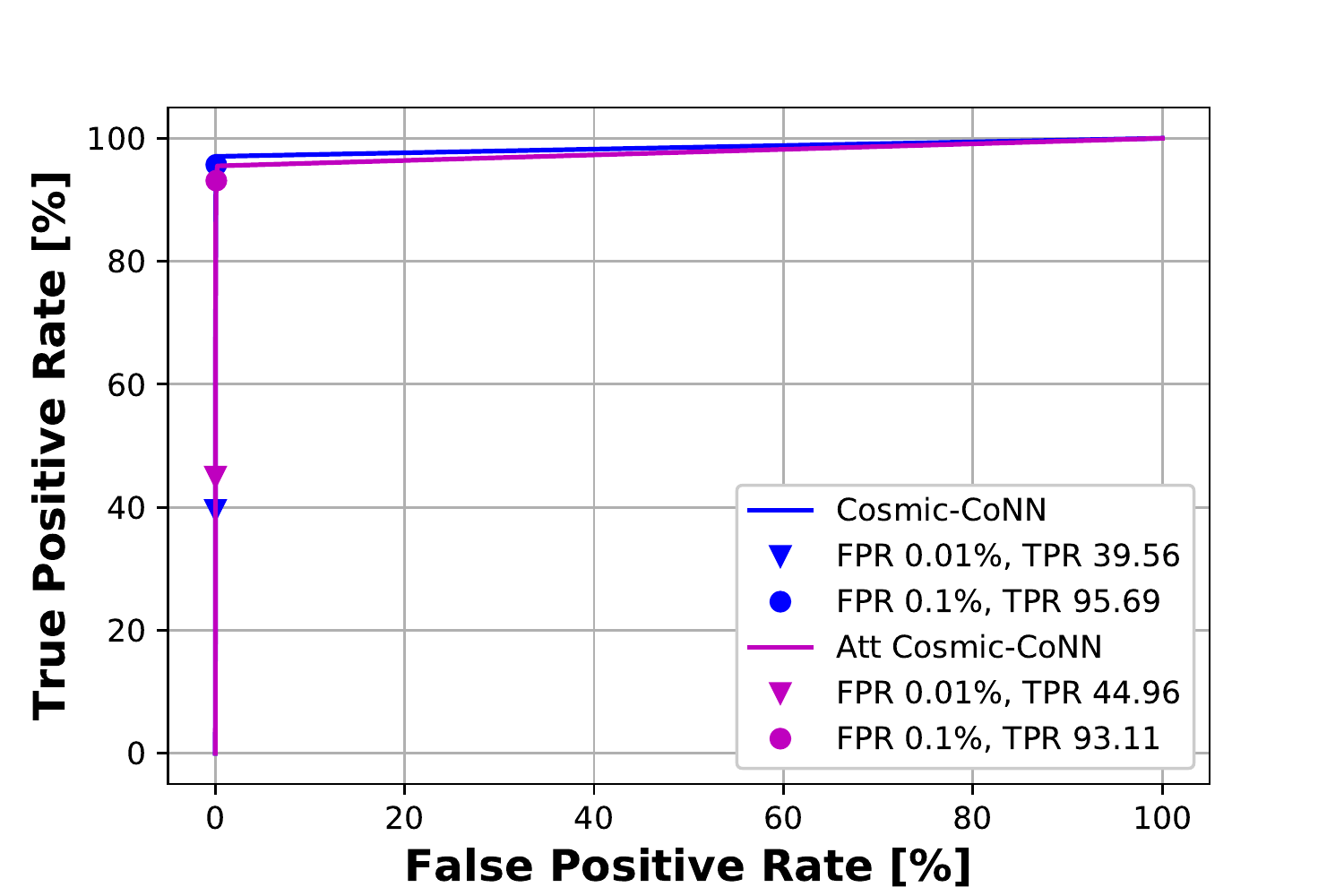}\label{fig:f83}}
  \hfill
  \subfloat[Cosmic-CoNN - PRC]{\includegraphics[width=0.25\textwidth,
  height=0.25\textwidth, keepaspectratio,]{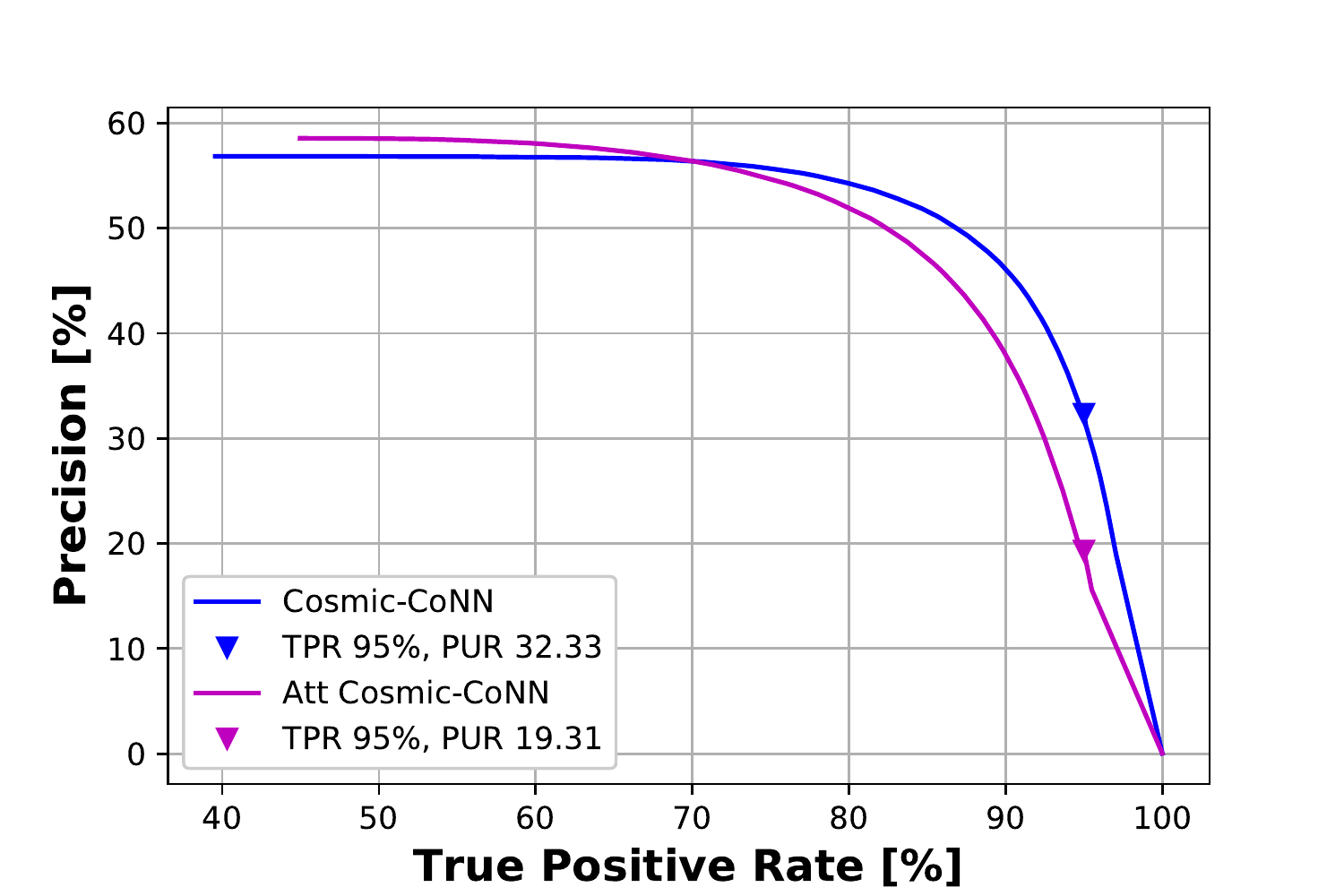}\label{fig:f84}}
  \caption{Performance of the proposed models on previously unseen data from the LCO CR test dataset using the 0.4-meter telescope data is presented here. (a) and (b) are the ROC and PRC plots on deepCR models with and without attention module insertion. Similar plots for the Cosmic-CoNN models are presented in (c) and (d). The deepCR model performs well on this data compared to the Cosmic-CoNN model in most cases. Also, the deepCR model benefits more from adding AGs than the Cosmic-CoNN model when testing on this data.}
  \label{fig:lco04m_dl}
\end{figure*}

\begin{figure*}[!tbp]
  \centering
  \subfloat[deepCR - ROC]{\includegraphics[width=0.25\textwidth,
  height=0.25\textwidth, keepaspectratio,]{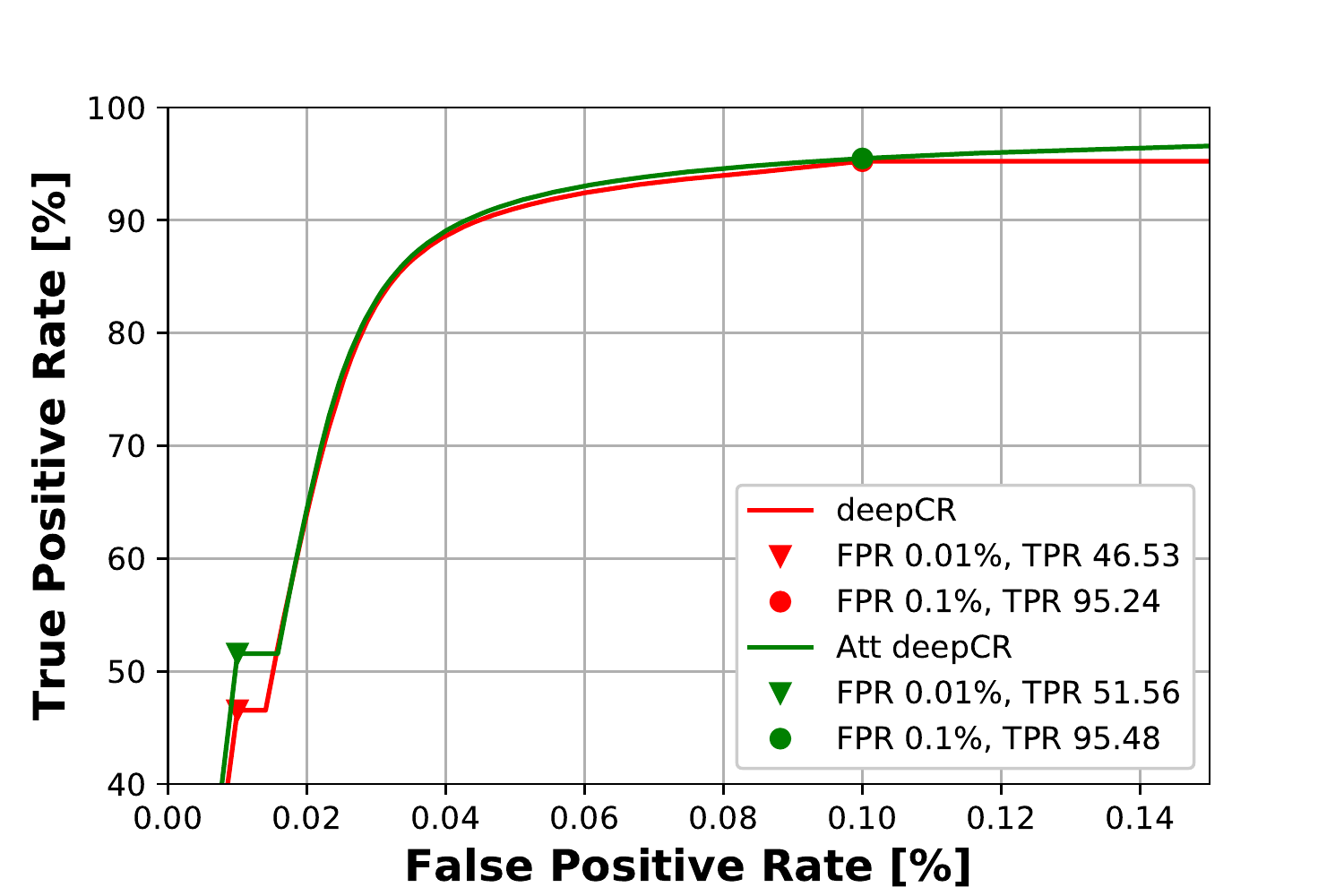}\label{fig:f81_ext}}
  \hfill
  \subfloat[deepCR - PRC]{\includegraphics[width=0.25\textwidth,
  height=0.25\textwidth, keepaspectratio,]{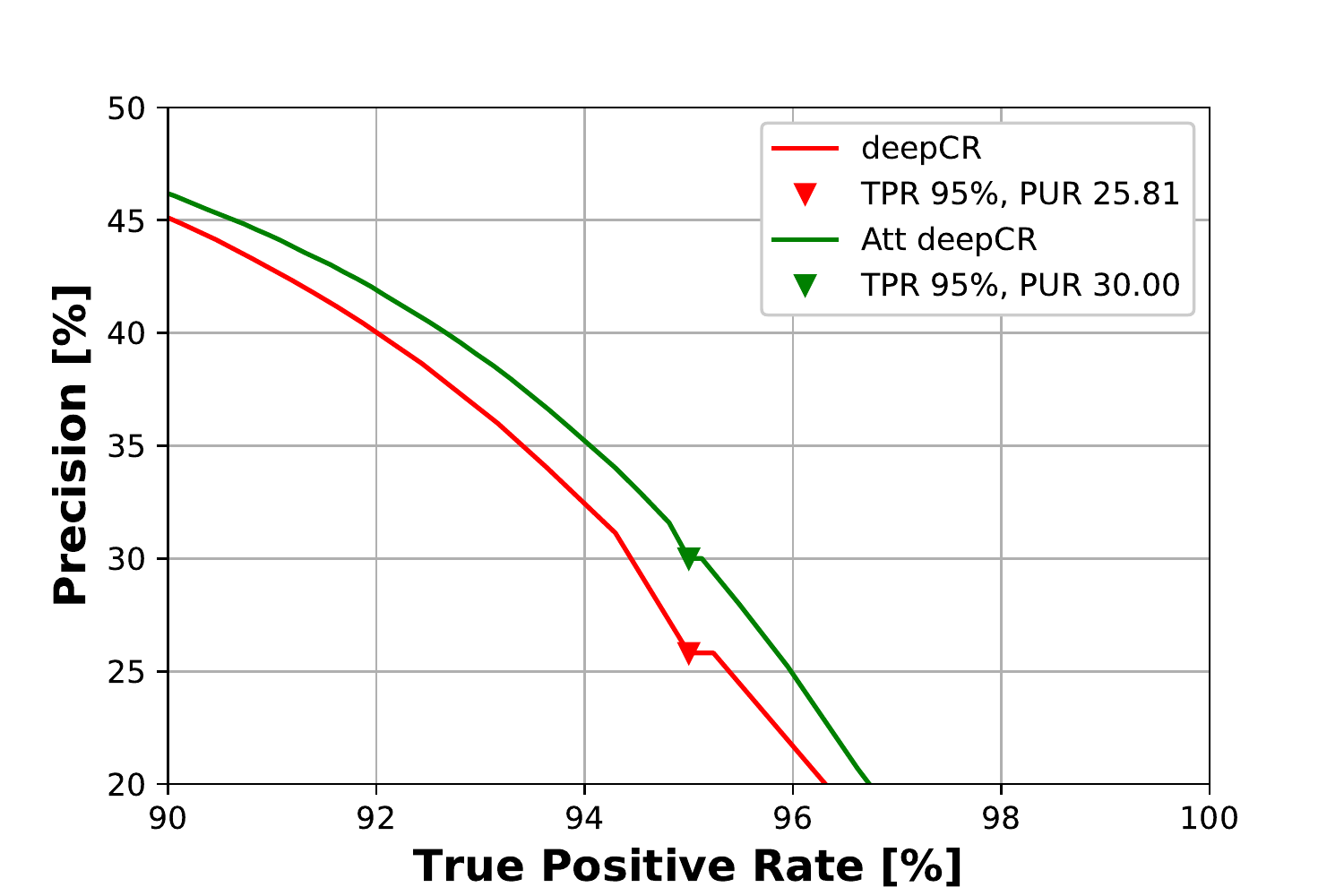}\label{fig:f82_ext}}
  \hfill
  \subfloat[cosmic-CoNN - ROC]{\includegraphics[width=0.25\textwidth,
  height=0.25\textwidth, keepaspectratio,]{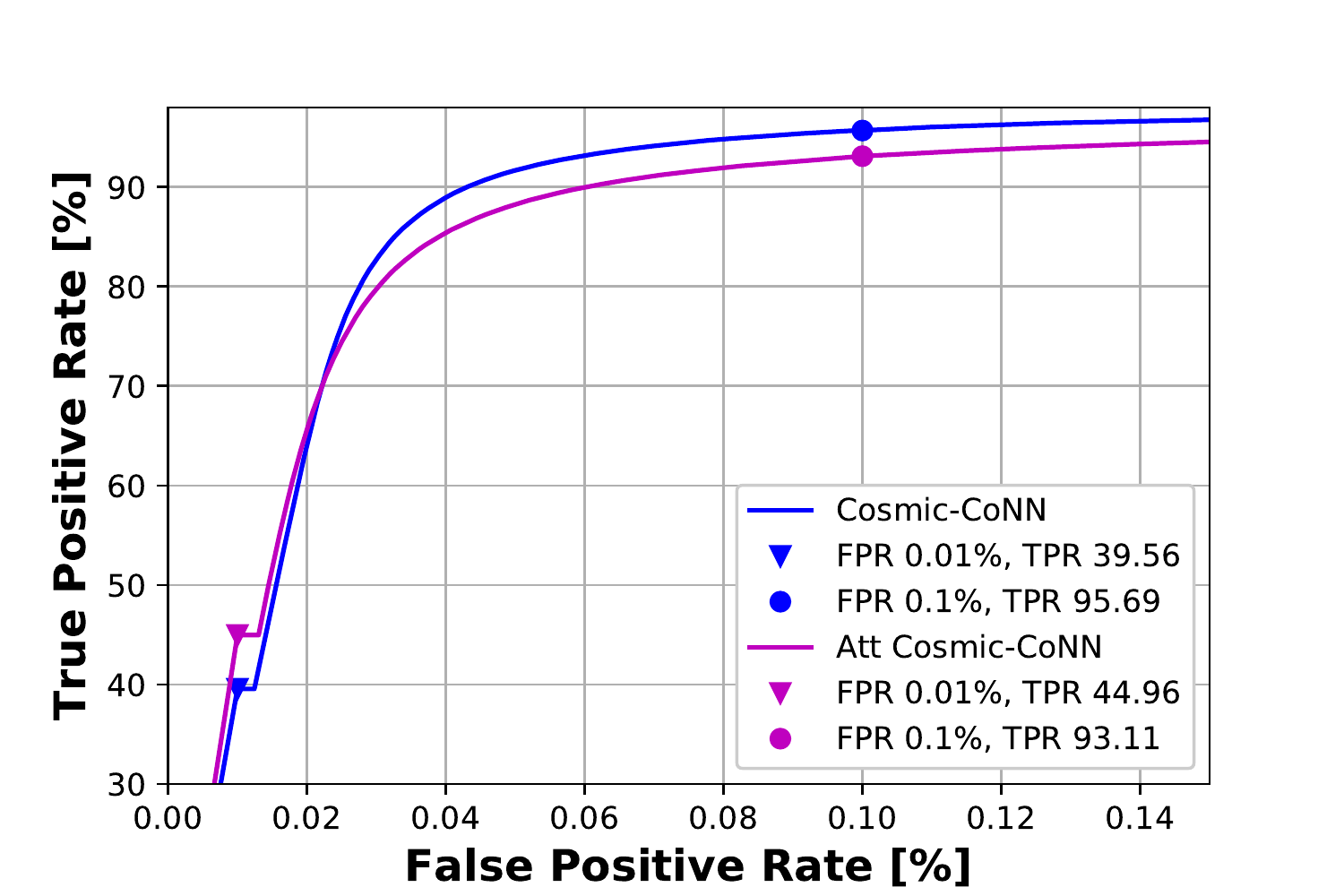}\label{fig:f83_ext}}
  \hfill
  \subfloat[Cosmic-CoNN - PRC]{\includegraphics[width=0.25\textwidth,
  height=0.25\textwidth, keepaspectratio,]{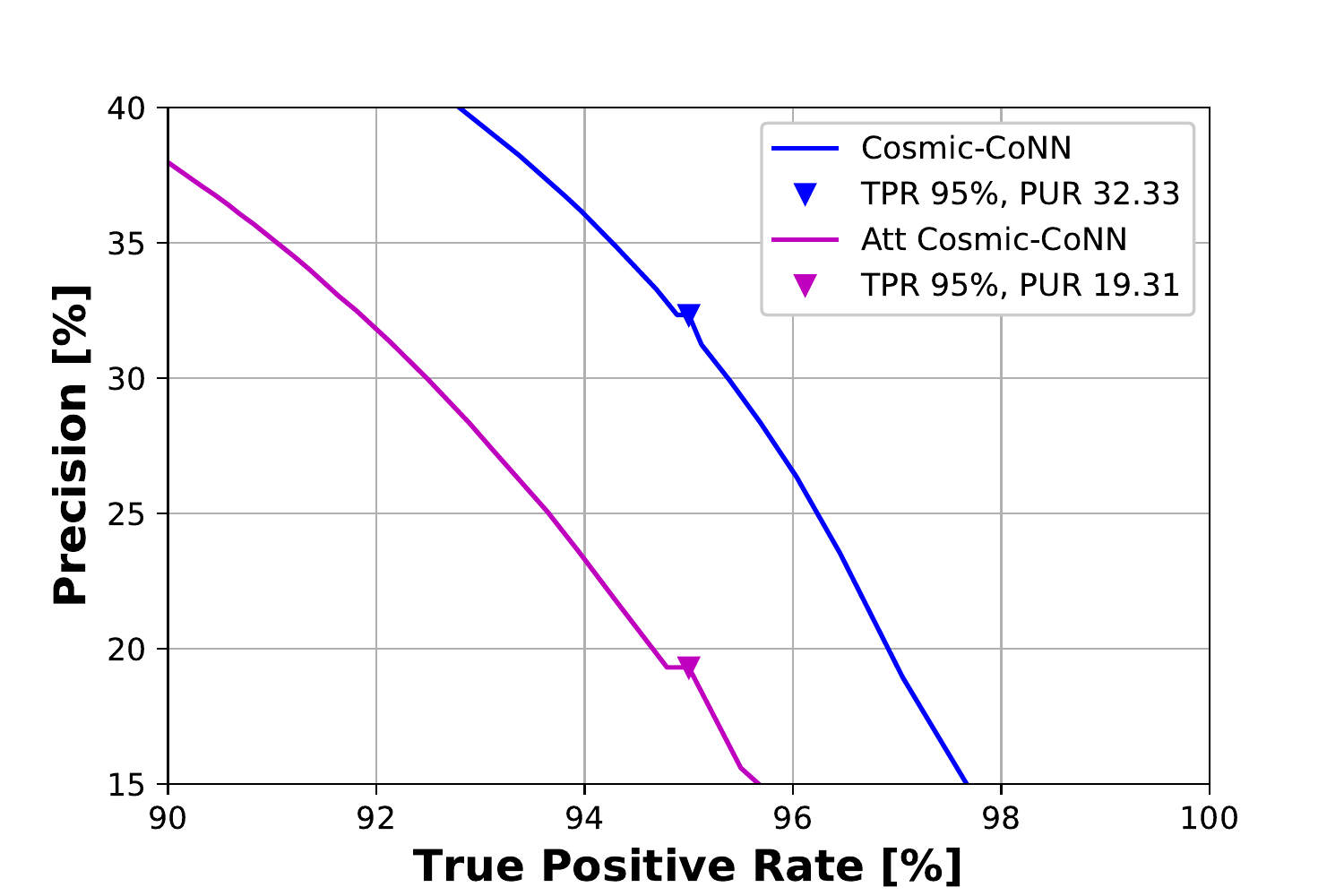}\label{fig:f84_ext}}
  \caption{The extended ROC and PRC plots for a better understanding of the proposed model performances from Fig.~\ref{fig:lco04m_dl}. The highest performance in CR detection can be noticed with the deepCR model at 0.01\% and 0.1\% FPR and the gains by adding AGs.}
  \label{fig:lco04m_dl_ext}
\end{figure*}

\begin{figure*}[!tbp]
  \centering
  \subfloat[LACosmic - ROC]{\includegraphics[width=0.25\textwidth,
  height=0.25\textwidth, keepaspectratio,]{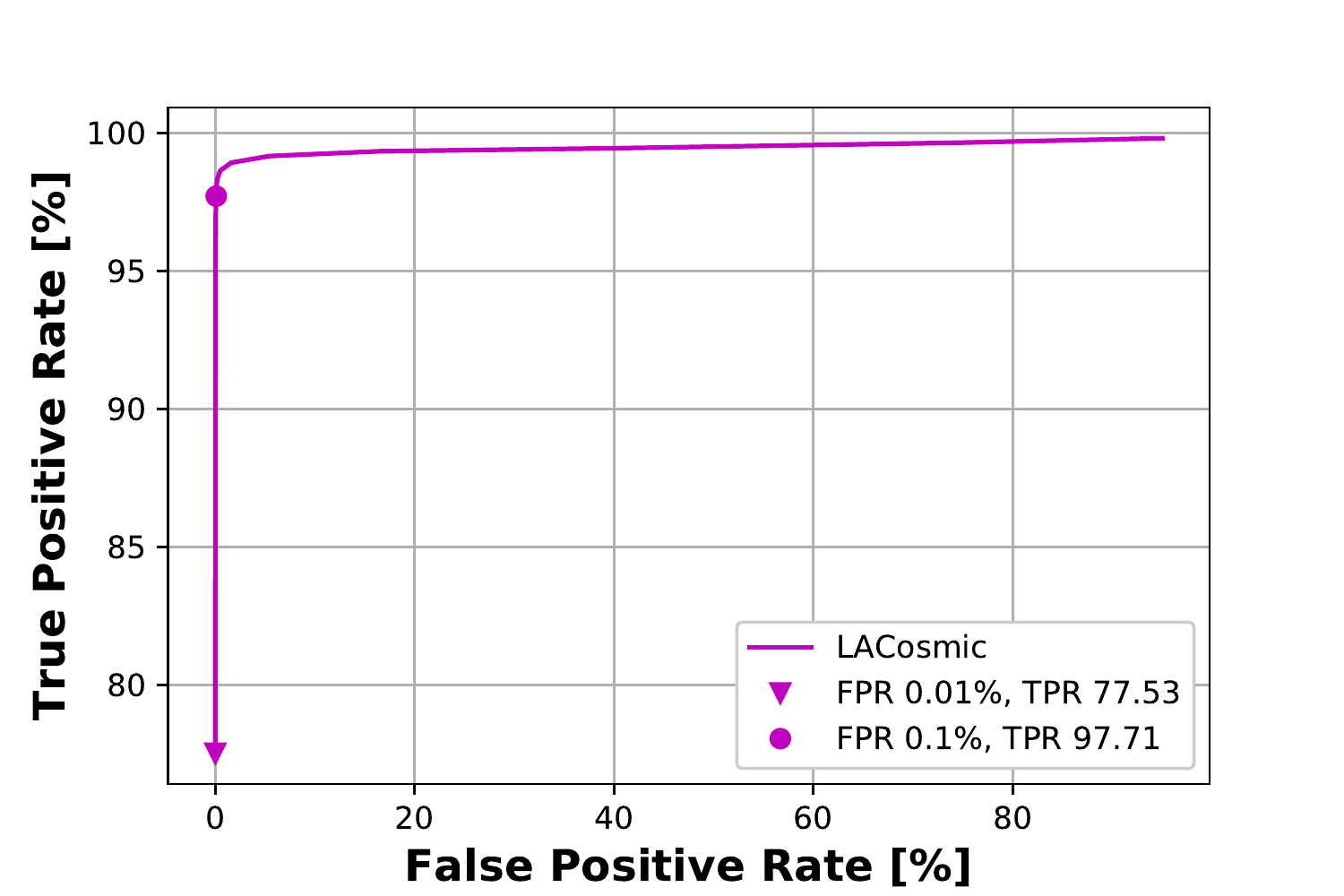}\label{fig:f31}}
  \hfill
  \subfloat[Astro-SCRAPPY - ROC]{\includegraphics[width=0.25\textwidth,
  height=0.25\textwidth, keepaspectratio,]{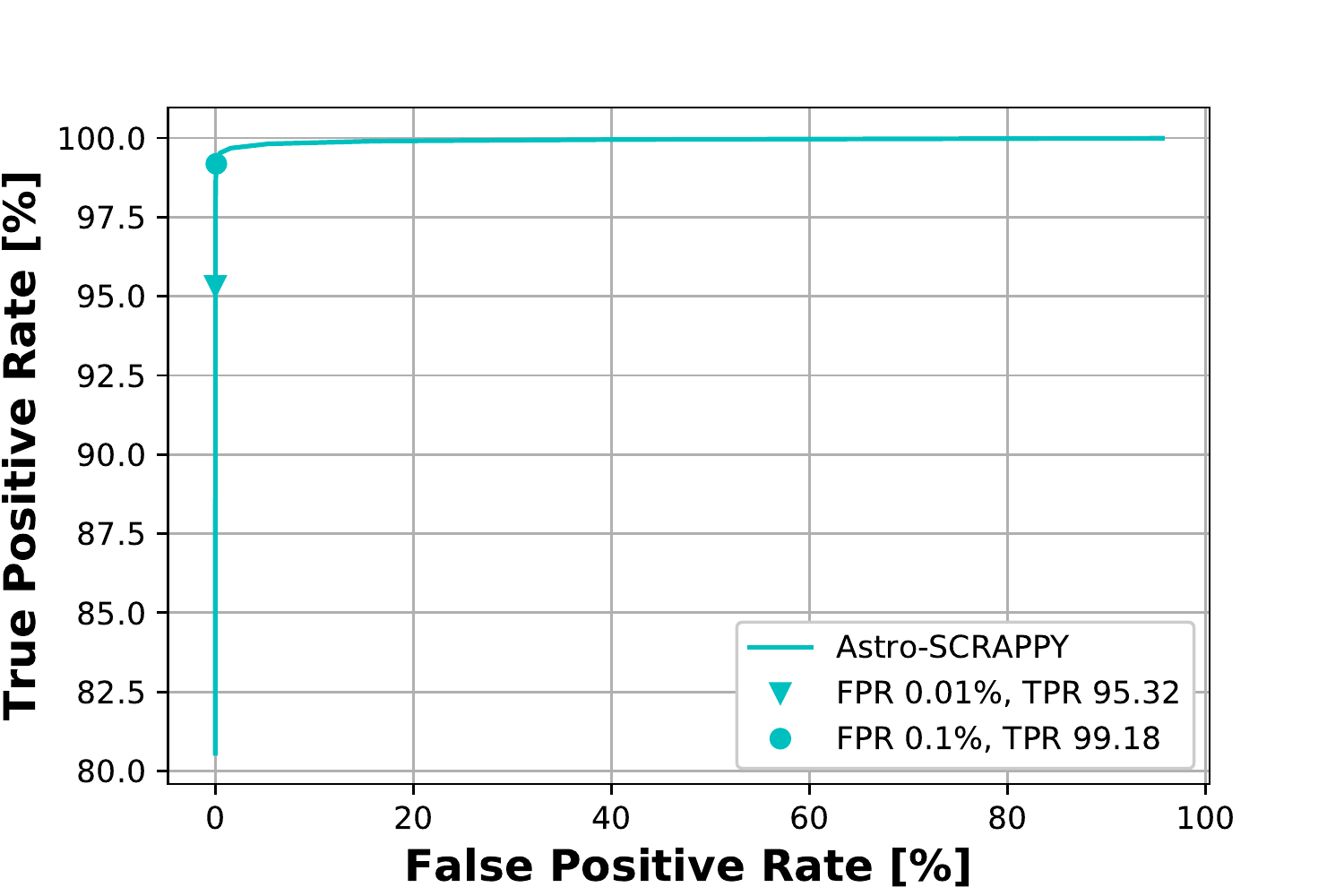}\label{fig:f32}}
  \hfill
  \subfloat[Pre-trained models - ROC]{\includegraphics[width=0.25\textwidth,
  height=0.25\textwidth, keepaspectratio,]{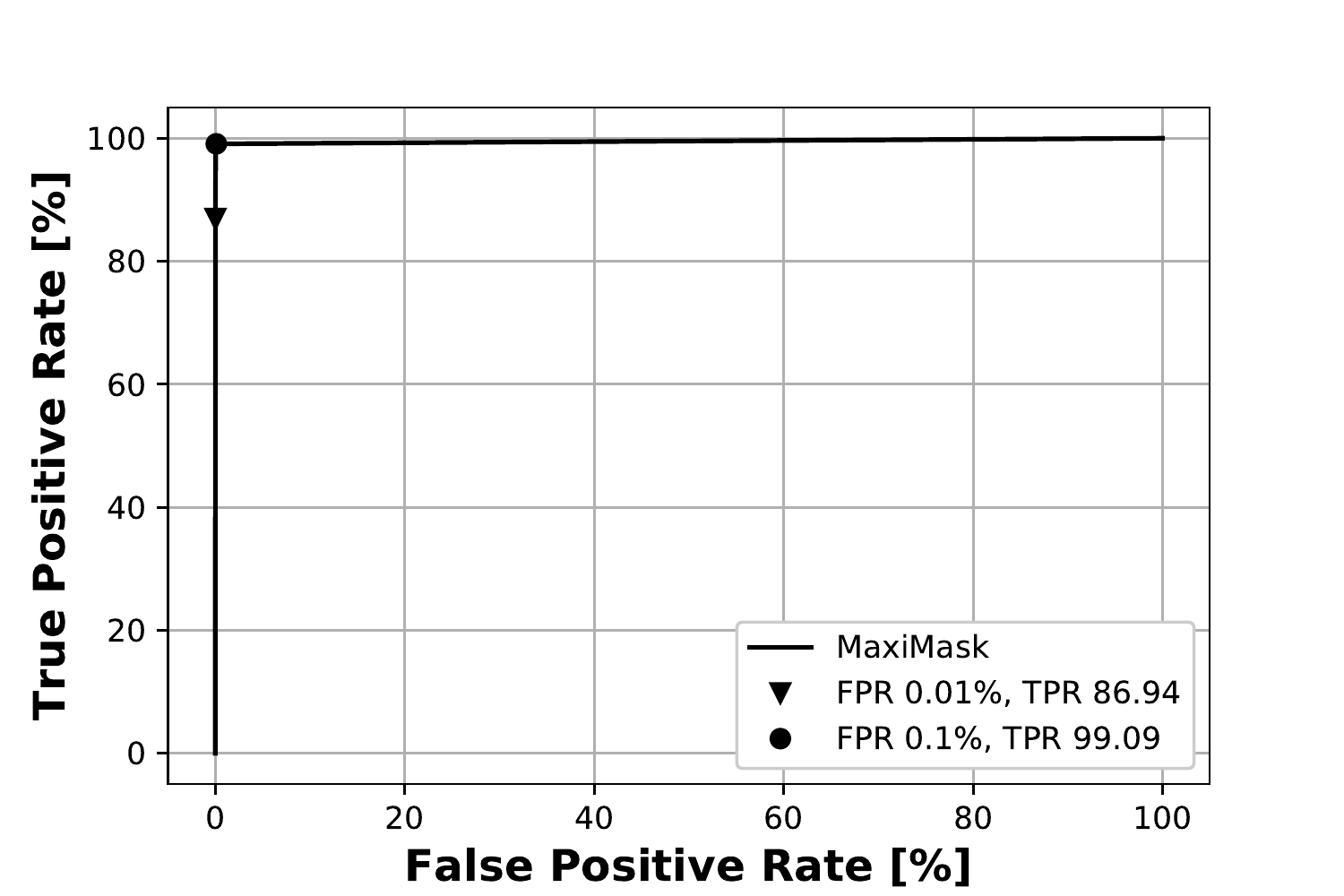}\label{fig:f33}}
  \hfill
  \subfloat[Pre-trained - PRC]{\includegraphics[width=0.25\textwidth,
  height=0.25\textwidth, keepaspectratio,]{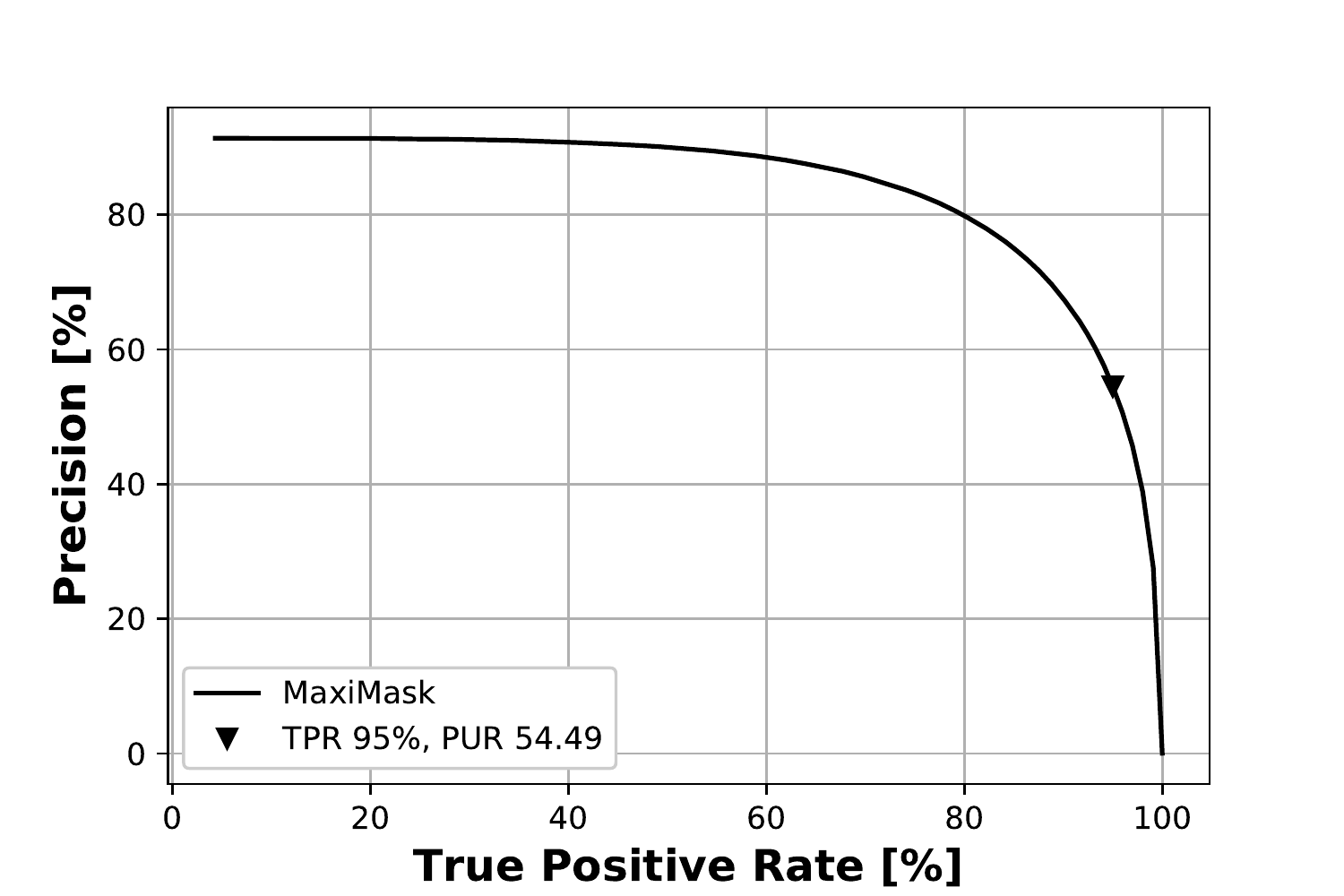}\label{fig:f34}}
  \caption{The ROC and PRC plots obtained with traditional CR detection algorithms, LACosmic and Astro-SCRAPPY on data with the 1-meter telescope from the LCO CR test dataset are presented in (a) and (b). The performance on the same data with pre-trained MaxiMask model is presented with ROC and PRC plots in (c) and (d), respectively. On this particular data, the Astro-SCRAPPY perform well than LACosmic and MaxiMask. The MaxiMask model also shows decent performance on this data.}
  \label{fig:lco1m_prev}
\end{figure*}

\begin{figure*}[!tbp]
  \centering
  \subfloat[LACosmic - ROC]{\includegraphics[width=0.25\textwidth,
  height=0.25\textwidth, keepaspectratio,]{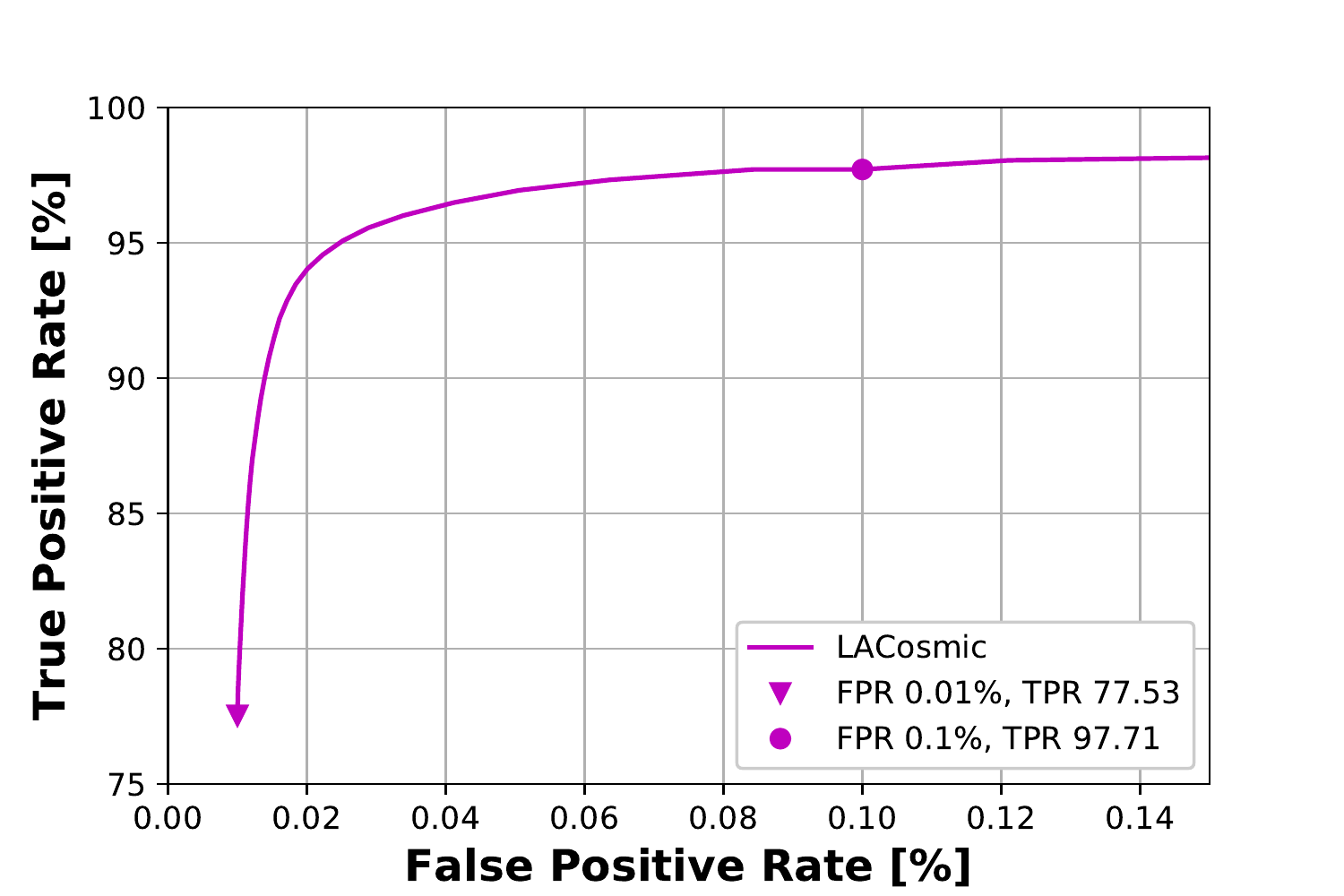}\label{fig:f31_ext}}
  \hfill
  \subfloat[Astro-SCRAPPY - ROC]{\includegraphics[width=0.25\textwidth,
  height=0.25\textwidth, keepaspectratio,]{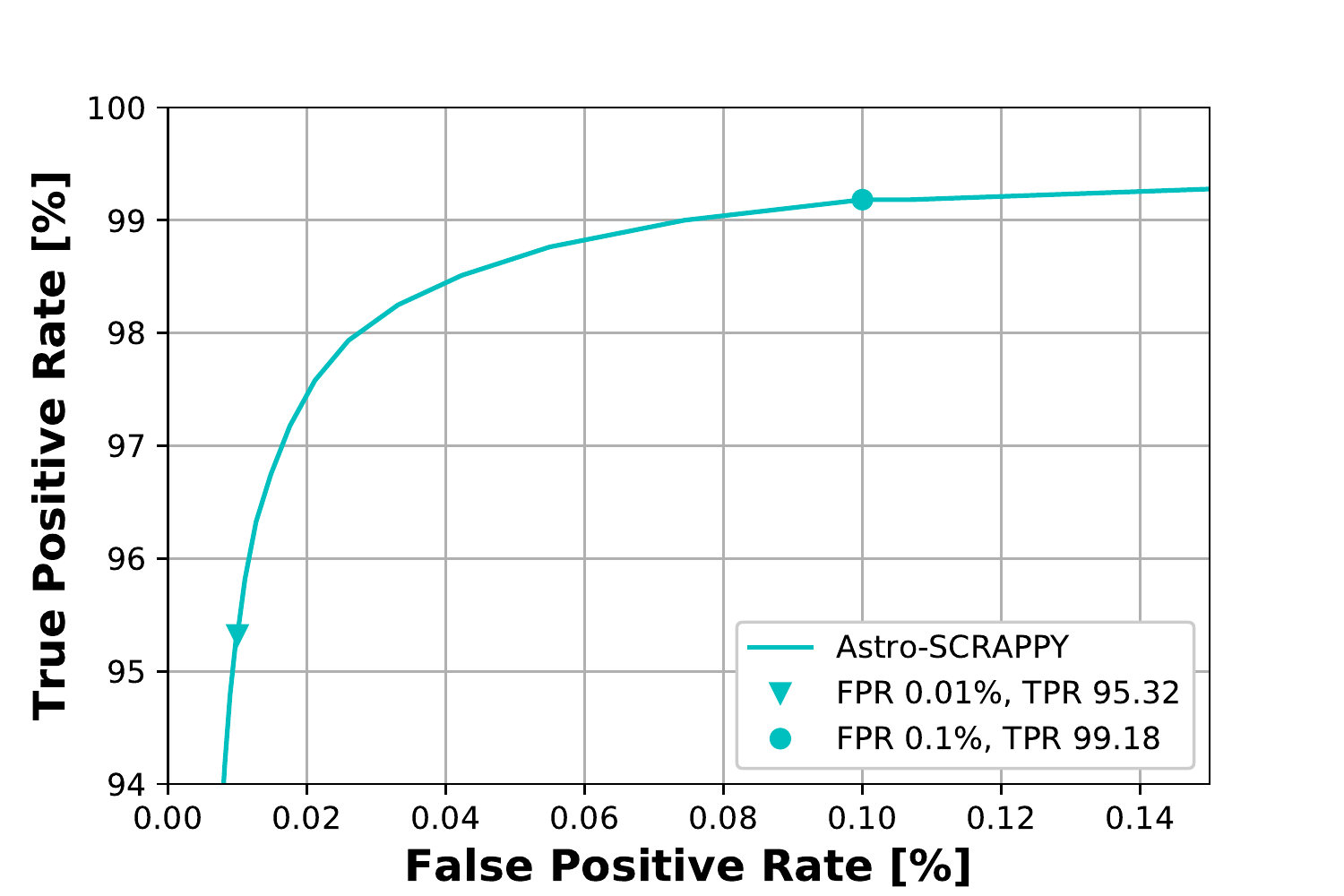}\label{fig:f32_ext}}
  \hfill
  \subfloat[Pre-trained models - ROC]{\includegraphics[width=0.25\textwidth,
  height=0.25\textwidth, keepaspectratio,]{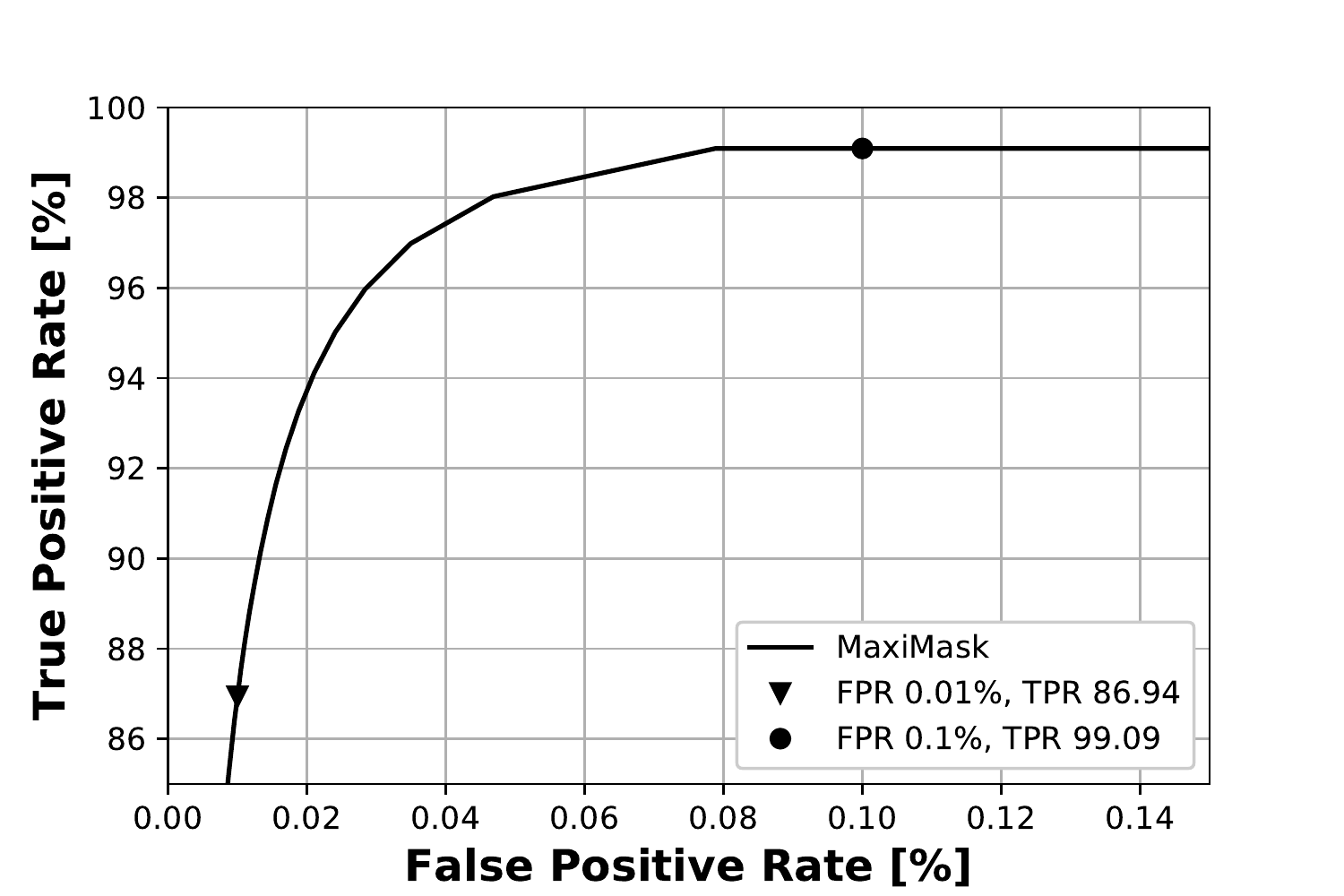}\label{fig:f33_ext}}
  \hfill
  \subfloat[Pre-trained - PRC]{\includegraphics[width=0.25\textwidth,
  height=0.25\textwidth, keepaspectratio,]{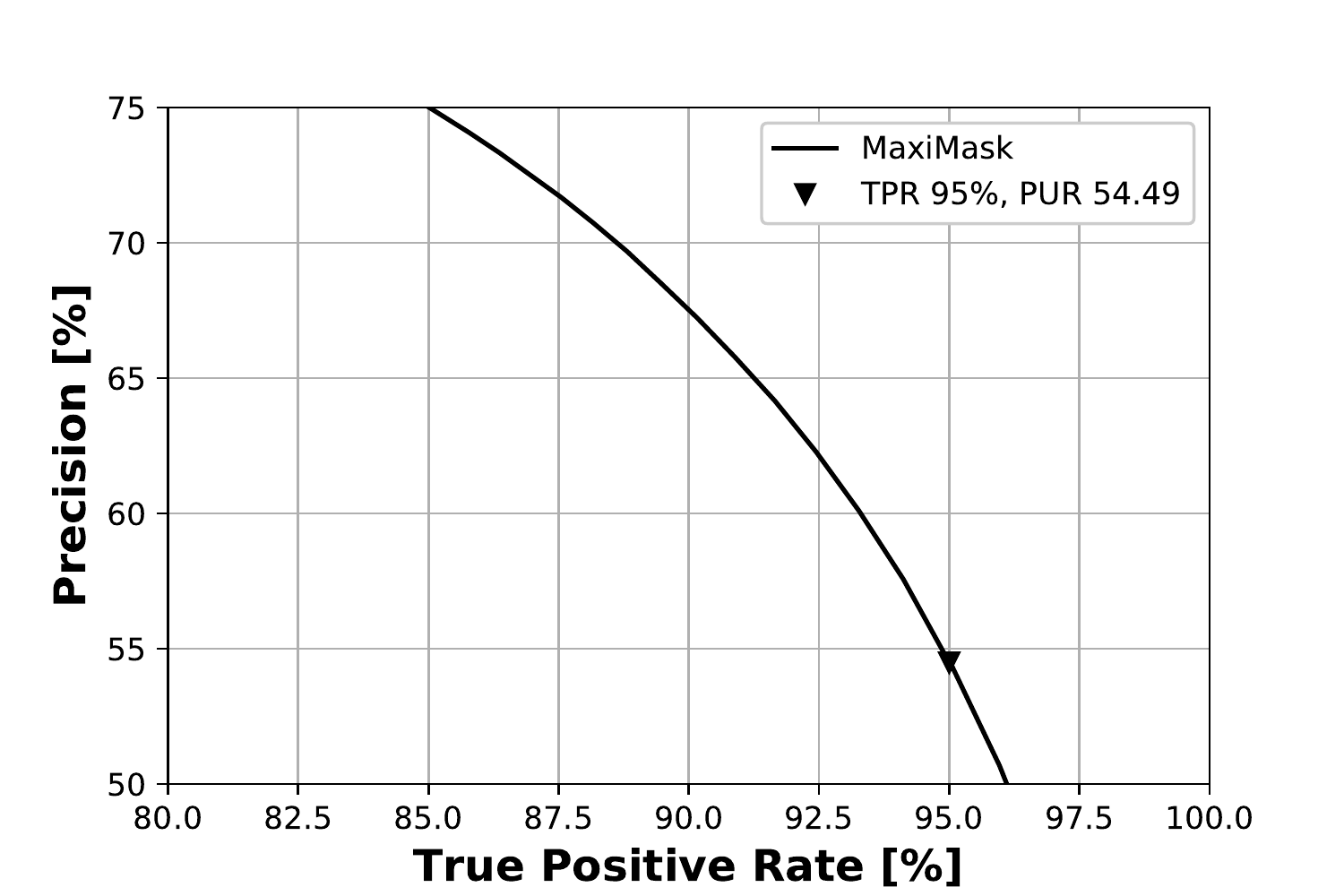}\label{fig:f34_ext}}
  \caption{The extended ROC and PRC plots for a better understanding of the model performances from Fig.~\ref{fig:lco1m_prev}. Both Astro-SCRAPPY and MaxiMask provide the highest CR detection performance on the LCO 1-meter data at 0.1\% FPR. The MaxiMask model also performs similar to Astro-SCRAPPY with a 1\% difference in TPR at 0.1\% FPR. However, at 0.01\% FPR, the TPR is significantly better with Astro-SCRAPPY than the MaxiMask model.}
  \label{fig:lco1m_prev_ext}
\end{figure*}

\begin{figure*}[!tbp]
  \centering
  \subfloat[deepCR - ROC]{\includegraphics[width=0.25\textwidth,
  height=0.25\textwidth, keepaspectratio,]{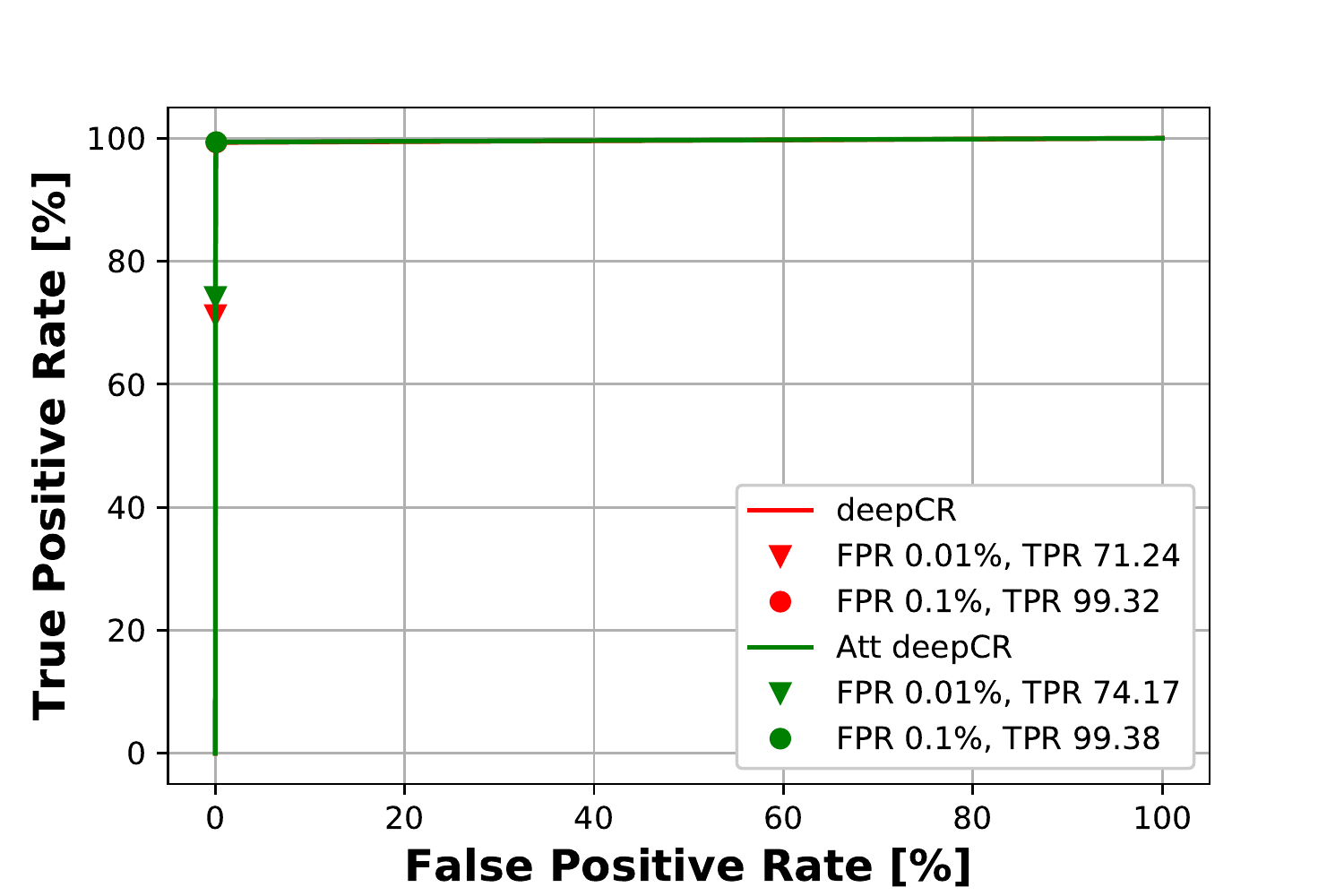}\label{fig:f41}}
  \hfill
  \subfloat[deepCR - PRC]{\includegraphics[width=0.25\textwidth,
  height=0.25\textwidth, keepaspectratio,]{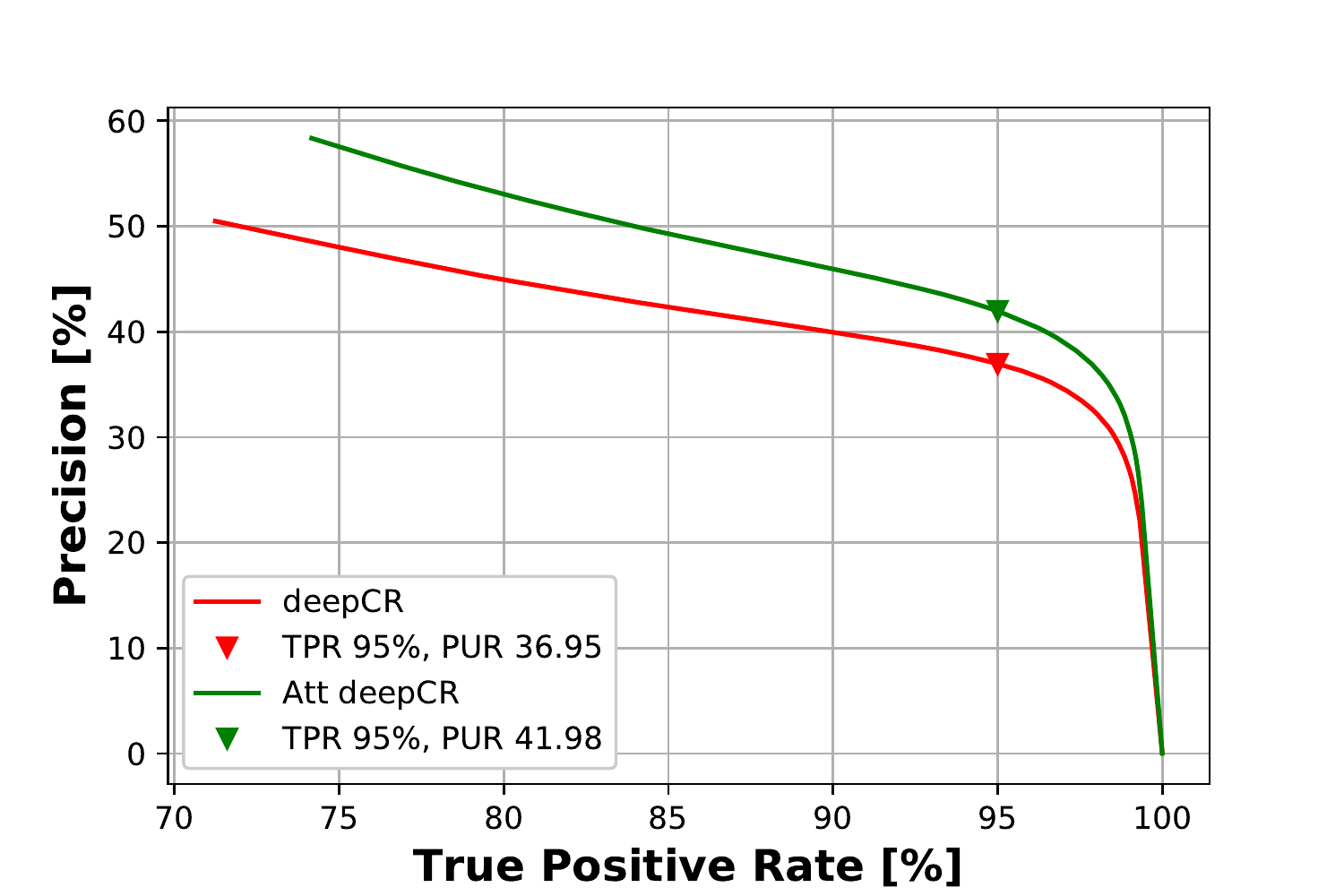}\label{fig:f42}}
  \hfill
  \subfloat[cosmic-CoNN - ROC]{\includegraphics[width=0.25\textwidth,
  height=0.25\textwidth, keepaspectratio,]{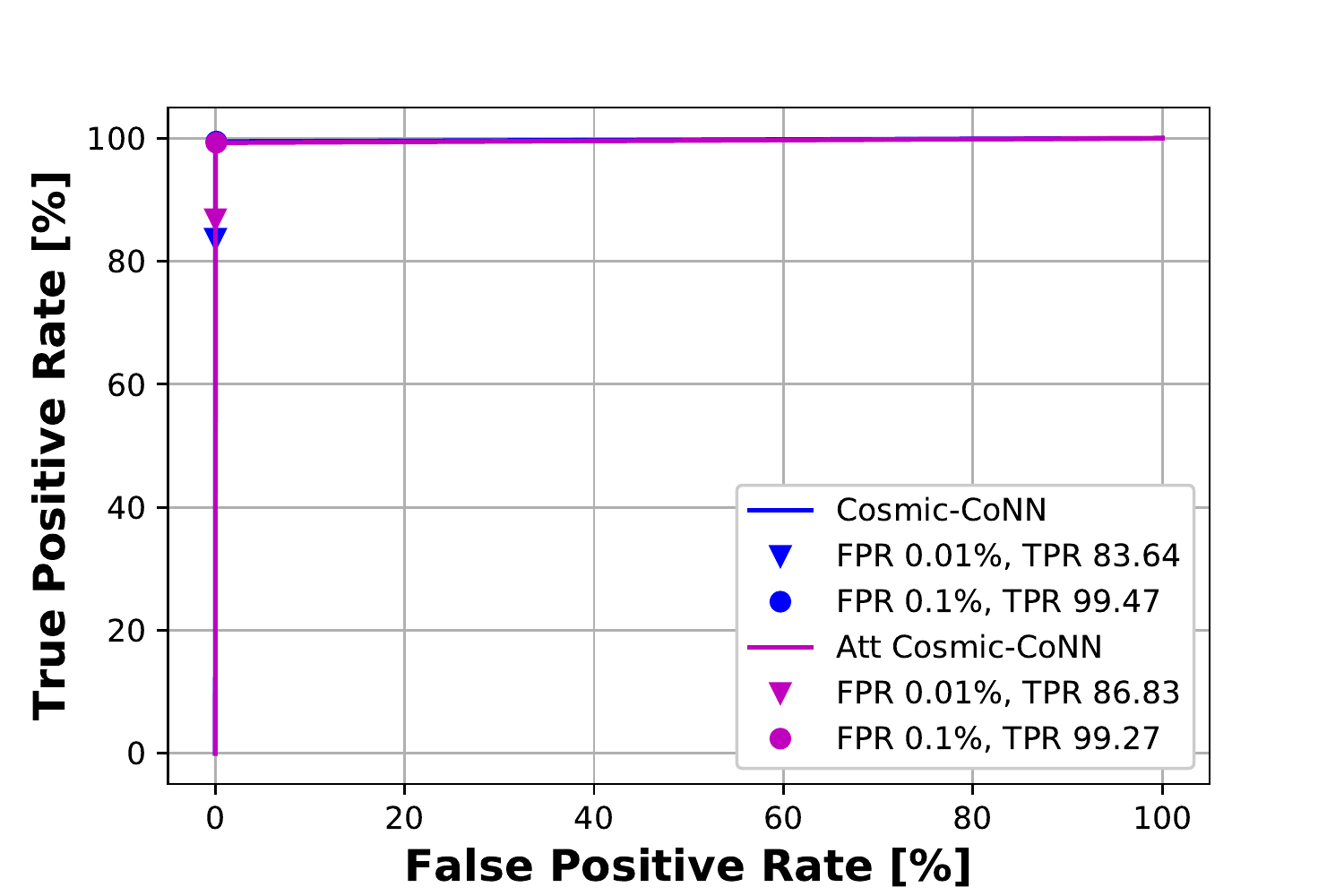}\label{fig:f43}}
  \hfill
  \subfloat[Cosmic-CoNN - PRC]{\includegraphics[width=0.25\textwidth,
  height=0.25\textwidth, keepaspectratio,]{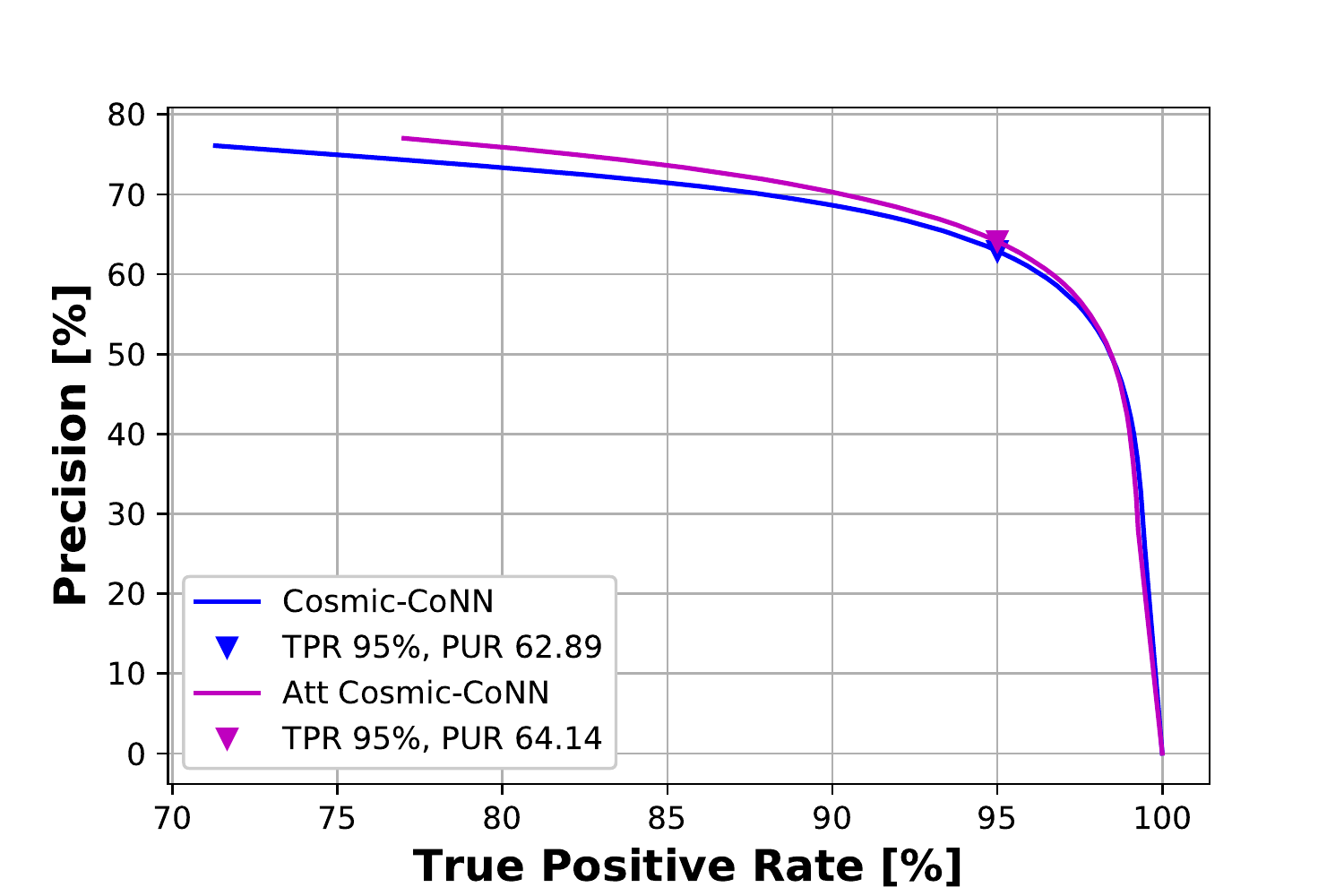}\label{fig:f44}}
  \caption{Performance of the proposed models on previously unseen data from the LCO CR test dataset using the 1-meter telescope data is presented here. (a) and (b) are the ROC and PRC plots on deepCR models with and without adding the attention gates. Similar plots for the Cosmic-CoNN models are presented in (c) and (d). The Cosmic-CoNN model performs better on this data than the deepCR model in most cases. Also, both the deepCR and Cosmic-CoNN models benefit from adding AGs more than the corresponding baselines on this data.}
  \label{fig:lco1m_dl}
\end{figure*}

\begin{figure*}[!tbp]
  \centering
  \subfloat[deepCR - ROC]{\includegraphics[width=0.25\textwidth,
  height=0.25\textwidth, keepaspectratio,]{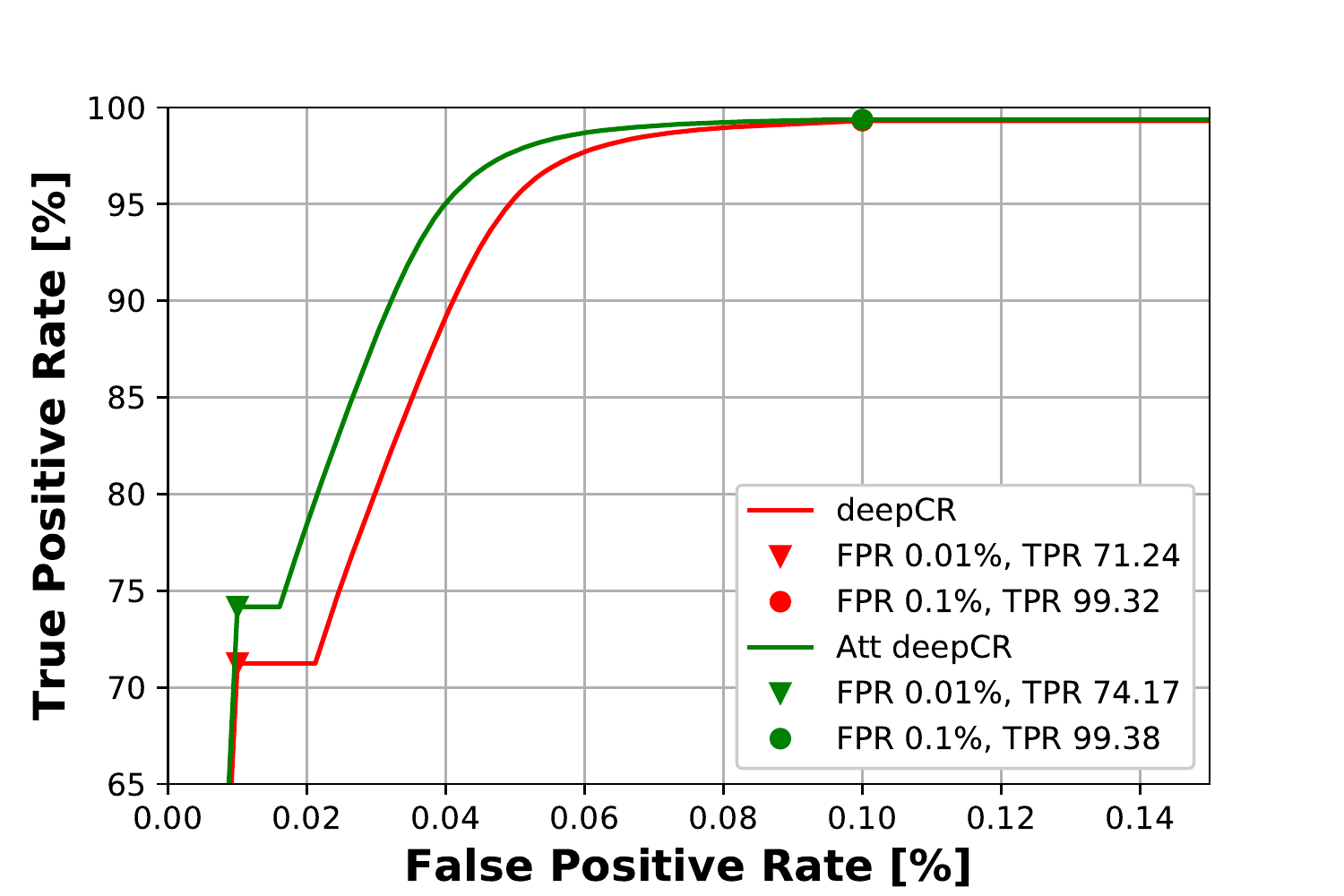}\label{fig:f41_ext}}
  \hfill
  \subfloat[deepCR - PRC]{\includegraphics[width=0.25\textwidth,
  height=0.25\textwidth, keepaspectratio,]{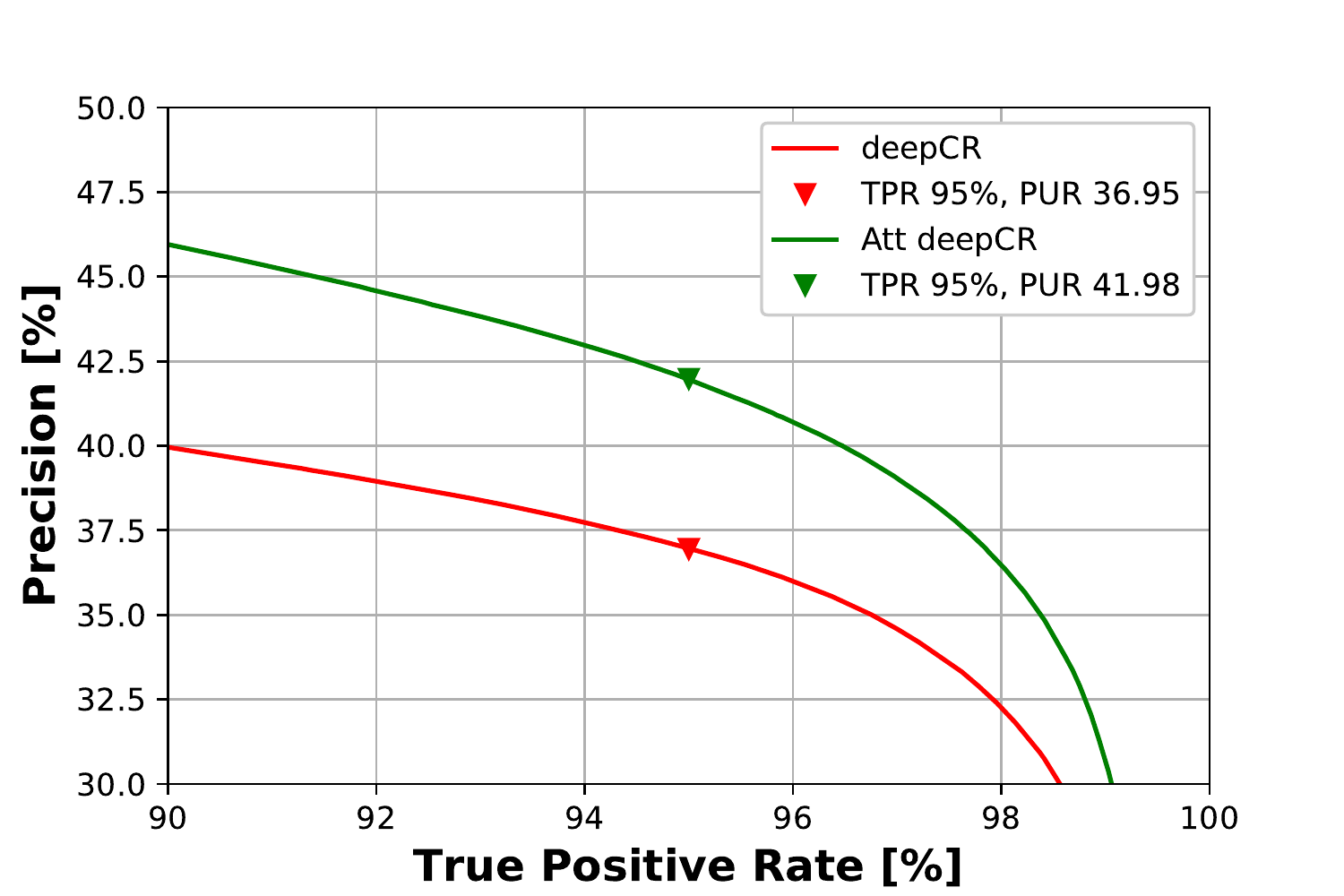}\label{fig:f42_ext}}
  \hfill
  \subfloat[cosmic-CoNN - ROC]{\includegraphics[width=0.25\textwidth,
  height=0.25\textwidth, keepaspectratio,]{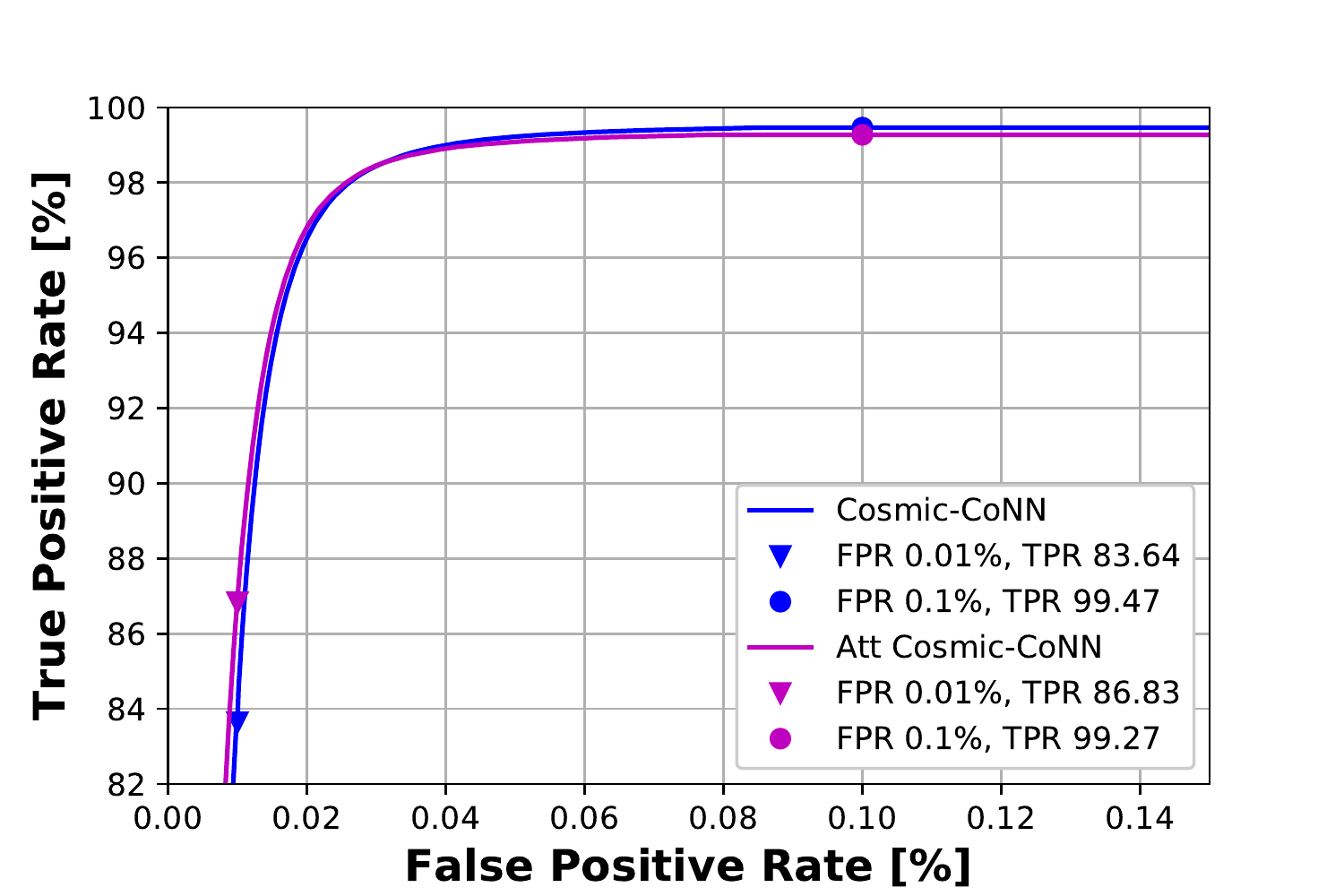}\label{fig:f43_ext}}
  \hfill
  \subfloat[Cosmic-CoNN - PRC]{\includegraphics[width=0.25\textwidth,
  height=0.25\textwidth, keepaspectratio,]{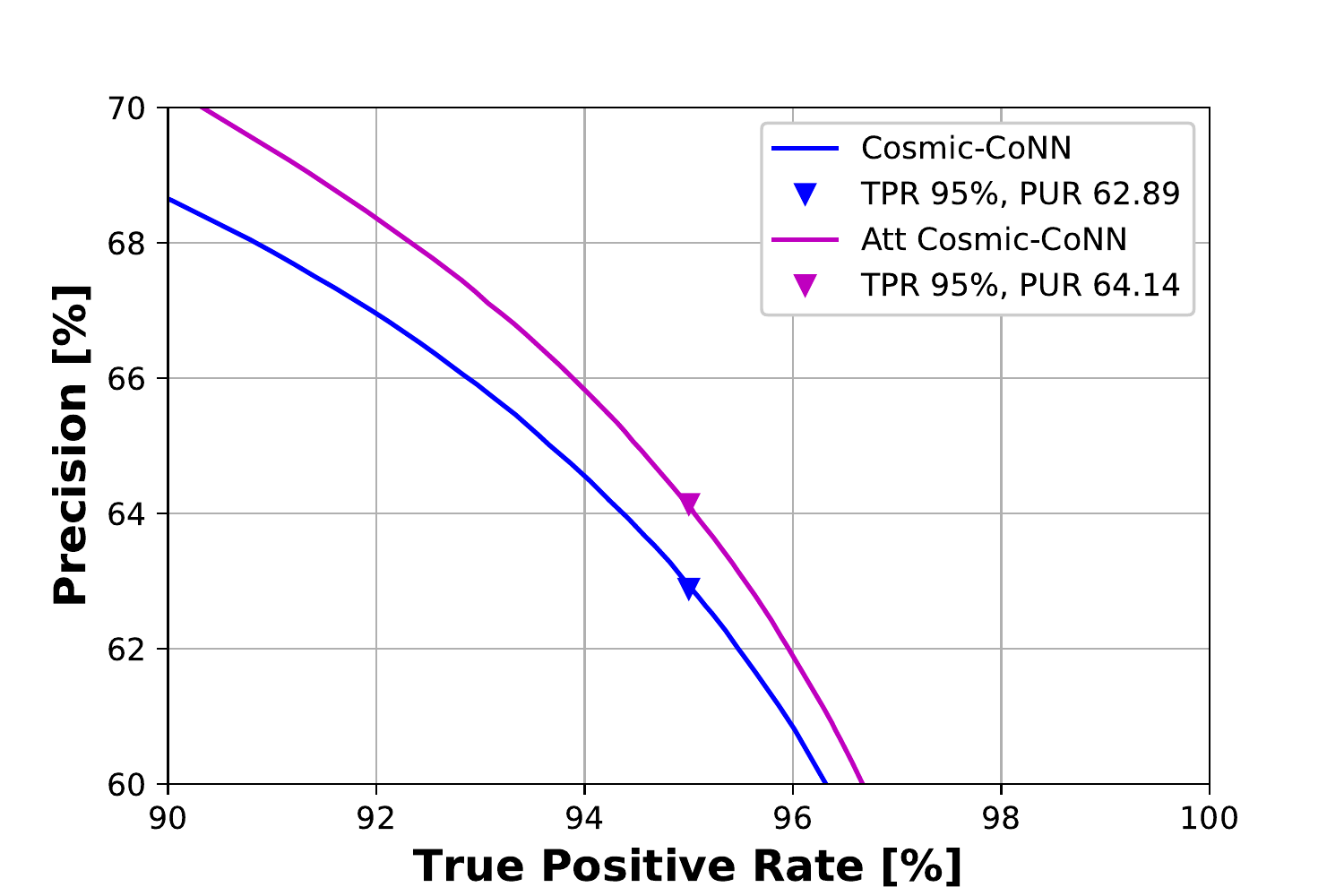}\label{fig:f44_ext}}
  \caption{The extended ROC and PRC plots for a better understanding of the proposed model performances from Fig.~\ref{fig:lco1m_dl}. The highest performance in CR detection can be noticed with the Cosmic-CoNN model both at 0.01\% and 0.1\% FPR, compared to baseline deepCR. The gains with adding AGs can be noticed with both baseline models.}
  \label{fig:lco1m_dl_ext}
\end{figure*}

\begin{figure*}[!tbp]
  \centering
  \subfloat[LACosmic - ROC]{\includegraphics[width=0.25\textwidth,
  height=0.25\textwidth, keepaspectratio,]{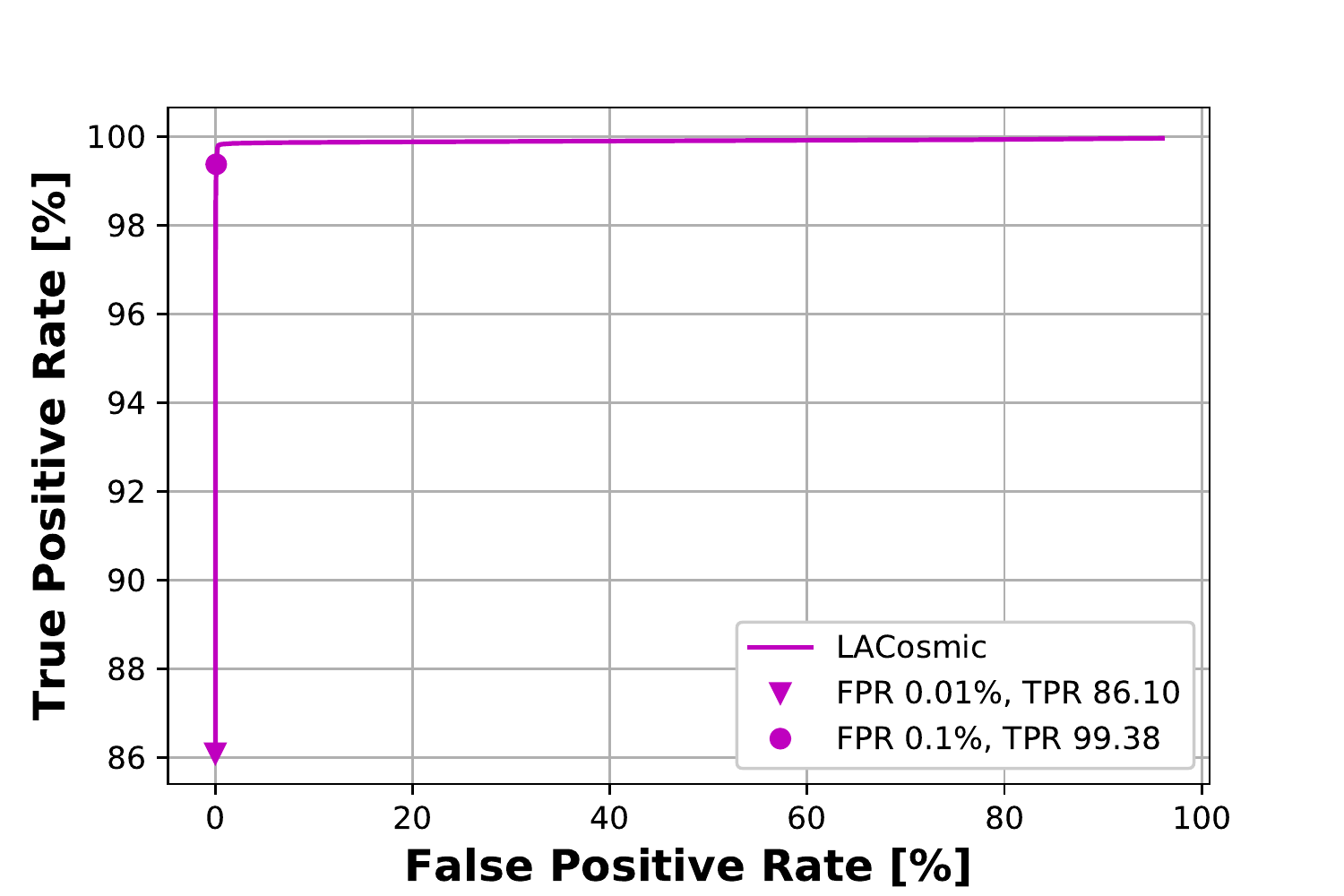}\label{fig:f51}}
  \hfill
  \subfloat[Astro-SCRAPPY - ROC]{\includegraphics[width=0.25\textwidth,
  height=0.25\textwidth, keepaspectratio,]{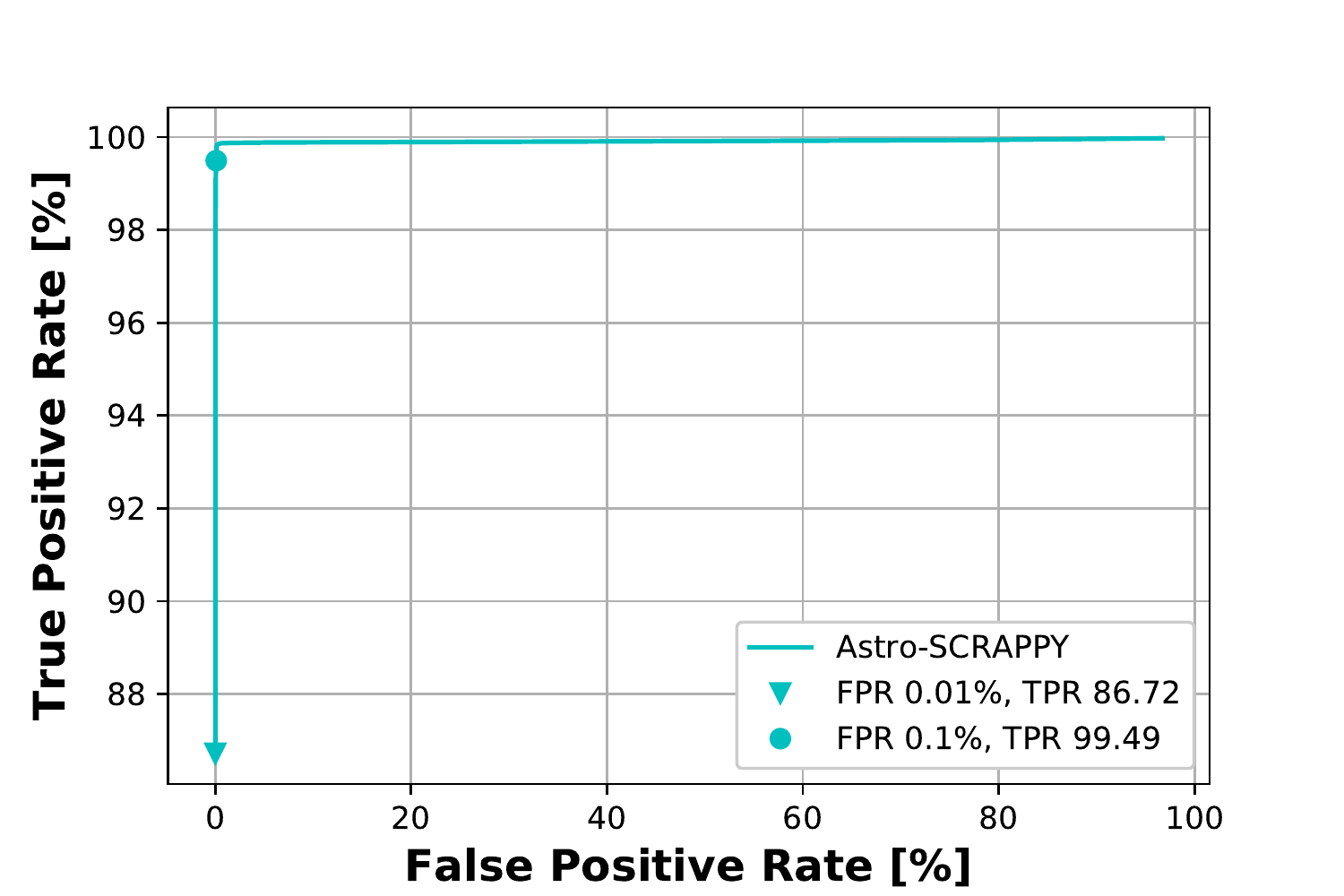}\label{fig:f52}}
  \hfill
  \subfloat[Pre-trained models - ROC]{\includegraphics[width=0.25\textwidth,
  height=0.25\textwidth, keepaspectratio,]{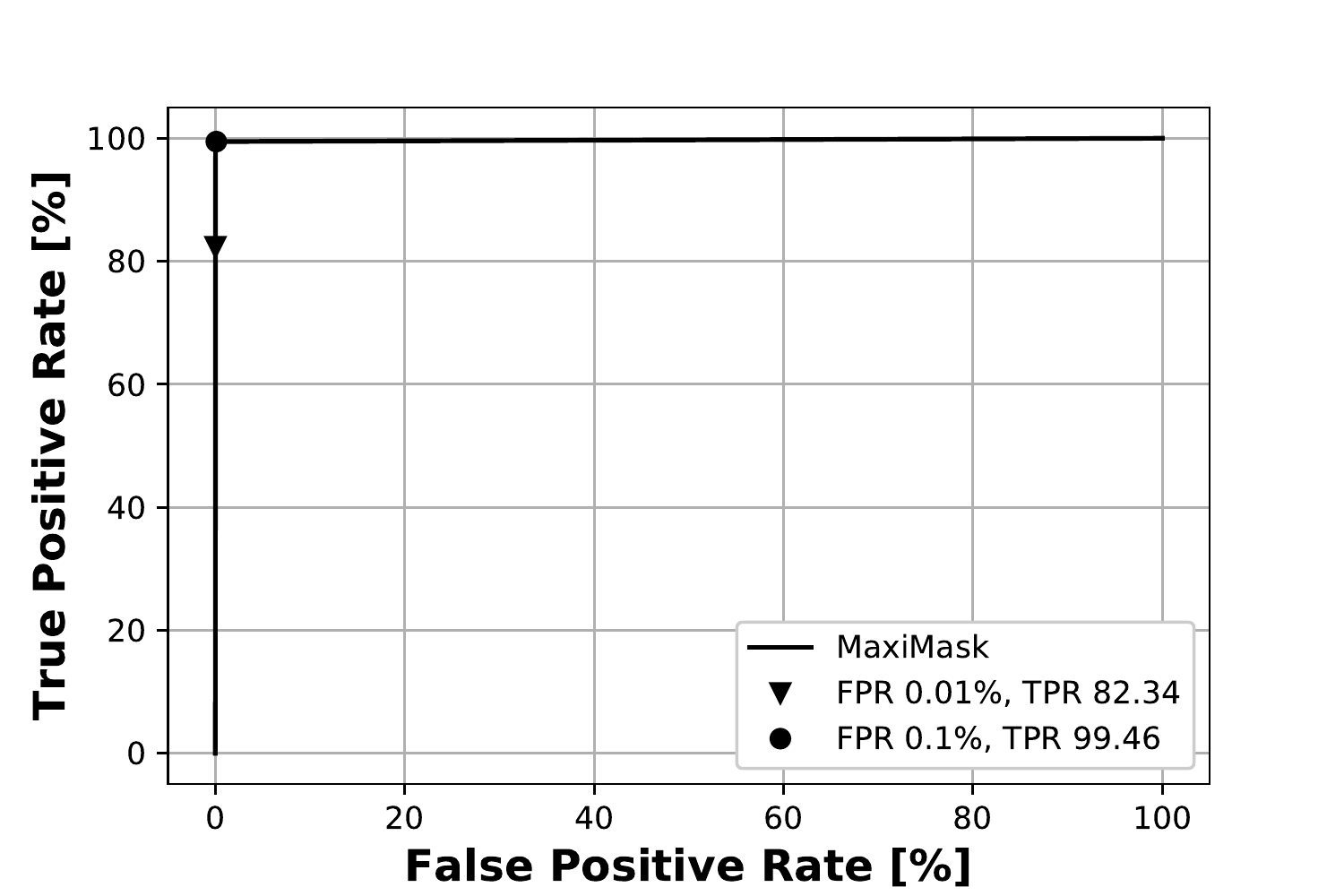}\label{fig:f53}}
  \hfill
  \subfloat[Pre-trained - PRC]{\includegraphics[width=0.25\textwidth,
  height=0.25\textwidth, keepaspectratio,]{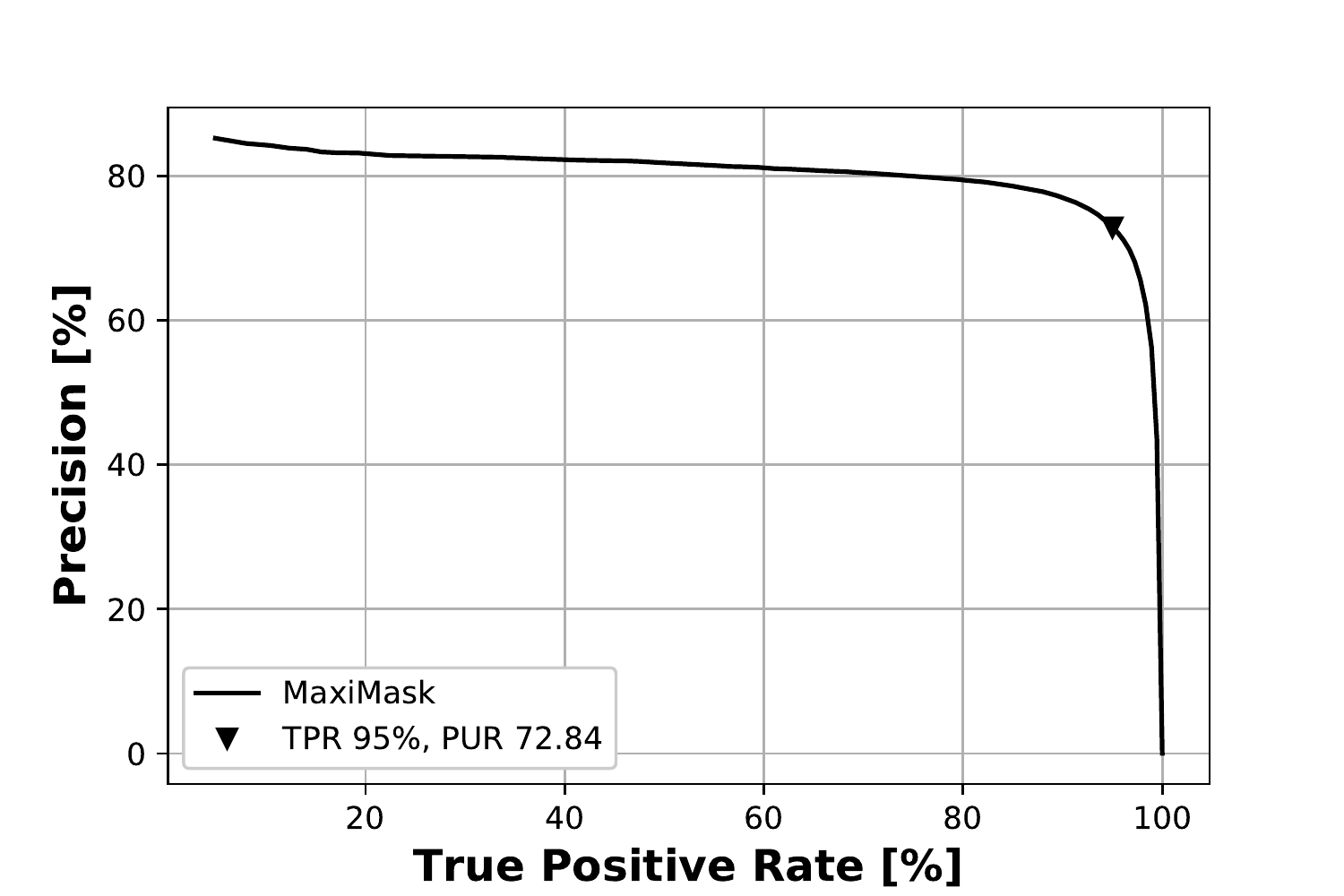}\label{fig:f54}}
  \caption{The ROC and PRC plots obtained with traditional CR detection algorithms, LACosmic and Astro-SCRAPPY on data with the 2-meter telescope from the LCO CR test dataset are presented in (a) and (b). The performance on the same data with pre-trained MaxiMask model is presented with ROC and PRC plots in (c) and (d), respectively. Both classical and pre-trained MaxiMask models show decent performance on this data.}
  \label{fig:lco2m_prev}
\end{figure*}

\begin{figure*}[!tbp]
  \centering
  \subfloat[LACosmic - ROC]{\includegraphics[width=0.25\textwidth,
  height=0.25\textwidth, keepaspectratio,]{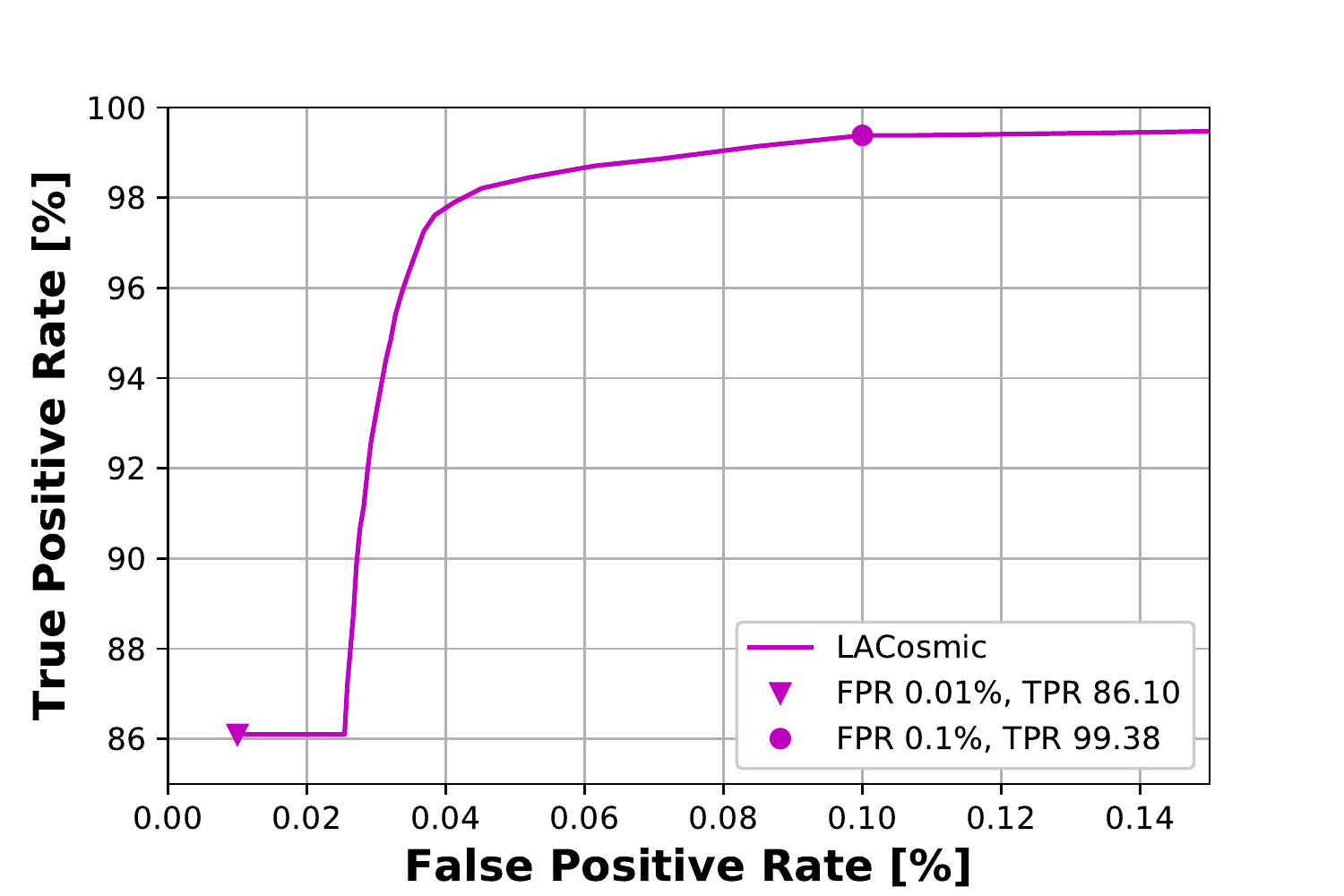}\label{fig:f51_ext}}
  \hfill
  \subfloat[Astro-SCRAPPY - ROC]{\includegraphics[width=0.25\textwidth,
  height=0.25\textwidth, keepaspectratio,]{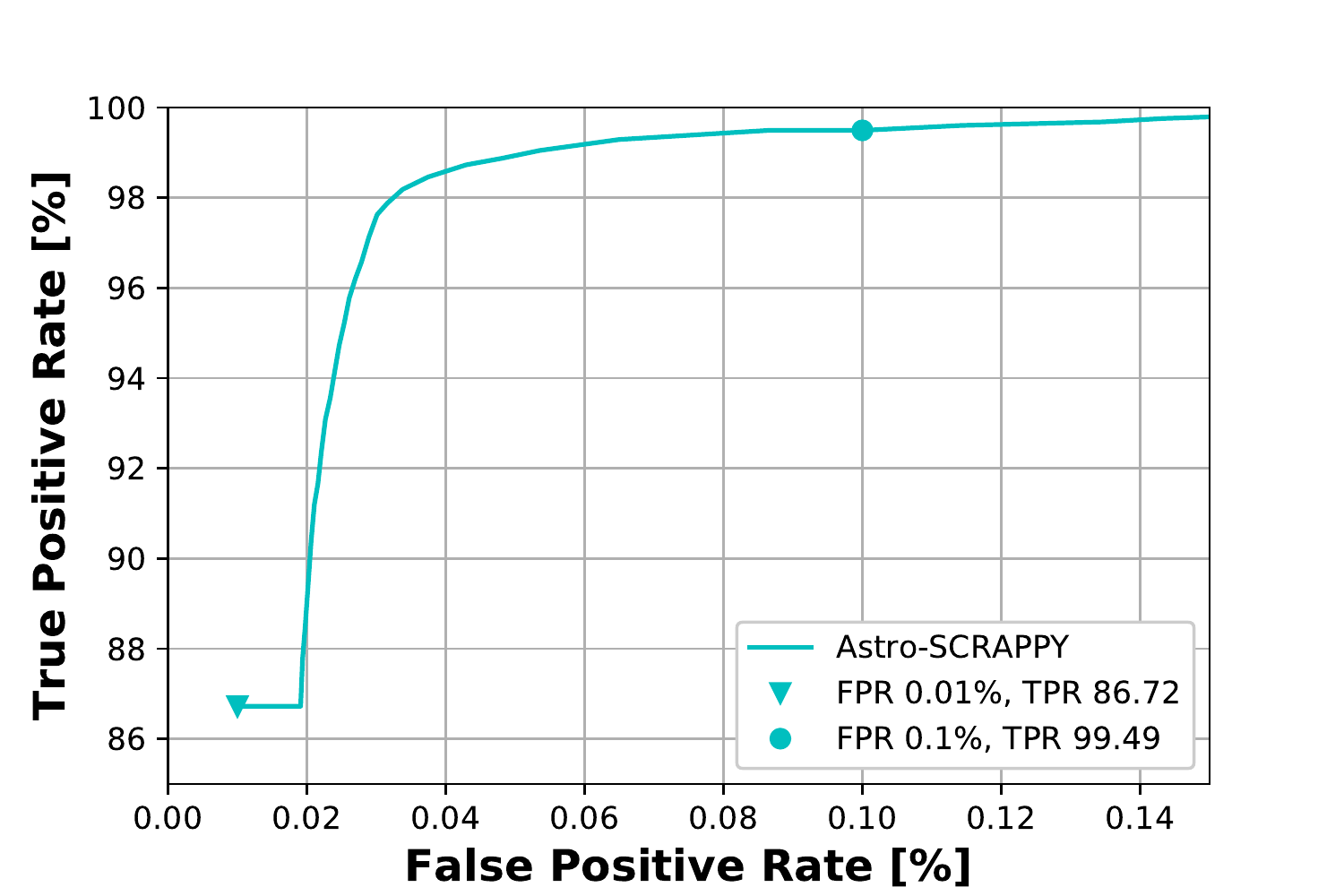}\label{fig:f52_ext}}
  \hfill
  \subfloat[Pre-trained models - ROC]{\includegraphics[width=0.25\textwidth,
  height=0.25\textwidth, keepaspectratio,]{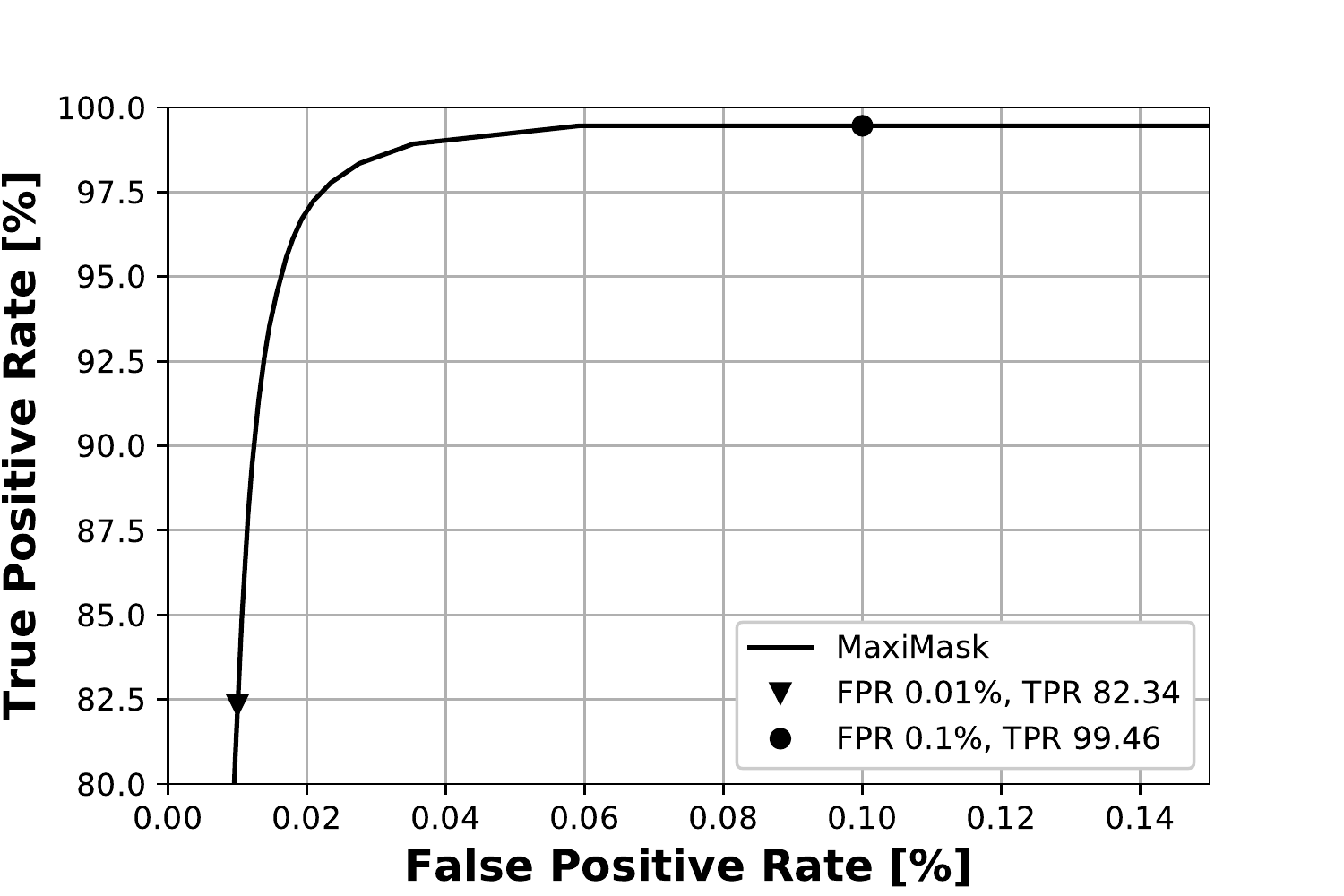}\label{fig:f53_ext}}
  \hfill
  \subfloat[Pre-trained - PRC]{\includegraphics[width=0.25\textwidth,
  height=0.25\textwidth, keepaspectratio,]{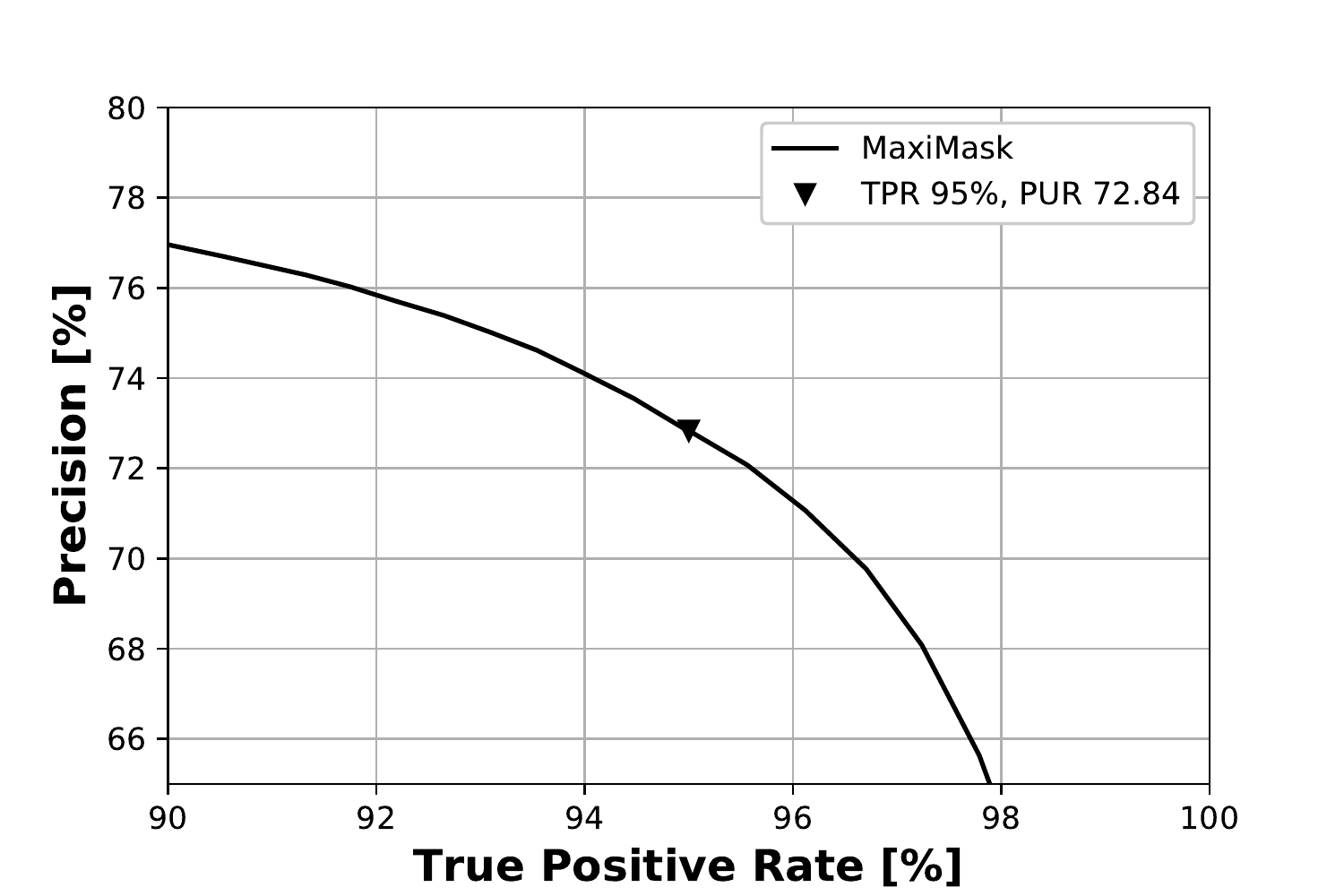}\label{fig:f54_ext}}
  \caption{The extended ROC and PRC plots for a better understanding of the model performances from Fig.~\ref{fig:lco2m_prev}. The cosmic-CoNN model provides the highest CR detection performance on the LCO 2-meter data at 0.01\% FPR. The detection rates are almost the same and above 99\% with LACosmic, Astro-SCRAPPY and Cosmic-CoNN models at 0.1\% FPR. The MaxiMask model also performs similar to these models with a 1\% difference in TPRs.}
  \label{fig:lco2m_prev_ext}
\end{figure*}

\begin{figure*}[!tbp]
  \centering
  \subfloat[deepCR - ROC]{\includegraphics[width=0.25\textwidth,
  height=0.25\textwidth, keepaspectratio,]{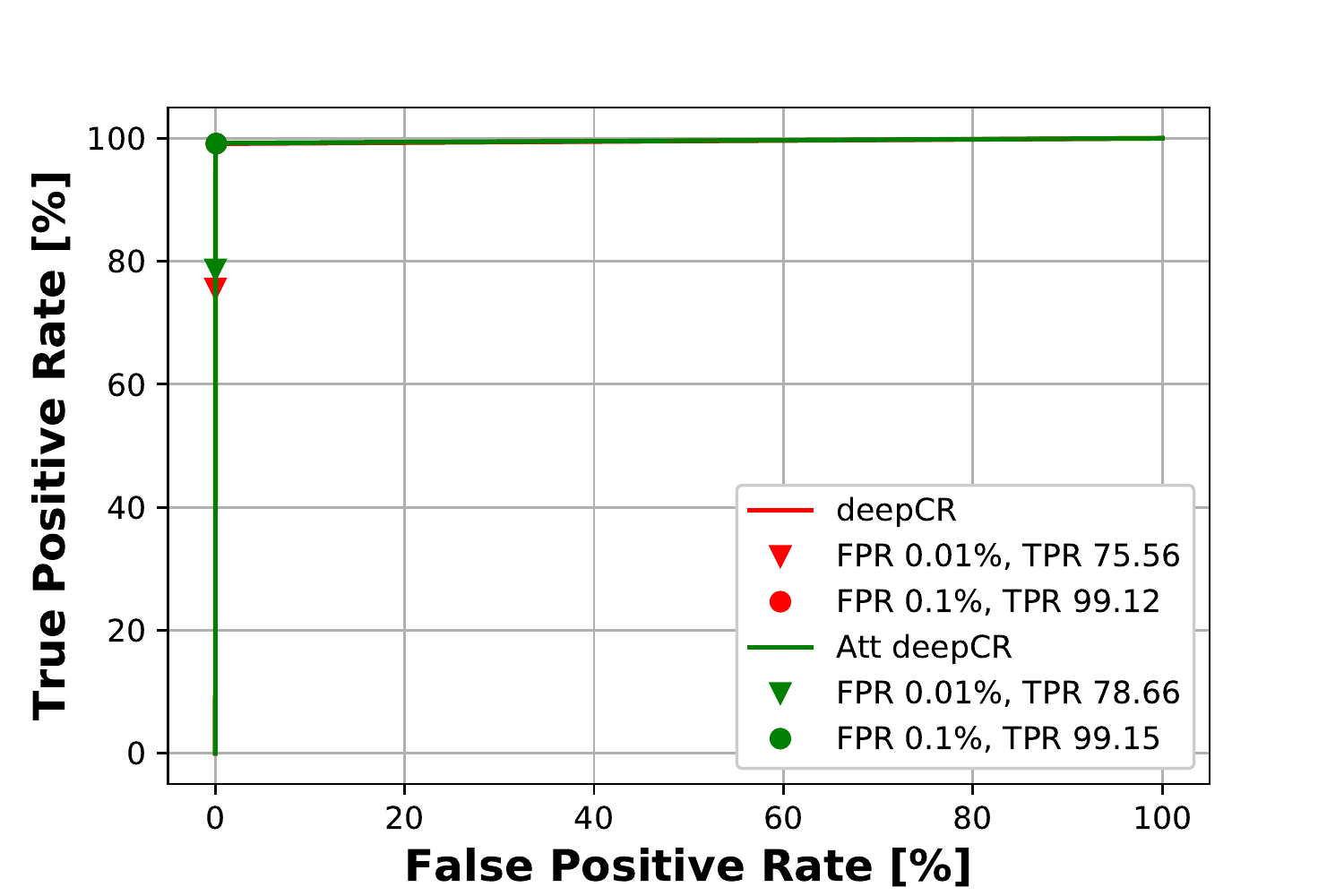}\label{fig:f61}}
  \hfill
  \subfloat[deepCR - PRC]{\includegraphics[width=0.25\textwidth,
  height=0.25\textwidth, keepaspectratio,]{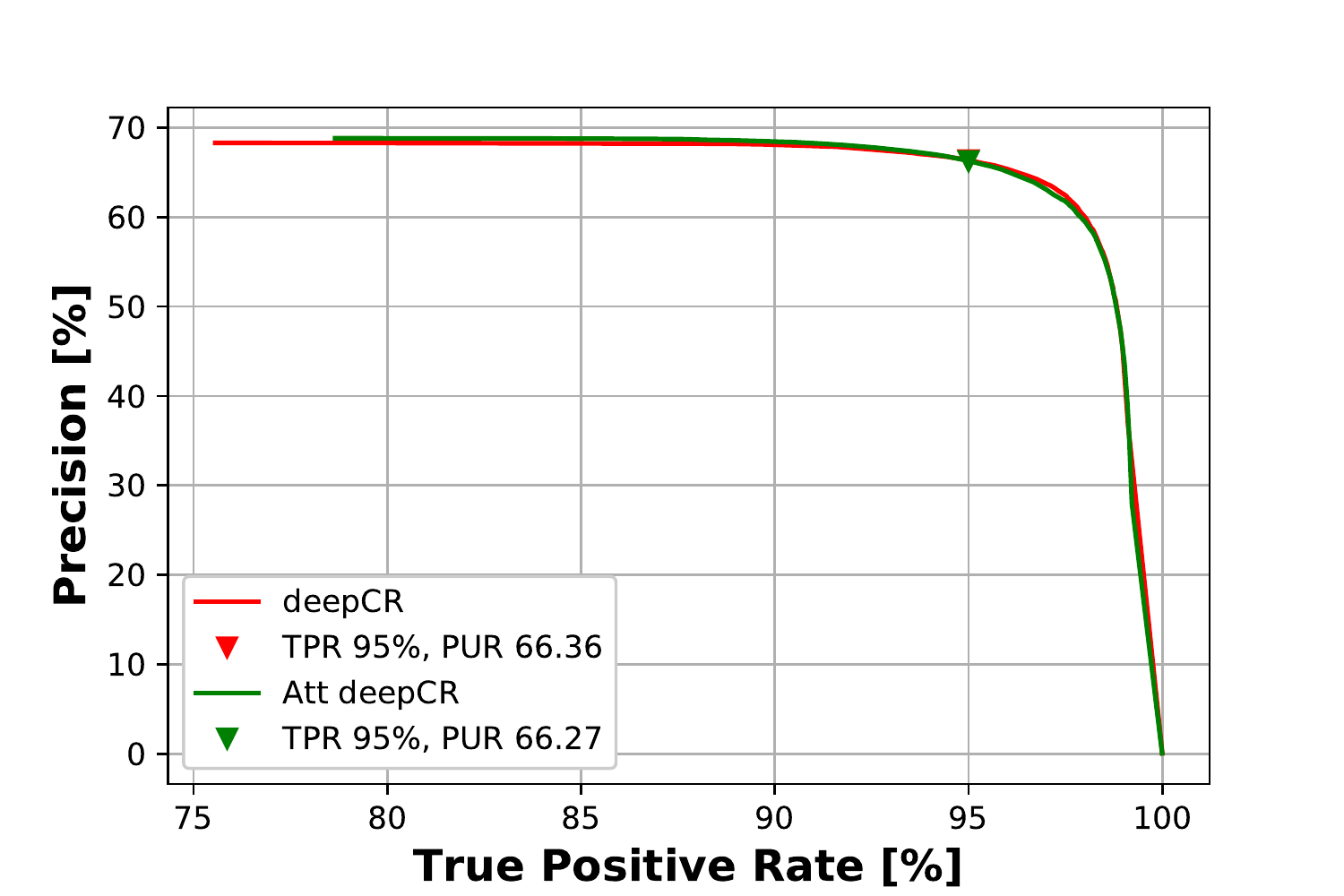}\label{fig:f62}}
  \hfill
  \subfloat[cosmic-CoNN - ROC]{\includegraphics[width=0.25\textwidth,
  height=0.25\textwidth, keepaspectratio,]{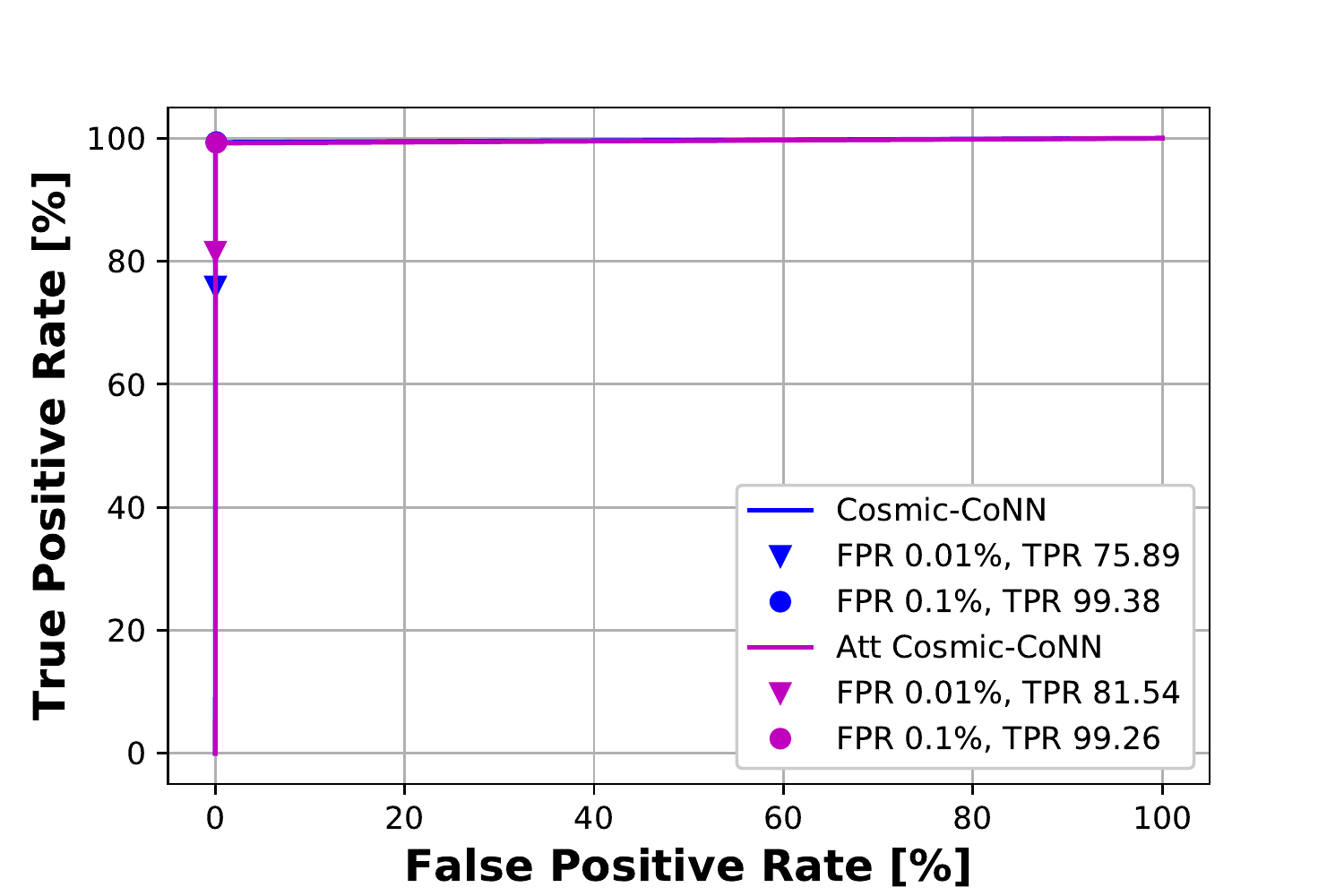}\label{fig:f63}}
  \hfill
  \subfloat[Cosmic-CoNN - PRC]{\includegraphics[width=0.25\textwidth,
  height=0.25\textwidth, keepaspectratio,]{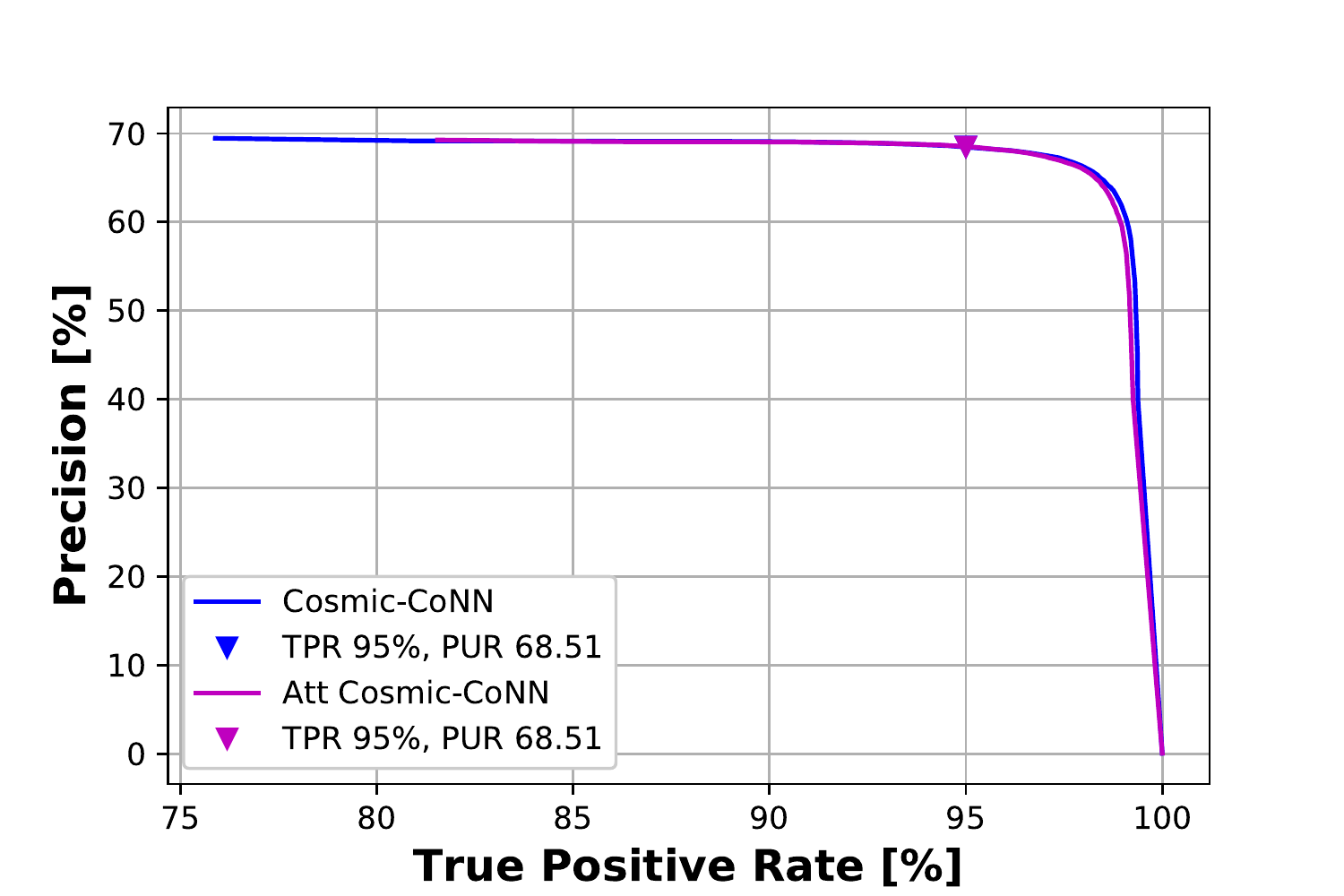}\label{fig:f64}}
  \caption{Performance of the proposed models on previously unseen data from the LCO CR test dataset using the 2-meter telescope data is presented here. (a) and (b) are the ROC and PRC plots on deepCR models with and without adding the attention gates. Similar plots for the Cosmic-CoNN models are presented in (c) and (d). The Cosmic-CoNN model performs well on this data compared to the deepCR model. Also, both the deepCR and Cosmic-CoNN models benefit from adding AGs more than the corresponding baselines when testing on this data.}
  \label{fig:lco2m_dl}
\end{figure*}

\begin{figure*}[!tbp]
  \centering
  \subfloat[deepCR - ROC]{\includegraphics[width=0.25\textwidth,
  height=0.25\textwidth, keepaspectratio,]{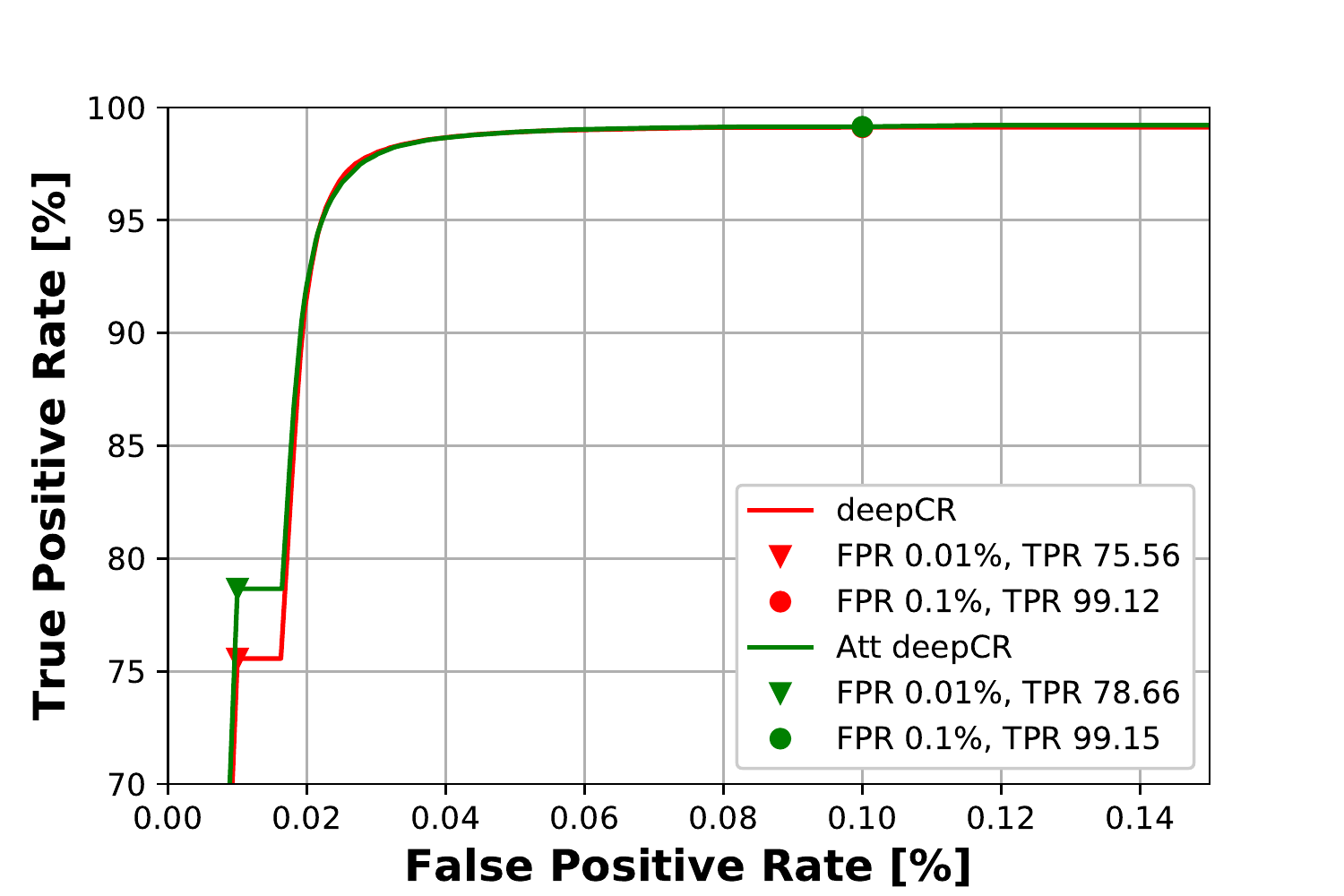}\label{fig:f61_ext}}
  \hfill
  \subfloat[deepCR - PRC]{\includegraphics[width=0.25\textwidth,
  height=0.25\textwidth, keepaspectratio,]{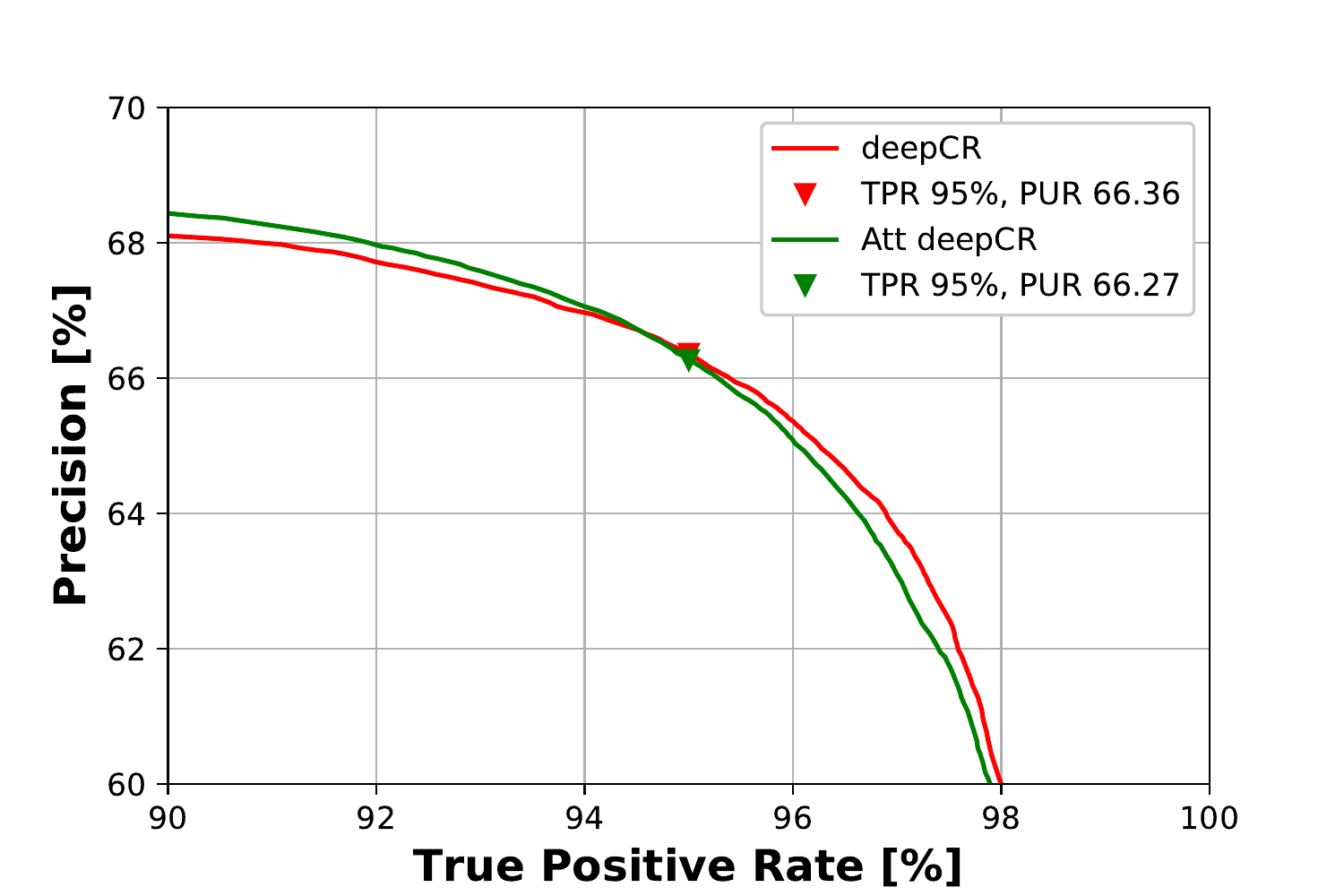}\label{fig:f62_ext}}
  \hfill
  \subfloat[cosmic-CoNN - ROC]{\includegraphics[width=0.25\textwidth,
  height=0.25\textwidth, keepaspectratio,]{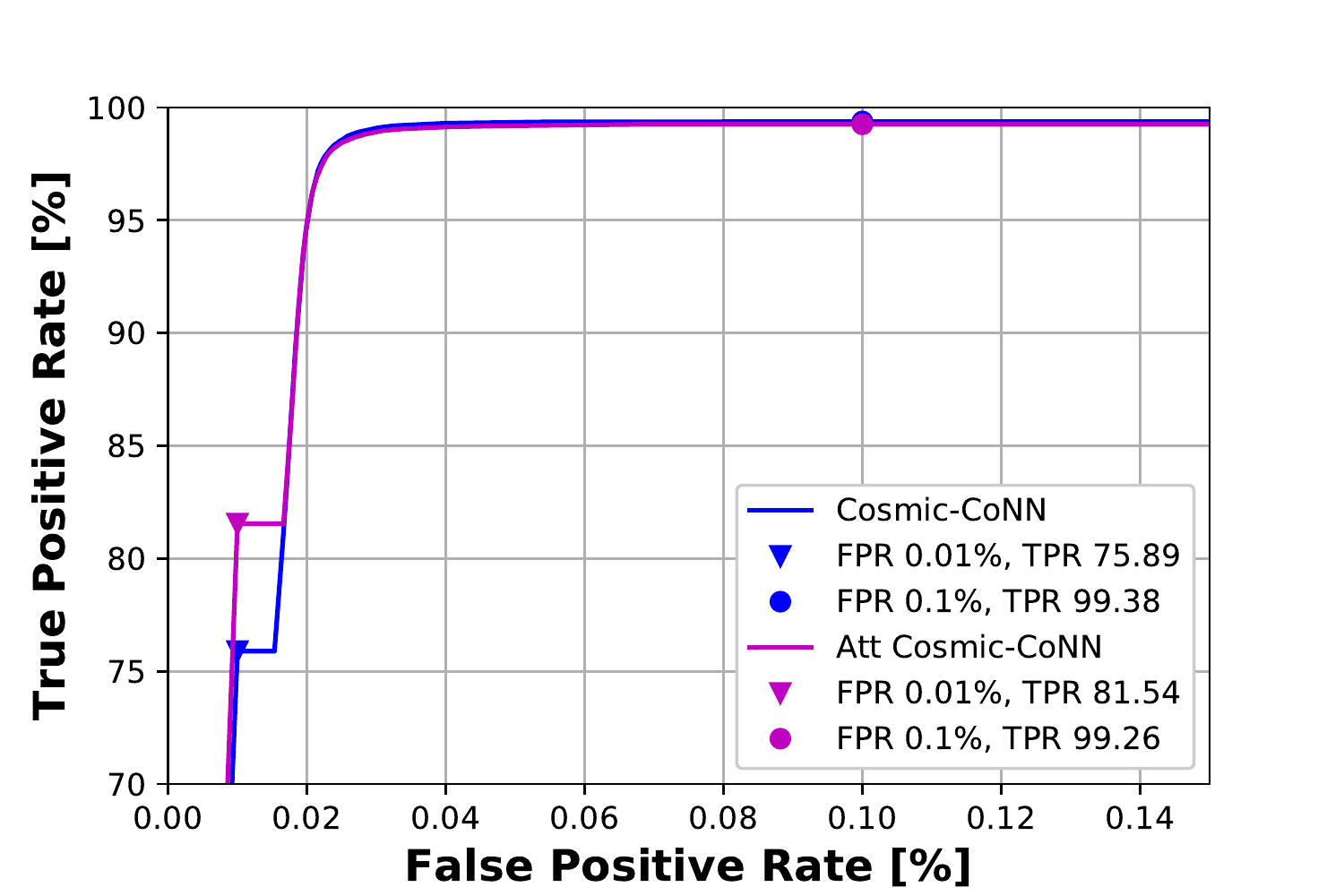}\label{fig:f63_ext}}
  \hfill
  \subfloat[Cosmic-CoNN - PRC]{\includegraphics[width=0.25\textwidth,
  height=0.25\textwidth, keepaspectratio,]{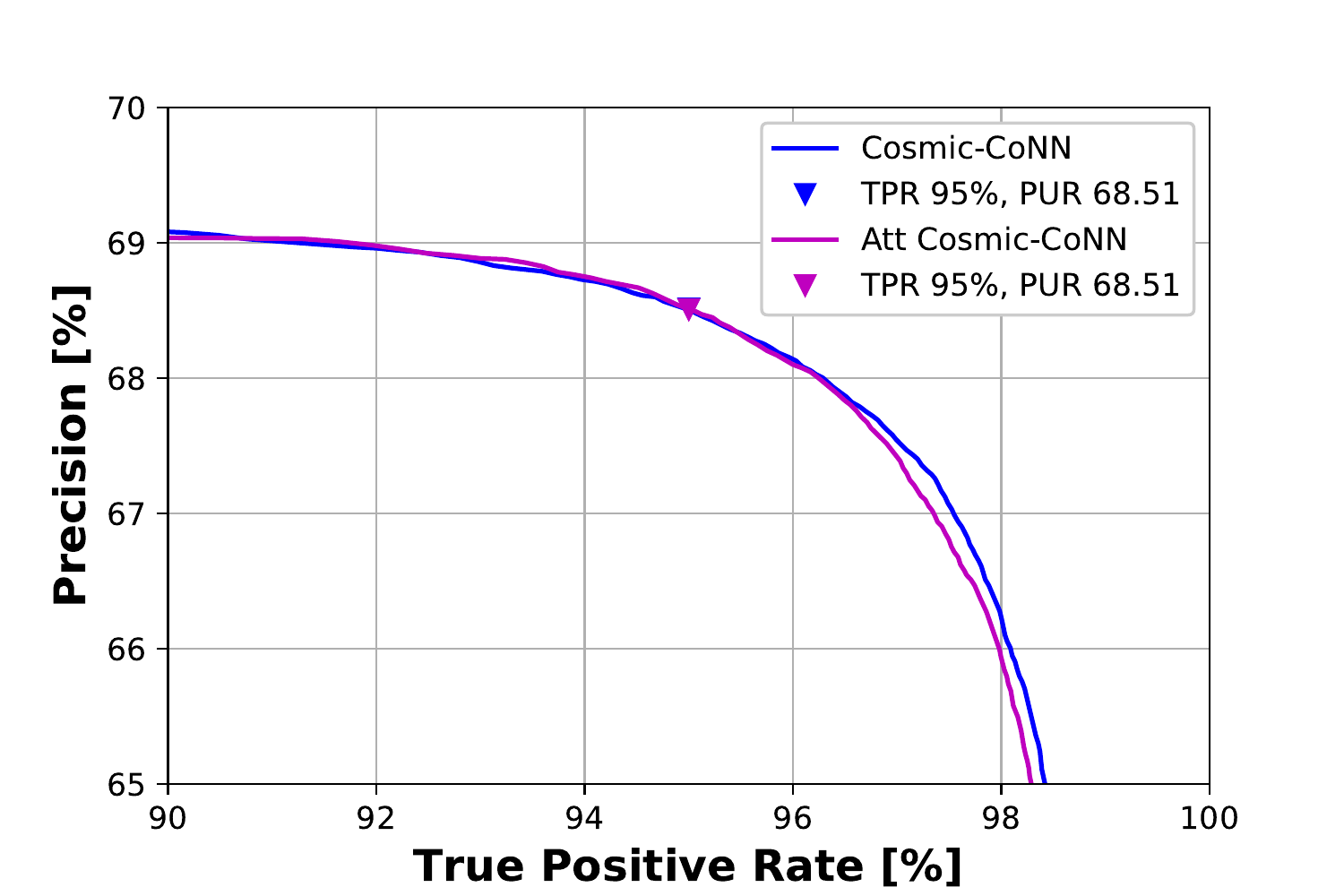}\label{fig:f64_ext}}
  \caption{The extended ROC and PRC plots for a better understanding of the proposed model performances from Fig.~\ref{fig:lco2m_dl}. The highest performance in CR detection can be noticed with the Cosmic-CoNN model both at 0.01\% and 0.1\% FPR. The gains with adding AGs can be noticed with both baseline models in all the cases.}
  \label{fig:lco2m_dl_ext}
\end{figure*}

\begin{figure*}
    \centering
    \includegraphics[width=18cm, height=30cm, keepaspectratio,]{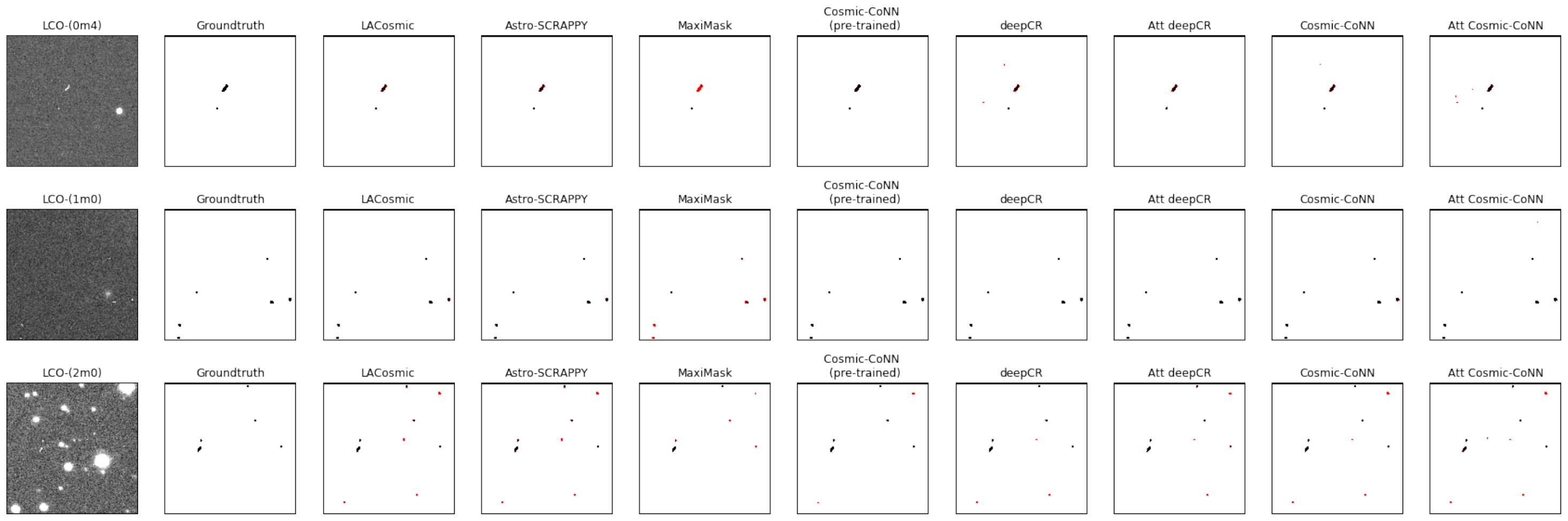}
    \caption{CR Detection discrepancy on LCOGT imaging data (one image from each LCO telescope class (0.4-meter, 1.0- and 2.0- meter)) with different CR detection algorithms. Incorrect or missing CR pixels are marked in red color. The pre-trained Cosmic-CoNN outperform all other CR detection models since it was initially trained using the LCO dataset. LACosmic and Astro-SCRAPPY give more False Positives and can be noticed clearly from a 2-meter telescope image. With the MaxiMask model, we can notice False Negatives and False Positives. The proposed models also detect most CR hits in images from all telescope classes but with a few False Positives. On 1-meter telescope data, the proposed models perform well compared to other data where we can notice more False Positives.}
    \label{fig:lco_cr}
\end{figure*}

\begin{figure*}
    \centering
    \includegraphics[width=18cm, height=20cm, keepaspectratio,]{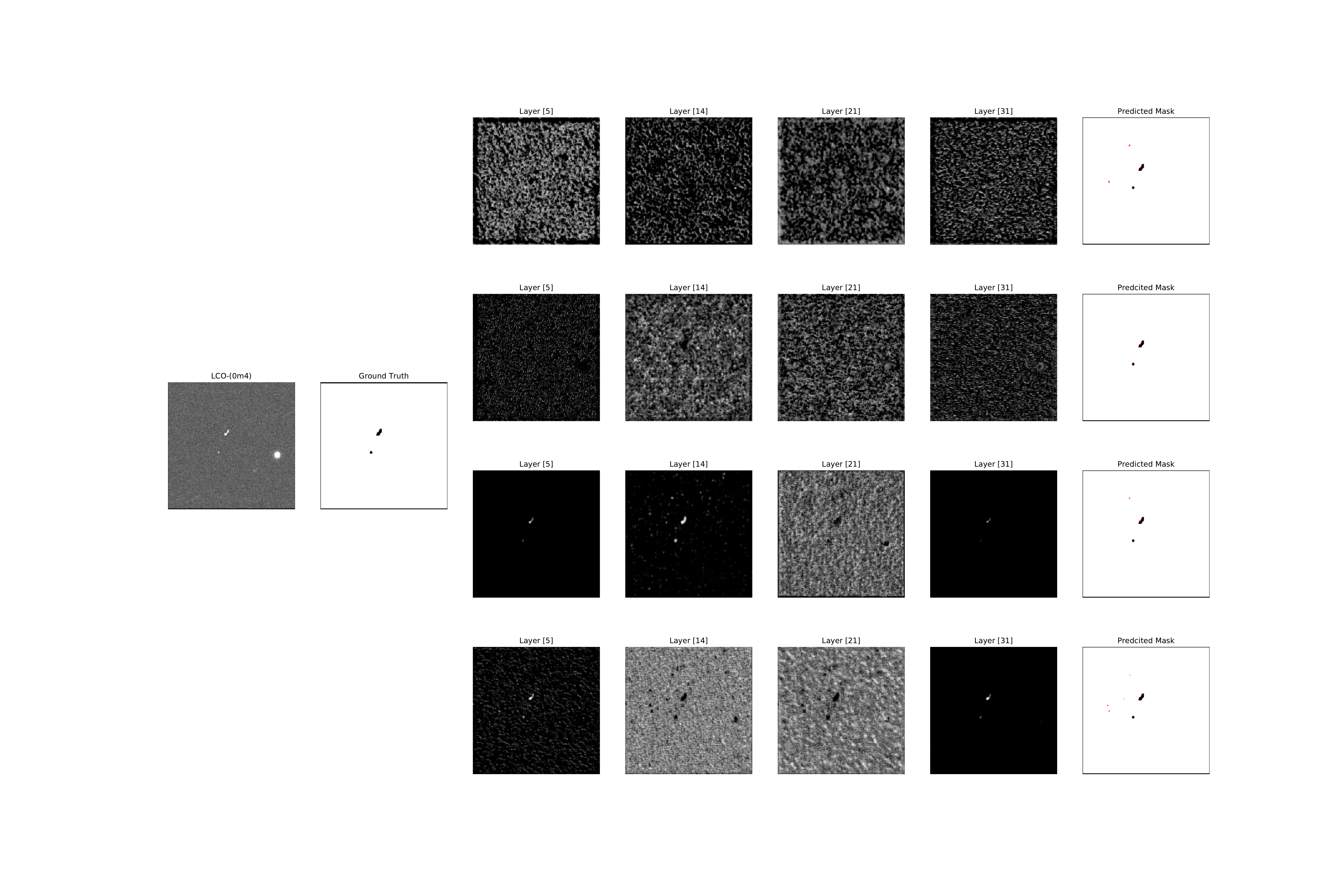}
    \caption{Feature maps were obtained on LCO image (0.4-meter Telescope) at different channels of deepCR and Cosmic-CoNN models with and without attention augmentation. The first row of each image is deepCR without AG augmentation, while the second row is deepCR with augmenting the AG module. Similarly, the third and fourth rows correspond to the Cosmic-CoNN models with and without attention augmentation. The AG models help in highlighting the CR induced pixels in an image better than the corresponding baseline models.}
    \label{fig:Feature_dc_lco}
\end{figure*}



\begin{table}
\begin{center}
\caption{Comparison of the number of trainable parameters.}
\begin{tabular}{ |c|c| } 
 \hline
 \textbf{Model} & \textbf{No. of Parameters} \\ 
 \hline
 deepCR & 467233  \\ 
 \hline
 Att deepCR & 472887 (1\% $\uparrow$) \\ 
 \hline
 Cosmic-CoNN & 465953  \\ 
 \hline
 Att Cosmic-CoNN & 471219 (1\% $\uparrow$) \\ 
 \hline
\end{tabular}
\label{tab:parameters}
\end{center}
\end{table}

\begin{table}
\begin{center}
\caption{Comparison of runtime complexity for a DECam image with 4K x 2K pixel size with different algorithms.}
\begin{tabular}{ |c|c| } 
 \hline
 \textbf{Model} & \textbf{CPU run time} \\ 
  & \textbf{(in sec.)}  \\ 
 \hline
 Maximask & 250.93 \\ 
 \hline
 deepCR & 24.98 \\ 
 \hline
 Att deepCR & 27.19 \\ 
 \hline
 Cosmic-CoNN & 26.04 \\ 
 \hline
 Att Cosmic-CoNN & 26.70 \\ 
 \hline
 Astro-SCRAPPY & 16.23  \\
 \hline
\end{tabular}
\end{center}
\label{tab:run time}
\end{table}

Since our models and the MaxiMask model are trained using the dilated CR masks as ground truth, for fair comparisons, we used the dilated CR masks as ground truth on the LCO test dataset as well. This is obtained by dilating the ground truth CR mask using a 3 $\times$ 3 kernel. Except for the MaxiMask and our proposed models, the output CR mask is dilated using a 3 $\times$ 3 dilation kernel for all other models. \rthis{Note that we are assessing the performance of the proposed models which are trained using images from a single imager (DECam), a single telescope (4-meter Blanco telescope) and even with a single exposure time (90 seconds). In order to understand the generalization capability of the proposed models with varying telescope diameters, we evaluated the proposed model's performance on each telescope class separately. The ROC and the PRC plots corresponding to the LCO test data are shown from Fig.~\ref{fig:lco04m_prev} to Fig.~\ref{fig:lco04m_dl_ext} for the 0.4-meter telescope data, from Fig.~\ref{fig:lco1m_prev} to Fig.~\ref{fig:lco1m_dl_ext} for the 1-meter telescope data, and from Fig.~\ref{fig:lco2m_prev} to Fig.~\ref{fig:lco2m_dl_ext} on data from the 2-meter telescope class respectively, with both classical and deep-learning-based models. The extended ROC and PRC plots are also presented to better compare the performance of different CR detection algorithms. Similarly, other performance metrics on the LCO data are reported in Table \ref{tab:perf1} and Table \ref{tab:perf2} again with individual telescope classes.} The conventional algorithms, LACosmic and Astro-SCRAPPY, perform decent CR detection on LCO test data across all telescope classes. However, these algorithms are limited by run time and hard-coded decision rules to obtain the optimal CR mask per image. 

\rthis{Since both MaxiMask and our trained models did not see the LCO data previously while training, these models did not generalize well when evaluated on the LCO test data. However, these models' performance is comparable to the classical LACosmic and Astro-SCRAPPY in most cases. The MaxiMask performance is inferior to our proposed models on the 0.4-meter LCO data, while it is superior with data from the 1-meter and 2-meter telescopes, across the majority of the metrics listed in Table~\ref{tab:perf1}. Unlike our models, the MaxiMask model was trained using images from multiple imagers, telescopes and even exposure times. From Table \ref{tab:perf1}, we demonstrate that the proposed baseline models and their attention-added variants perform closer to the classical algorithms on previously unseen LCO test data across all telescope classes. However, a high TPR is achieved at high FPR over all the telescope data. Between the two baseline models considered, the Cosmic-CoNN model shows better performance at 0.01\% and 0.1\% FPR than the deepCR model for all data classes. In all the cases, we infer that the attention models provide significant CR detection performance compared to their corresponding baselines at 0.01\% FPR for all classes of telescope data. Further, if we look at the TPR at 0.01\% FPR, the baseline deepCR and Cosmic-CoNN models deliver lower performance than the MaxiMask model on 1-meter and 2-meter telescope data. Nevertheless, the proposed attention augmentation is helping the baseline models improve their CR detection rates. Thus, the proposed AG augmentation idea helps improve the baseline models' generalization capability. Comparatively, our proposed models work better on the 1-meter and 2-meter telescope data than data from the 0.4-meter telescope. Further, Table~\ref{tab:perf2} lists the performance metrics on the LCO test data from the second row for each telescope class. While computing the listed metrics for LACosmic and Astro-SCRAPPY, we used \texttt{sigclip=15.0} and \texttt{sigfrac=0.1}. As discussed before, the \texttt{objlim} parameter is changed based on the telescope class. We demonstrate that the AG models offer consistent and marginal gains compared to the corresponding baseline models with the 1-meter data. However, for the data from the 0.4-meter and 2-meter telescopes, the performance with both the attention-based and baseline models is equivalent. The lowest FDR is noticed with the MaxiMask model for the LCO test data for all telescope classes.}

The CR detection performance of each model is compared qualitatively with individual images from each telescope class from the LCO test data, and the corresponding results are illustrated in Fig.~\ref{fig:lco_cr}. The conventional LACosmic, Astro-SCRAPPY, and Cosmic-CoNN (pre-trained) detect more CR peripherals than the MaxiMask and our proposed models. With the MaxiMask model, we can notice either missing CR hits (False Negatives) or detecting source or background pixels as CR hits (False Positives). Whereas in the proposed models, we are mainly seeing False Positives than the False Negatives. Hence, the proposed models perform well on previously unseen data when tested using the LCO CR data. However, high CR detection is achieved at high false-positive rates. This indicates that our models can easily capture the unique spatial signatures of the CR hits through the convolutional and attention layers. Similarly the attention maps are illustrated in Fig.~\ref{fig:Feature_dc_lco}, from the 0.4-meter telescope image. From here, we can infer that the attention gate augmented models pay more attention to the CR hits than the baseline models and thus help improve the CR detection performance and the generalization performance.


Finally, we compare the run time complexity of the proposed models with MaxiMask and Astro-SCRAPPY algorithms using an Intel Core i3 processor running at 2.30GHz. The baseline models with and without attention augmentation are running up to 10 times faster than the MaxiMask model when checked with an image of $4K\times2K$ pixel size. This is because the proposed models are more lightweight than the MaxiMask model. Since the MaxiMask was developed to detect several other contaminants in the image along with the CR hits, this model has many parameters and thus takes a long time to flag the contaminants. We noticed that the run time for CR mask computation using Astro-SCRAPPY depends on the parameters used. Furthermore, Astro-SCRAPPY is an iterative algorithm that depends on hard-coded decision rules to obtain the optimal CR mask per image. The reported time for Astro-SCRAPPY is with \texttt{sigclip=3.0}, \texttt{objlim=1.0}, and \texttt{sigfrac=0.1} while making all other parameters set to default.

\section{Conclusions}

In conclusion, we have demonstrated the efficacy of the baseline deep learning CR detection models on the ground-based imaging data from two different imagers: DECam (mounted on the 4-meter Blanco telescope) and LCOGT \rthis{(with data from three different telescope classes).} We have constructed a dataset for the DECam consisting of image pairs with CR hits and the corresponding ground truth CR mask. Specifically, these images and the CR masks have been taken from the DECam observations. This dataset is used to train the baseline, and the attention augmented models. Furthermore, we have shown that adding an attention gate module at each decoder layer of the baseline models consistently improves qualitative and quantitative performance on the DECam data for most cases with. Specifically, the imoprovement with attention models is noticed with TPR at 0.01\% FPR and Precision at 95\% TPR from ~\ref{tab:perf1}. However, the gains with attention models are marginal. \rthis{Qualitatively, the gain using the proposed attention models is obtained  with far fewer False Positives (cf. Fig.~\ref{fig:decam_cr}).
We noticed that deepCR and cosmic-CoNN benefit more from adding attention gates irrespective of the source normalization technique used while training the models.} The performance gained with attention augmented models is with a small added computational cost of increasing the number of trainable parameters by 1\%. Also, the proposed attention-based models outperform state-of-the-art methods such as LACosmic, Astro-SCRAPPY, MaxiMask and Cosmic-CoNN (pre-trained) on the DECam test data. The paired dataset constructed in this work using DECam images could also be used to train other CR detection algorithms on the DECam data and made available online \href{https://drive.google.com/file/d/1zDHN6zZ7wRCU08-GHKvdKtJqdd46TF2P/view?usp=sharing}{here}. We have also made our training codes and models available at our \href{https://github.com/lfovia/Cosmic-Ray-Detection-in-Astronomical-Images}{GitHub repository}.

We evaluated four of our models (including baseline deepCR, Cosmic-CoNN and their attention added variants) on previously unseen LCOGT test data to check the generalization capability of the proposed CR detection models. We demonstrated the efficacy of the proposed deep-learning models, which are trained using images from a single camera/instrument (DECam), a single telescope (4-meter Blanco telescope) and even with a single exposure time (90 seconds for all the images) on LCO test data. \rthis{The LCO CR dataset has been obtained from three distinct telescope imagers with different diameters, namely 0.4-meter, 1-meter, and 2-meter telescopes. Even though these models do not perform as well as with the DECam data, the majority of the CR streaks are detected on LCO test data. However, this gain is at the cost of a few false-positive rates. The Maximask model also performs close to the classical LACosmic and Astro-SCRAPPY algorithms on the LCO data, given that this model also did not use the LCO data while training. Nevertheless, the MaxiMask model gives more False Positives and False Negatives. MaxiMask achieves the lowest FDR across the DECam and LCO data. The highest performance with the proposed models can be seen with the LCO 1-meter and 2-meter telescope data, whereas the 0.4-meter telescope data has the least performance. With the proposed attention augmentation, the baseline models can generalize well on the LCO data with significant gains in TPR at 0.01\% FPR across all the telescope data. Further, attention helps the baseline models perform similarly to the classical algorithms. Between the two baseline models studied, we noticed that the Cosmic-CoNN model marginally performs better than the deepCR model on DECam data. However, this performance is significant on LCO data with high detection rates at 0.01\% FPR over all classes of telescopes.} The CR hits in the LCOGT images have different signatures than those observed in the DECam images and can be visually noticed. The difference in CR morphology between DECam and LCOGT stems from the differences in CCD thickness between DECam and LCOGT. The DECam has long CR streaks as the significant contamination, and CRs with an area of one or two pixels are less. On the other hand, dots with one or two pixels are the primary CR contamination in the LCOGT images, whereas long worms are the minor. Even the morphology of the data will change between DECam and LCO data. All these factors made our models not perform very well on the unseen data. In the future, we plan to improve the generalization capabilities of the models to make them independent of the underlying image generation process.

\label{sec:conclusion}

\section{Acknowledgments}
SR was supported by Tata Consultancy Services (TCS) and Department of Science and Technology - Interdisciplinary Cyber Physical Systems (DST-ICPS) (under the grant T-641).
We are grateful to the anonymous referees for several constructive comments and feedback on our manuscript.
This project used data obtained with the Dark Energy Camera (DECam), which was constructed by the Dark Energy Survey (DES) collaborating institutions: Argonne National Lab, University of California Santa Cruz, University of Cambridge, Centro de Investigaciones Energeticas, Medioambientales y Tecnologicas-Madrid, University of Chicago, University College London, DES-Brazil consortium, University of Edinburgh, ETH-Zurich, University of Illinois at Urbana-Champaign, Institut de Ciencies de l'Espai, Institut de Fisica d'Altes Energies, Lawrence Berkeley National Lab, Ludwig-Maximilians Universitat, University of Michigan, National Optical Astronomy Observatory, University of Nottingham, Ohio State University, University of Pennsylvania, University of Portsmouth, SLAC National Lab, Stanford University, University of Sussex, and Texas A\&M University. Funding for DES, including DECam, has been provided by the U.S. Department of Energy, National Science Foundation, Ministry of Education and Science (Spain), Science and Technology Facilities Council (UK), Higher Education Funding Council (England), National Center for Supercomputing Applications, Kavli Institute for Cosmological Physics, Financiadora de Estudos e Projetos, Fundacao Carlos Chagas Filho de Amparo a Pesquisa, Conselho Nacional de Desenvolvimento Científico e Tecnologico and the Ministerio da Ciencia e Tecnologia (Brazil), the German Research Foundation-sponsored cluster of excellence ``Origin and Structure of the Universe" and the DES collaborating institutions. This work also makes use of observations from the Las Cumbres Observatory Global Telescope Network~\citep{brown2013cumbres} having 25 telescopes at seven sites around the world.

\section{Software} 
Astropy \citep{robitaille2013astropy}, LACosmic \citep{van2001cosmic}, Astro-SCRAPPY \citep{mccully2019astro}, Numpy \citep{harris2020array}, Scipy \citep{virtanen2020scipy}, Matplotlib \citep{hunter2007matplotlib}, Jupyter \citep{kluyver2016jupyter}, scikit-image \citep{van2014scikit}, SExtractor \citep{bertin1996sextractor}, and Pytorch \citep{paszke2019pytorch}.

\bibliographystyle{model2-names}
\bibliography{mybibfile}

\section*{Appendix}
We have implemented our deep-learning models in {\tt PyTorch 1.9.0} with Adam optimizer \citep{kingma2014adam}. While training our models using the DECam dataset, we randomly sampled and withheld 1$\%$ of the training set for validation. We trained the deepCR models with identical networks and adopted the two-phase training procedure used in ~\citep{zhang2020deepcr}. The network is first trained in ``training mode," in which batch normalization layers record running statistics of layer activations with a momentum of 0.005 and use training batch statistics for normalizing for 40 \% of epochs. Following the initial training phase, the network is switched to ``evaluation mode" for the next 60 \% of the epochs, in which the running statistics are frozen and used in both forward and backward batch normalization passes. Similarly, while training Cosmic-CoNN ~\citep{xu2021cosmic} models, which use group normalization, we adopt a fixed number of groups (equal to eight) for all feature layers.

An initial learning rate of 0.005 was used for both the baseline and their attention augmented variant models. During training, we monitor the validation loss for each model and automate to decay the learning rate by 0.1 when the validation loss does not improve by 0.1 \% for four consecutive epochs. Both deepCR and Cosmic-CoNN models and their attention variants are converged within 60 epochs of training. In order to make a fair comparison, both Cosmic-CoNN and deepCR models were carefully tuned, and the best models were used for evaluation. Each training epoch took approximately 235 seconds (or 4 minutes) for all the models on Nvidia Titan Xp with 12GB GPU.

\end{document}